\documentclass{jfm}

\usepackage{hyperref}
\usepackage{graphicx}
\usepackage{fancyhdr}
\usepackage{epsfig}
\usepackage{rotating}
\usepackage{amssymb}
\usepackage{mathrsfs}
\usepackage[round]{natbib}
\usepackage{epsf}
\usepackage{amsmath}
\usepackage{color}
\usepackage{lscape}
\usepackage{setspace}
\usepackage{enumerate}
\usepackage{subcaption}
\usepackage[latin1]{inputenc}
\usepackage{textcomp}
\usepackage[english]{babel}
\usepackage[super]{nth}

\hypersetup{
	colorlinks=false,
	pdfborder={0 0 0},
	pdfborderstyle={/S/U/W 0},
}

\newcommand{\f}{\frac}

\newcommand{\be}{\begin{equation}}
\newcommand{\ee}{\end{equation}}
\newcommand{\ba}{\begin{eqnarray}}
\newcommand{\ea}{\end{eqnarray}}
\newcommand{\md}{\mathrm{d}}
\newcommand{\mi}{\mathrm{i}}
\newcommand{\me}{\mathrm{e}}
\newcommand{\mF}{\mathcal{F}}

\DeclareMathOperator{\Lapl}{\mathcal{L}}

\newcommand{\ssA}{\mathsfbi{A}} 
\newcommand{\ssB}{\mathsfbi{B}} 
\newcommand{\ssE}{\mathsfbi{E}} 

\newcommand{\bF}{\boldsymbol{F}}
\newcommand{\bH}{\boldsymbol{H}}

\def\Xint#1{\mathchoice
   {\XXint\displaystyle\textstyle{#1}}%
   {\XXint\textstyle\scriptstyle{#1}}%
   {\XXint\scriptstyle\scriptscriptstyle{#1}}%
   {\XXint\scriptscriptstyle\scriptscriptstyle{#1}}%
   \!\int}
\def\XXint#1#2#3{{\setbox0=\hbox{$#1{#2#3}{\int}$}
     \vcenter{\hbox{$#2#3$}}\kern-.5\wd0}}
\def\dashint{\Xint-}

\shorttitle{Unsteady aerodynamics of membrane wings}
\shortauthor{S. Tiomkin and J. W. Jaworski}

\title{Unsteady aerodynamic theory for membrane wings}

\author{Sonya Tiomkin\aff{1}\corresp{\email{sot220@lehigh.edu}} \and Justin  W. Jaworski\aff{1}}

\affiliation{\aff{1}Department of Mechanical Engineering and Mechanics, Lehigh University, Bethlehem,
PA 18015, USA}

\begin{document}

\maketitle

\begin{abstract}
We study analytically the dynamic response of membrane aerofoils subject to arbitrary, small-amplitude chord motions and transverse gusts in a two-dimensional inviscid incompressible flow. 
The theoretical model assumes linear deformations of an extensible membrane under constant tension, which are coupled aeroelastically to external aerodynamic loads using unsteady thin aerofoil theory. 
The structural and aerodynamic membrane responses are investigated for harmonic heave oscillations, an instantaneous change in angle of attack, sinusoidal transverse gusts, and a sharp-edged gust. The unsteady lift responses for these scenarios produce aeroelastic extensions to the Theodorsen, Wagner, Sears, and K\"{u}ssner functions, respectively, for a membrane aerofoil.
These extensions incorporate for the first time membrane fluid-structure interaction into the expressions for the unsteady lift response of a flexible aerofoil.
The indicial responses to step changes in the angle of attack or gust profile are characterised by a slower lift response in short times relative to the classical rigid-plate response, while achieving a significantly higher asymptotic lift at long times due to aeroelastic camber. 
The unsteady lift for harmonic gusts or heaving motions follows closely the rigid plate lift responses at low reduced frequencies but with a reduced lift amplitude and greater phase lag.
However, as the reduced frequency approaches the resonance of the fluid-loaded membrane, the lift response amplitude increases abruptly and is followed by a sharp decrease. This behaviour of the unsteady lift response function is visualised as circular paths in the complex plane. 
Each circle in the complex plane representation of the lift response functions corresponds to a different dominant mode of the membrane dynamic response, and the inflection points between these circles identify a shift in dominance between two consecutive membrane modes.
This behaviour reveals a frequency region, controlled by the membrane tension coefficient, for which the classical Theodorsen and Sears functions underestimate the load on the aerofoil, followed by a reduced frequency regime where a sizeable lift reduction is obtained through passive membrane oscillations.
These results suggest that membrane aerofoils with appropriately tuned pretension could possess substantial aerodynamic benefits over rigid aerofoils in unsteady flow conditions.

\end{abstract}

%

\section{Introduction}

The growing industrial interest in small-scale unmanned aerial vehicles (SUAVs) for sensing, reconnaissance, and parcel delivery continues to spur scientific interest into novel aerodynamic design solutions for low-speed flows, inspired by biological fliers \citep{Hassanalian2017}. 
A special focus has been given to the membrane wings of bats, who possess impressive maneuvering and gliding abilities without relying upon high-frequency wing flapping for lift \citep{Hedenstrom2015}. These membrane wings are lightweight and are therefore appealing for SUAV applications. However, the compliance of membrane wings couples their geometrical shape and dynamics to the surrounding fluid mechanics and complicates the prediction of their aerodynamic performance. 
Several computational and experimental studies have examined the aerodynamics of these flexible membrane wings in steady flow conditions over the last two decades \citep[e.g.,][]{Song2008,Gordnier2009,Rojratsirikul2009,Arbos-Torrent2013,Serrano-Galiano2018}. \cite{Gordnier2009} and \cite{Rojratsirikul2009} showed that membrane wings in low Reynolds number flows delay stall and enhance the mean lift coefficient due to the onset of flow-membrane oscillations. The oscillations of the compliant membrane are essential to these aerodynamic benefits, as \cite{Gordnier2009} found no significant advantage for a static membrane wing when compared to an equivalent (cambered) rigid aerofoil. Thus, the unsteady behaviour of membrane wings is a principal source of interest to discover mechanisms for aerodynamic performance improvement.
While many studies investigated the membrane wing response to steady flow conditions \citep[see][for a recent literature survey]{Tiomkin2021}, few have studied its response to unsteady flow conditions or prescribed motions, where aeroelastic membrane deformation may yield further aerodynamic benefits. 

The pursuit of novel mechanisms to improve the unsteady aerodynamic performance of membrane wings and understand their associated fluid mechanics has led to a recent research focus on flapping membrane wings.
The combination of aerofoil flexibility with flapping motions can potentially eliminate flow separation along the aerofoil altogether and improve the aerodynamic maneuverability of the vehicle, as demonstrated in the context of bat flight \citep{Muijres2008,Chin2016}.
Several computational \citep{Gopalakrishnan2010,Jaworski2015} and experimental \citep{Tregidgo2013} studies investigated the membrane wing response to prescribed flapping motions, which are generally described as pitch or heave oscillations, or as a combination of the two.
\cite{Gopalakrishnan2010} used coupled large eddy simulations of a rectangular membrane to show that induced camber enhances both the lift and the thrust during a flapping pitching motions. 
These simulations identified the movement of the leading-edge vortex along the membrane aerofoil surface to be the main source of the increased lift and thrust relative to a flapping rigid wing, for which the leading-edge vortex detaches and moves away of the wing which causes a drop in the lift coefficient \citep{Eldredge2019ARFM}. These results are supported by the computations of \cite{Jaworski2015} that focused on the role of prestress and elastic modulus in the propulsion of a flapping membrane aerofoil, which is enhanced by the interaction of the leading-edge vortex with the local elastic deformation. Experiments of \cite{Tregidgo2013} focused on the membrane dynamic response to a transient sinusoidal pitch maneuver of reduced frequency of $k=0.022$ and amplitude of $10^\circ$. Different vibrational modes were identified that depended on the stationary angle of attack about which the unsteady maneuver was carried out. For small stationary angles of attack $(0^\circ\le\alpha\le4^\circ)$, first mode oscillations were observed with a small lag in the membrane dynamic response relative to the prescribed motion. This delay was more pronounced for a larger stationary angle of attack of $\alpha=10^\circ$, for which hysteresis was identified between the pitch-up and the pitch-down sections of the motion, which were accompanied by different vibrational modes.

The above studies collectively emphasize the complexity of the flapping membrane wing problem. However, due to their computational and experimental nature, their scope is limited to a few specific points in the parameter regime of flapping membrane wings, i.e., specific values of reduced frequency, mass ratio, and membrane elasticity. An analytical solution of a simplified model problem is therefore desired to shed light on the role of each dimensionless group in the wide parameter space of flapping membrane wings. 
Physical insights from such an analytical solution are expected to inform future computational and experimental studies en route to obtaining a more complete understanding of the physics of flapping membrane wings.

Several analytical studies have been carried out that focus on propulsive thrust and efficiency predictions for flapping flexible wings.
\cite{Alben2008} presented an analytical solution for a flapping inextensible elastic sheet (with a free trailing edge), utilizing unsteady thin aerofoil theory coupled to a beam structural model. Their work identified an optimal thrust condition at the resonance peaks for small pitching amplitudes. 
More recently, \citet{Alon2019} showed via analytical solution that a heaving membrane wing transitions between thrust and drag near the membrane resonance frequency, as the reverse von K\'{a}rm\'{a}n wake transitions to a traditional von K\'{a}rm\'{a}n wake.
In their reviews on flapping wing aerodynamics of biological and bio-inspired flyers, \cite{Shyy2013book,Shyy2016} highlighted the importance of using a time-domain approach to predict accurately the aerodynamic performance of flapping wings at the scale of bats and birds due to the inseparable flapping and body time-scales of these flyers, which is not the case for smaller insect-scale flyers.
Thus, while a quasi-steady model can make accurate predictions for insect-scale vehicles, this model assumption is not recommended for SUAV applications, where a time-dependent approach is essential to address vehicle stability and control.
Furthermore, whilst the studies of \cite{Alben2008} and \cite{Alon2019} elucidate the propulsive potential of flapping flexible wings, a theoretical basis to understand the membrane wing aerodynamic performance in prescribed flapping motions remains underdeveloped, specifically in terms of the ability to predict its unsteady lift and structural dynamic response.

In addition to the unsteady lift and thrust enhancement mechanisms engendered by membrane wings under prescribed flapping motions, 
an understanding of the response of these wings to flow disturbances such as gusts is important to the design of membrane wing SUAVs.
Due to their small size and slow flight speed, SUAVs are especially susceptible to flight disruption from small gusts typical of urban environments \citep[][]{Watkins2006atmospheric,elbanhawi2017enabling,Jones2022}. 
Classical linear unsteady aerodynamic theory \citep[see][]{vonKarman1938,Sears1940} predicts the transient lift response of a rigid aerofoil to transverse gusts of small gust ratios, where the gust ratio is the transverse gust amplitude divided by the freestream flow speed.
This theory has long been utilized to predict the unsteady load on rigid wings in terms of lift amplitude and phase lag.
However, when compliant membrane wings are considered, the lift response is composed of both the local change in angle of attack and the resulting deformation of the aerofoil. The membrane deformation couples aeroelastically to the aerodynamic load, which may amplify or attenuate the unsteady lift response.
Initial results by \cite{Berci2013} from a semi-analytical state-space model indicate the appearance of structural oscillations in the massless membrane response to a sharp-edged transverse gust. However, these oscillations were described only in terms of the mid-chord membrane deformation and without consideration of the structural mode of oscillation and the lift response of the aerofoil.
A complete analysis of the membrane response to unsteady flow is currently lacking in the literature. 

The current study aims to fill this knowledge gap by presenting an unsteady analytical model and its solution for a membrane wing in inviscid incompressible flow, under the unsteady conditions of prescribed motions or transverse gust profiles. 
The transient membrane response is determined in the Laplace domain; steady-state harmonic oscillations of the membrane deformation and the unsteady lift response are investigated using a simplified solution in the frequency (Fourier) domain, which is convenient to compare against established rigid aerofoil theory. 
These solutions yield novel extensions to the classical unsteady aerodynamic functions for flexible membrane wings. 

The remainder of this paper is organised as follows. 
Section~\ref{sec:Formulation} presents the mathematical problem for the generalised case of a membrane wing in arbitrary motion or gust, and for specific canonical unsteady flow scenarios.
In \S~\ref{sec:results}, the results of the theoretical model are presented in terms of membrane wing deformation and aerodynamic performance, as represented by extensions to the classical unsteady aerodynamic functions by Theodorsen, Wagner, Sears, and K\"{u}ssner.
Section~\ref{sec:conclusions} closes with concluding remarks.

\section{Formulation}\label{sec:Formulation}

\subsection{Membrane wing}\label{sec:Formulation-membrane}

Consider an \emph{extensible} membrane aerofoil of thickness $h$ and density $\rho_m$, which is held by simple supports at a distance $2b$ from one edge to the other. The membrane is initially still and taut, and is immersed in a uniform and inviscid incompressible freestream of density $\rho$ and speed $U$, aligned parallel to the membrane chord (see figure~\ref{fig:MemGustSkatch_t0}).
Assuming small deformations of the membrane, the membrane dynamic equation is
\be
\label{eq:DynamicMem-linear}
\rho_m h \,\tilde{y}_{\tilde{t}\tilde{t}} = T \,\tilde{y}_{\tilde{x}\tilde{x}} + \Delta p ,
\ee
where $\tilde{y}$ denotes the membrane profile, $\tilde{t}$ represents time, $\tilde{x}$ is a coordinate along the chord, and $T$ and $\Delta p$ are the tension and pressure difference along the membrane, respectively.
While the membrane is extensible, we note that \cite{Tiomkin2017} showed that the tension can be considered constant to leading order for the small angles of attack and deformations assumed in the current study. 

The non-dimensional form of the dynamic equation is
\be
\label{eq:DynamicSail-LinearNorm}
4\mu \,y_{tt}=2C_T\,y_{xx}+\Delta C_p ,
\ee
in which $b, b/U, \rho, U, \f{1}{2}\rho U^2,$ and $\rho U^2 b$ are used as the units of length, time, density, circulation (per unit length), pressure, and force (per unit span).
Note that $b$ is used as the unit of length throughout the dynamic equation, but the mass ratio is normalised with $c$ as the unit of length, namely $\mu=\rho_m h/\rho c$, following the convention in previous membrane wing studies \citep[][]{Jaworski2012,Alon2019}.
The mass ratio $\mu$ and tension coefficient $C_T$ are fixed parameters in the present analysis, and the unsteady membrane deformation and pressure coefficient profiles, $y$ and $\Delta C_p$, respectively, are part of the solution.
A schematic drawing of the membrane geometry in the non-dimensional form is presented in figure~\ref{fig:MemGustSkatch_t} for the gust response case; this coordinate system is used to describe the membrane deformation in all of the considered cases.

\subsection{Incompressible potential flow}

The extensible membrane aerofoil may encounter or produce an unsteady flow field that superposes on the uniform background flow. Inviscid, incompressible potential flow is considered with an initial angle of attack of $\alpha=0^\circ$, which isolates the effects of unsteady angle of attack variations or transient gusts on the membrane dynamic response; this approach is similar to the traditional formulation available for the arbitrary motion of rigid aerofoils \citep[e.g.,][]{Bisplinghoff_book1996}.
For completeness, this section outlines the formulation of \cite{Tiomkin2017} for a membrane aerofoil in steady flow and extends it to include the dynamic membrane response to an unsteady flow.

The standard coordinate transformation
\be
\label{eq:Tet-transformation}
x=-\cos\theta
\ee
places the profile leading edge at $x=-1$ ($\theta=0$) and the trailing edge at $x=1$ ($\theta=\upi$).
This coordinate transformation permits the membrane slope, $y_x$, to be expressed as a Fourier cosine series expansion per \cite{Nielsen1963}, which is augmented here by allowing the Fourier coefficients to be time-dependent:
\be
\label{eq:SailSlope-Fourier}
y_x(t,\theta)=\f{1}{2}F_0(t)+\sum_{n=1}^\infty F_n(t)\cos n\theta .
\ee
We proceed with expressing the membrane dynamic equation \eqref{eq:DynamicSail-LinearNorm} in terms of the new coordinate $\theta$, which yields a system of differential equations for the Fourier coefficients.

Integration of \eqref{eq:SailSlope-Fourier} along the horizontal coordinate, from the leading edge to a point $x$ along the chord, yields the membrane profile:
\ba
\label{eq:SailShape_Fourier}
y\left(t,\theta\right)&=& \f{1}{2}F_0(t)\left(1-\cos\theta\right)+\f{1}{2}F_1(t)\sin^2\theta  \nonumber\\
&& \mbox{}- \f{1}{2}\sum_{n=2}^\infty F_n(t) \left(\f{1}{n^2-1}\right)
\,\left[2+(n-1)\cos \left(n+1\right)\theta - (n+1)\cos \left(n-1\right)\theta\right] , \quad\qquad
\ea
which has to sustain the fixed boundary conditions of the membrane edges. The leading-edge boundary condition is automatically satisfied by \eqref{eq:SailShape_Fourier}. However, the fixed trailing-edge boundary condition imposes the constraint
\be
\label{eq:F_0_Time}
F_0(t)=2\sum_{\substack{m=1}}^\infty \f{F_{2m}(t)}{(2m)^2-1} .
\ee
In addition, the assumption of an initially still and taut membrane yields zero-valued initial conditions for the Fourier coefficients and their first time derivative.

The pressure difference across the membrane in \eqref{eq:DynamicSail-LinearNorm} is obtained by using the unsteady vortex sheet method \cite[][p. 274]{Bisplinghoff_book1996},
\be
\label{eq:DP-def}
\Delta C_p(t,x)=2 \gamma(t,x) + 2\f{\p}{\p t} \int_{-1}^x \gamma \left(t,\zeta\right) \md \zeta ,
\ee
where $\gamma$ is the normalised vortex sheet strength per unit length along the profile.
The vorticity distribution along the aerofoil is determined by the fundamental equation of thin aerofoil theory, 
\be
\label{eq:ThinAirfoil-SmallCamber_norm}
\f{1}{2\upi}\dashint_{-1}^1\f{\gamma(t,\xi)}{x-\xi}\md\xi= w_a(t,x) - \f{1}{2\upi}\int_{1}^{1+t}\f{\gamma_w\left(t,\eta\right)}{x-\eta}\md\eta, \qquad x\in \left(-1,1\right),
\ee
where the dashed integral denotes the Cauchy principal value.
Here $w_a(t,x)$ is the normal velocity on the membrane surface (normalised by $U$), and $\gamma_w (t,\eta)$ describes the normalised vorticity per unit length at location $\eta$ along the wake, $\eta\in(1,\infty)$, at time $t$.
Wake vortices are assumed to be continuously shed from the trailing edge at the freestream velocity into a flat wake and have a fixed strength, which asserts that the wake vorticity distribution, $\gamma_w (t,\eta)$, is equivalent to the vorticity at the trailing edge at time $t-\eta+1$:
\be
\gamma_w (t,\eta)=\gamma_w (t-\eta+1,1)\triangleq \gamma_{_{T\!E}}(t-\eta+1).
\ee
Application of S\"{o}hngen's inversion formula to \eqref{eq:ThinAirfoil-SmallCamber_norm} and enforcement of Kelvin's theorem \citep[cf.,][p. 289]{Sohngen1939,Bisplinghoff_book1996} leads to
\be
\label{eq:GammaInt-step5}
2\int_{-1}^{1} \sqrt{\f{1+\xi}{1-\xi}}\,w_a(t,\xi)\md\xi = -\int_{1}^{1+t}\sqrt{\f{\eta+1}{\eta-1}}\gamma_{_{T\!E}} (t-\eta+1) \md\eta .
\ee
\cite{Tiomkin2017} showed that the application of the Laplace transform to \eqref{eq:GammaInt-step5} yields a closed-form expression for the wake vorticity distribution in the Laplace plane.

Provided that a solution for $\gamma_{_{T\!E}}$ is obtainable in the time domain, the method of \cite{Schwarz1940} \citep[see also][]{Iosilevskii2007} produces a general expression for the pressure difference along the aerofoil:
\ba
\label{eq:DP-GeneralSol}
\Delta C_p(t,x)&=& -\f{4}{\upi}\sqrt{\f{1-x}{1+x}}\dashint_{-1}^{1} \sqrt{\f{1+\xi}{1-\xi}}\,\f{w_a(t,\xi)}{x-\xi}\md\xi + \f{4}{\upi}\dashint_{-1}^{1}\Lambda_1 (x,\xi) w_{a_t}(t,\xi)\md\xi
\nonumber\\
&&\mbox{} + \f{2}{\upi}\sqrt{\f{1-x}{1+x}} \int_{1}^{1+t} \f{\gamma_{_{T\!E}} (t-\eta+1)}{\sqrt{\eta^2-1}}\md\eta ,
\ea
where $\Lambda_1$ is an auxiliary function expressed in \eqref{eq:a4_Lam1} of appendix~\ref{a4:Math}.
The first integral term describes the quasi-steady pressure difference, the second term is the apparent mass contribution (non-circulatory term), and the third term describes the contribution of the wake.

The contributions to expression \eqref{eq:DP-GeneralSol} for the aerodynamic load along the membrane may be further separated and analyzed by describing the normal velocity on the membrane surface as a superposition:
\be
\label{eq:wa_def}
w_a(t,x)= w_{a_d}(t,x)+w_{a_f}(t,x),
\ee
where $w_{a_\mathit{f}}(t,x)$ is the contribution of the unsteady flow (i.e., prescribed chord motion or a traveling gust) to the normal flow velocity on the membrane, and $w_{a_{d}}(t,x)$ is the respective contribution of the membrane deformation,
\be
\label{eq:wa_s_def}
w_{a_d}(t,x)= -y_x(t,x)-y_t(t,x) . 
\ee
Substitution of \eqref{eq:wa_def} into \eqref{eq:DP-GeneralSol} permits a separation of the effect of the membrane deformation, $w_{a_d}$, from the effect of the unsteady flow, $w_{a_f}$, on the aerodynamic load, namely
\be
\label{eq:Dcp_tot}
\Delta {C_p} (t,\theta) = \Delta C_{p_d} (t,\theta)+\Delta C_{p_f} (t,\theta) ,
\ee
where the subscripts $d$ and $f$ denote terms due to membrane deformation and unsteady flow, respectively.
Details of the analytical expressions for $\Delta C_{p_d}$ (in the Laplace plane) are available in appendix A of \cite{Tiomkin2017}. We develop in \S~\ref{sec:UnsteadyFlow} the closed-form expressions for $\Delta C_{p_f}$ that are necessary to complete the description of the aerodynamic load on a membrane undergoing prescribed chord motion or encountering a gust.
Note that the membrane Fourier coefficients appear only in the expression for $\Delta C_{p_d}$, whilst $\Delta C_{p_f}$ depends only on the prescribed motion or gust. 

The next section combines the terms obtained for the membrane deformation and the resulting aerodynamic load to produce a set of equations for the coupled aeroelastic problem for any arbitrary prescribed chord motion or gust.

\subsection{Aeroelastic coupling and methods of solution}

The coupled aeroelastic equation that describes the membrane response to unsteady flow conditions is obtained by substituting \eqref{eq:SailSlope-Fourier}, \eqref{eq:SailShape_Fourier}, and \eqref{eq:Dcp_tot} into \eqref{eq:DynamicSail-LinearNorm}. 
This procedure yields a matrix equation in which the unknowns are the Fourier coefficients that describe the membrane deformation.
The aeroelastic equation is described and solved in the Laplace domain for generalised time-dependent cases, or in the frequency domain for harmonic motions or gusts. Details of these two methods are given next in \S\S~\ref{sec:Laplace} and \ref{sc:FD_solution}, respectively.
Note that the overbar and hat symbols are used throughout to denote variables in the Laplace and frequency domains, respectively.

\subsubsection{Laplace domain}\label{sec:Laplace}

The membrane dynamic equation \eqref{eq:DynamicSail-LinearNorm} is expressed in the Laplace domain by applying the Laplace transform to \eqref{eq:SailSlope-Fourier}, \eqref{eq:SailShape_Fourier}, and \eqref{eq:Dcp_tot} and substituting the resulting expressions into the Laplace transform of \eqref{eq:DynamicSail-LinearNorm}.
We then multiply the resulting equation by $\sin\theta$ and use the mathematical relations \eqref{eq:a4_trigo3} and \eqref{eq:a4_trigo4} to construct a matrix system of equations,
\be
\label{eq:MatrixEqUS-Lap}
\left\{\ssA s^2 + \ssB s + \ssE\right\}\bar{\bF}= \bH ,
\ee
where $\bar{\bF}$ is the vector of Fourier coefficients $\bar{F}_n (s) , n=1\ldots N$, and $N$ is the number of coefficients chosen to represent the membrane-profile slope in \eqref{eq:SailSlope-Fourier}, taken here as $N=24$ following the numerical convergence studies of \cite{Nielsen1963} and \cite{Tiomkin2017}. Here, the overbar denotes the Laplace transform of the variable, $\bar{F}_n(s) = \Lapl \left\{F_n(t);s\right\}$. The matrices $\ssA, \ssB, \ssE$, and the vector $\bH$ are obtained by matching the coefficients of the harmonics of $\sin\theta$ in the dynamic equation \eqref{eq:DynamicSail-LinearNorm}. It is noted here that matrices $\ssB$ and $\ssE$ and vector $\bH$ depend on the Laplace variable $s$, while $\ssA$ is constant.
We further note that the matrices $\ssA,\ssB,$ and $\ssE$ are obtained from the steady flow solution under zero angle of attack, i.e., by applying $\Delta {C_p} = \Delta C_{p_d}$ to the dynamic equation; these matrices are detailed in \cite{Tiomkin2017}. 
The effect of the unsteady flow appears only in $\bH$, on the right hand side of the resulting dynamic equation, and is determined by $\Delta C_{p_f}$. Thus, $\Delta C_{p_f}$ acts as an excitation force that is applied to the membrane.

The Fourier coefficients, $\bar{\bF}$, can now be computed from \eqref{eq:MatrixEqUS-Lap}, and their substitution into the Laplace transform of \eqref{eq:SailShape_Fourier} produces the membrane dynamic solution in the Laplace domain. 
This approach predicts the membrane dynamic response to any arbitrary motion or gust. 
However, a numerical Laplace inversion is required to obtain a solution in the time domain, as no analytical expression is available for the inverse Laplace transform of our problem. We apply the numerical scheme of \cite{Valsa1998} to carry out this inversion, which is robust and reliable for both oscillatory and non-oscillatory functions.

A solution can alternatively be determined 
in the frequency (Fourier) domain by setting $s=\mi k$, where $k$ is the reduced frequency \citep[][p. 292]{Bisplinghoff_book1996}. This approach computes readily the steady-state response of the membrane wing to harmonic gusts or motions. However, this method cannot obtain the transient response of the membrane and will therefore only be used here for the harmonic cases and as a means of verification of the indicial lift responses obtained in the Laplace domain.
Details of the application of the frequency-domain method are presented in the next section.

\subsubsection{Frequency domain}\label{sc:FD_solution}

The assumption of harmonic motion for all variables converts the membrane dynamic solution to the frequency domain, where, for example $y(t,x)=\hat{y}(k,x)\, \me^{\mi kt}$, and the hat denotes a complex-valued amplitude. Assignment of $s=\mi k$ into \eqref{eq:MatrixEqUS-Lap} yields this equation in the frequency domain,
\be
\label{eq:MatrixEqUS-Freq}
\left\{-\ssA\, k^2 + \hat{\ssB}\, \mi k + \hat{\ssE}\right\}\hat{\bF}= \hat{\bH} ,
\ee
where $\hat{\ssB}=\ssB(s=\mi k),  \hat{\ssE}=\ssE(s=\mi k), \hat{\bH}=\bH(s=\mi k)$, and $\hat{\bF}$ is the vector of complex amplitudes of the Fourier coefficients $\hat{F}_n (s) , n=1\ldots N$. 
Once determined by \eqref{eq:MatrixEqUS-Freq}, these Fourier coefficients produce the resulting membrane deformation through \eqref{eq:SailShape_Fourier}. 
Note that the constant matrix $\ssA$ is unaffected by the shift from the Laplace \eqref{eq:MatrixEqUS-Lap} to the frequency \eqref{eq:MatrixEqUS-Freq} domain.

Frequency domain analysis is a natural approach to study the canonical unsteady aerodynamic scenarios of Theodorsen (harmonic oscillations) and Sears (sinusoidal gust). For these two scenarios, the frequency-domain solution yields the membrane dynamic response and the aerodynamic lift response; these results are compared against the Laplace domain results for verification. 
Integration of the Theodorsen and Sears harmonic functions over the entire frequency domain yields the indicial lift responses to a step change in angle of attack (Wagner's function) and to a sharp-edged gust (K\"{u}ssner's function), respectively (\citeauthor{Bisplinghoff_book1996}, \citeyear{Bisplinghoff_book1996}, pp.~284-287; \citeauthor{Baddoo2021unsteady}, \citeyear{Baddoo2021unsteady}).
Thus, the frequency-domain solution can generate all four canonical functions for the membrane aerofoil, whilst the transient dynamic response of the membrane can only be studied through the Laplace-domain solution. 
Case-specific technical details for this approach are further discussed in \S\S~\ref{sc:Formulation-EquivTheodorsen} and \ref{sc:Formulation-EquivSears} for harmonic heave motions and sinusoidal gusts, respectively.

\subsection{Unsteady flow conditions}\label{sec:UnsteadyFlow}

We next describe the unsteady flow conditions that the membrane aerofoil encounters for two cases of prescribed motion: generalised and harmonic heave motions, and a step change in angle of attack.
We describe the generalised problem of a membrane aerofoil that encounters a small-amplitude transverse gust, and then focus on two canonical gust profiles of sinusoidal or sharp-edged geometry. 
For each of these cases, the aerodynamic load due to the unsteady flow, $\Delta C_{p_f}$, is derived by applying $w_a=w_{a_f}$ to \eqref{eq:DP-GeneralSol}, where the normal velocity on the aerofoil, $w_{a_f}$, is defined according to \cite{vonKarman1938} and \cite{wagner1925}. $\Delta C_{p_f}$ forms the $\bH$ vector in \eqref{eq:MatrixEqUS-Lap} for the Laplace-domain solution, or the $\hat{\bH}$ vector in \eqref{eq:MatrixEqUS-Freq} for the frequency-domain solution, which concludes the formulation of our problem.

\subsubsection{Prescribed heave motion}\label{sec:Heave_def}

Consider a membrane aerofoil that performs a prescribed translatory motion, $h(t)$, that is normal to the flight direction, where $h$ is normalised by $b$ and is positive downwards.
Under these conditions, the aerodynamic load on the membrane due to heave motion is obtained by substituting $w_a(t,x)=\dot{h}(t)$ into \eqref{eq:GammaInt-step5}, where the overdot denotes a time derivative.
The solution of \eqref{eq:GammaInt-step5} in the Laplace domain yields a closed-form expression for the wake vorticity distribution, which is substituted into the Laplace transform of \eqref{eq:DP-GeneralSol}. Subsequent application of the coordinate transformation \eqref{eq:Tet-transformation} yields the Laplace transform of the aerodynamic load due to harmonic heave oscillations: 
\be
\label{eq:DCp_h-Laplace}
\Delta \bar{C}_{p_h} (s,\theta ) = 4 s^2 \bar{h}(s)\left[\bar{\Phi}(s) \cot{\f{\theta}{2}}+\sin{\theta}\right] ,
\ee
where $\bar{h}(s)$ is the Laplace transform of the prescribed heave motion, and $\bar{\Phi}(s)$ is the Laplace transform of Wagner's function \citep[e.g.,][]{Sears1940},
\be
\label{eq:Wagner_Lap}
\bar{\Phi}(s)=\f{{C}(s)}{s}.
\ee
Here ${C}(s)$ is the generalised Theodorsen function \citep[e.g.,][]{Edwards1979},
\be
\label{eq:C_Th-Laplace}
C(s) = \f{K_1(s)}{K_0(s)+K_1(s)},
\ee
and $K_0$ and $K_1$ are modified Bessel functions of the second kind.

In the frequency domain, assuming $h(t) = h_0\, \me^{\mi kt}$, the amplitude of the effective angle of attack is $\alpha_0=\mi k h_0$, and the aerodynamic load due to harmonic heave oscillations is 
\be
\label{eq:DCp_h-Freq}
\hat{\Delta {C_p}_h} (k,\theta) = 4 (\mi k)h_0 C(k)\, \cot{\f{\theta}{2}} -4 k^2 h_0 \sin{\theta} ,
\ee
where $C(k)$ is the frequency-domain Theodorsen's function,
\be
\label{eq:C_Th-FreqDomain}
C(k) = \f{H_1^{(2)}(k)}{H_1^{(2)}(k)+\mi H_0^{(2)}(k)},
\ee
and $H_0^{(2)}$ and $H_1^{(2)}$ are Hankel functions of the second kind.
The corresponding lift response is 
\be
\label{eq:CL_h-Freq}
{C_l}_{_h}(t) = 2\pi C(k)\,\dot{h}(t) + \pi\ddot{h}(t) ,
\ee
which is in fact the rigid plate response to heave oscillations \citep[][p. 272]{Bisplinghoff_book1996}. 
The aerodynamic load expressions \eqref{eq:DCp_h-Laplace} and \eqref{eq:DCp_h-Freq} form the term $\Delta C_{p_f}$ in \eqref{eq:Dcp_tot} for prescribed heave motions in the Laplace and frequency domains, respectively.

\subsubsection{Step angle of attack}

The canonical unsteady aerodynamics problem for the indicial lift response of an aerofoil to a step in angle of attack was originally solved by \citet{wagner1925} for a rigid flat plate. 
The aerodynamic load on a membrane due to a step change in angle of attack may be computed using $w_a(t,x)=~\alpha(t)$, where
\be
\label{eq:alpha_t-AoA_step}
\alpha(t) = \alpha_0 \,\mathcal{H}(t)\; \Rightarrow \; \bar{\alpha}(s) = \f{\alpha_0}{s} ,
\ee
and $\mathcal{H}(t)$ is the Heaviside function.
The procedure detailed in \S~\ref{sec:Heave_def} is repeated to derive an expression for the pressure difference coefficient,
\be
\label{eq:DCp_a_AoA_step-Laplace}
\Delta \bar{C}_{p_{\alpha_0}}\left(s,\theta\right) = 4 \alpha_0\left[\bar{\Phi}(s)\,\cot{\f{\theta}{2}}+\sin{\theta}\right].
\ee
The corresponding lift coefficient due to a step change in angle of attack is
\be
\label{eq:CL_a_AoA_step-Laplace}
\bar{C}_{l_{_{\alpha_0}}}(s) = 2\pi\alpha_0\left[\bar{\Phi}(s)+\f{1}{2}\right],
\ee
which recovers the indicial lift response found by \cite{wagner1925}.

\subsubsection{Generalised transverse gust}\label{sc:Formulation-Gust}

\begin{figure}
	\begin{center}
		\begin{subfigure}[]{0.46\textwidth}
			\includegraphics[width=\textwidth]{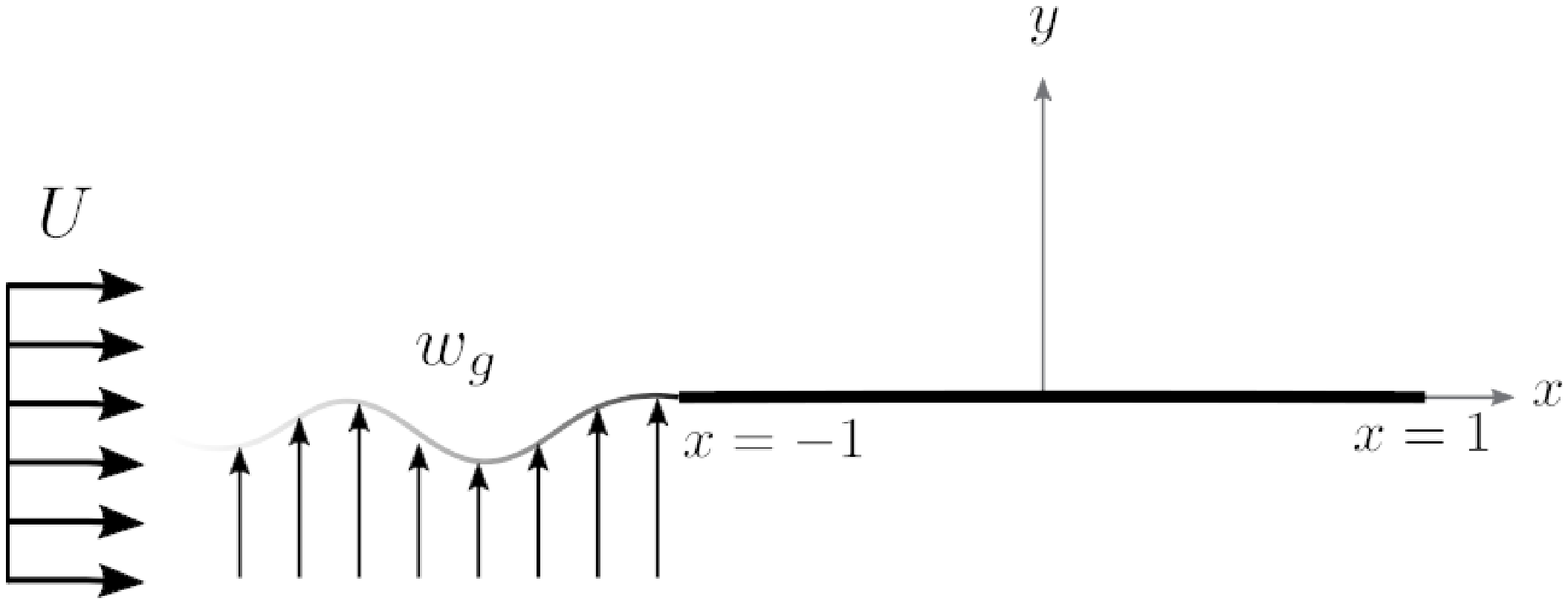}
			\caption{$t=0$}
			\label{fig:MemGustSkatch_t0}
		\end{subfigure}
		\quad
		\begin{subfigure}[]{0.46\textwidth}
			\includegraphics[width=\textwidth]{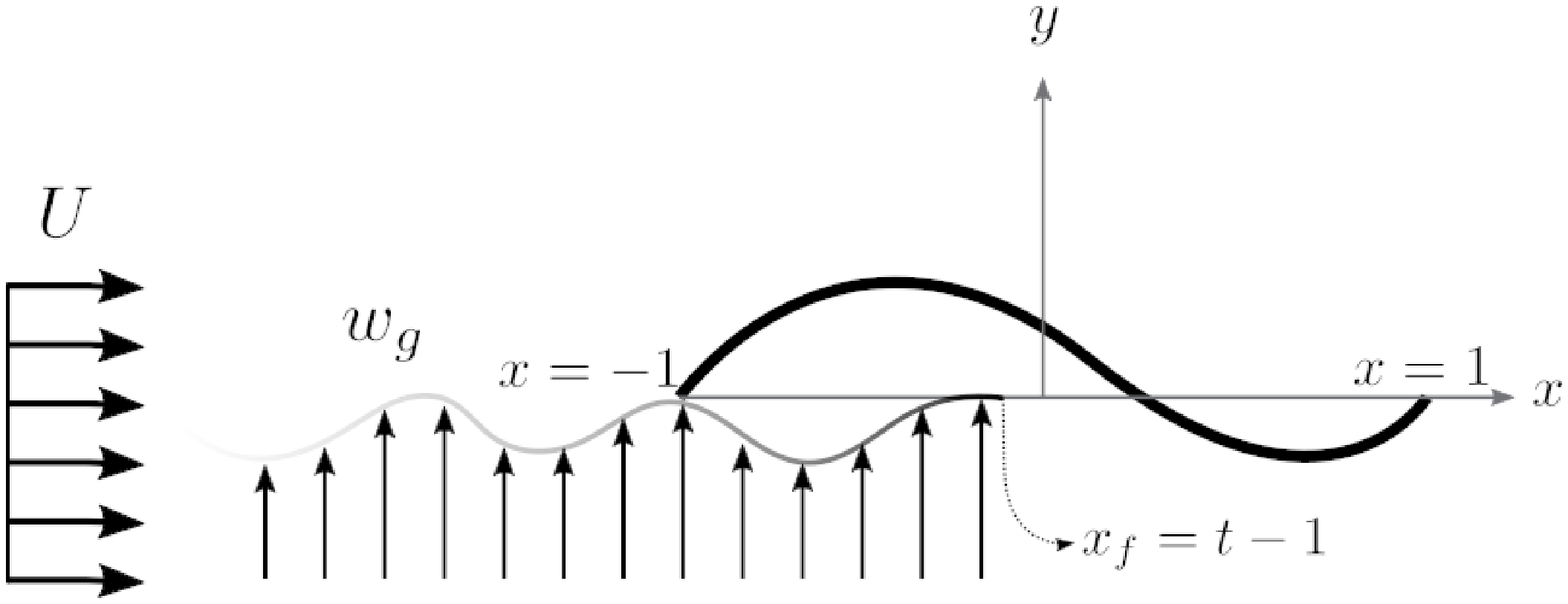}
			\caption{$t>0$}
			\label{fig:MemGustSkatch_tg0}
		\end{subfigure}
		\caption{Sketch of the membrane aerofoil gust problem: (a) initial time $t=0$; (b) later time $t>0$. Note that the membrane is initially taut, at zero angle of attack, and deforms under transient (gust) fluid loads.}
		\label{fig:MemGustSkatch_t}
	\end{center}
\end{figure}

We consider a membrane aerofoil that encounters a vertically-oriented gust with an arbitrary profile. The leading edge of the aerofoil encounters the gust front at time $t=0$ (figure~\ref{fig:MemGustSkatch_t}), and the gust amplitude is assumed to be small with respect to the freestream velocity. 
The aerodynamic load due to the imposed unsteady flow, $\Delta C_{p_f}$, is briefly detailed here to complete the formulation of the problem, as it is in fact the rigid aerofoil response to a transverse gust presented in appendix B of \cite{Iosilevskii2007}. 
Note that $\Delta C_{p_f}$ in \eqref{eq:Dcp_tot} is denoted $\Delta C_{p_g}$ in the present context of gusts.

The gust model assumptions permit the substitution of $w_a(t,x)=\alpha_g(t-x-1)$ into \eqref{eq:GammaInt-step5} to produce the gust effect on the vorticity distribution along the wake, which leads to a closed-form expression for the aerodynamic load along the aerofoil using \eqref{eq:DP-GeneralSol}: 
\be
\label{eq:DCp_g-Laplace}
\Delta \bar{C}_{p_g}\left(s,\theta\right) = 4s\,\bar{\Psi}(s)\, \bar{\alpha}_g(s) \,\cot{\f{\theta}{2}},
\ee
where 
\be
\label{eq:Kussner-Laplace}
\bar{\Psi}(s)=\f{\me^{-s}}{s^2}\f{1}{K_0(s)+K_1(s)}
\ee
is the Laplace transform of K\"{u}ssner's function, $\Psi(t)$ \citep[e.g.,][]{Sears1940}. Note that the aerodynamic load due to an arbitrary transverse gust is obtained by a convolution of K\"{u}ssner's function and the time derivative of the gust angle of attack,
\be
\label{eq:DCp_g-Time}
\Delta C_{p_g}\left(t,\theta\right) = 4\cot{\f{\theta}{2}}\int_0^t \Psi\left(t-\tau\right)\f{\md\alpha_g(\tau)}{\md \tau}\md\tau .
\ee
Equation \eqref{eq:DCp_g-Laplace} forms an expression for $\Delta C_{p_f}$ in the Laplace domain, for an arbitrary, small-amplitude gust profile.

\subsubsection{Sinusoidal gust}\label{sc:Formulation-Sinus_gust}
A sinusoidal gust encountered at the leading edge at time $t=0$ can be expressed as
\be
\label{eq:alpha_g-sinus}
\alpha_g\left(t-x-1\right) = \alpha_0\sin\left(k\left(t-x-1\right)\right) \, \mathcal{H}\left(t-x-1\right).
\ee
The substitution of the Laplace transform of \eqref{eq:alpha_g-sinus} into \eqref{eq:DCp_g-Laplace} yields the aerodynamic load along the aerofoil due to a sinusoidal gust,
\be
\label{eq:DCp_g-sinus}
\Delta \bar{C}_{p_g}\left(s,\theta\right) = 4k\alpha_0 \f{s}{s^2+k^2}\,\bar{\Psi}(s)\, \cot{\f{\theta}{2}}.
\ee 

It is natural to solve the steady-state problem for harmonic gusts in the frequency domain by assuming
\be
\label{eq:alpha_g-exp_ikt}
\alpha_g\left(t-x-1\right) = \alpha_0\, \me^{\mi k\left(t-x-1\right)},
\ee
which yields
\be
\label{eq:DCpg_exp_ikt-FD}
\Delta \hat{C}_{p_g}\left(k,\theta\right) = 4\alpha_0\mi k\,\hat{\Psi}(k)\cot{\f{\theta}{2}} .
\ee
Here 
\be
\label{eq:Kussner-FD}
\hat{\Psi}(k)=\f{2}{\pi}\f{\me^{-\mi k}}{k^2}\left(\f{1}{H_1^{(2)}(k)+\mi H_0^{(2)}(k)}\right) 
\ee
is the Fourier transform of the K\"{u}ssner function, which is obtained by assigning $s=\mi k$ in \eqref{eq:Kussner-Laplace}.
The lift response to sinusoidal gusts in the frequency domain is
\be
\label{eq:L_g-Time-sinus}
\hat{C}_{l_g}(k) = 2\pi\alpha_0\mi k\, \hat{\Psi}(k) = 2\pi \alpha_0\, S(k),
\ee
where the modified Sears function, $S(k)$, is given by \citep[][p. 287]{Bisplinghoff_book1996}
\be
\label{eq:Sears-FD}
S(k) = \mi k\,\hat{\Psi}(k) = \left\{C(k)\left[J_0(k)-\mi J_1(k)\right]+\mi J_1(k)\right\}\me^{-\mi k} = \tilde{S}(k)\,\me^{-\mi k} .
\ee
$\tilde{S}(k)$ is the classical Sears function whose gust front is at the mid-chord location at time $t=0$, and $J_0$ and $J_1$ are Bessel functions of the first kind.

\subsubsection{Sharp-edged gust}
A sharp-edged gust is similarly described by
\be
\label{eq:alpha_g-SEG}
\alpha_g\left(t-x-1\right) = \alpha_0 \,\mathcal{H}\left(t-x-1\right).
\ee
The substitution of \eqref{eq:alpha_g-SEG} into \eqref{eq:DCp_g-Laplace} yields the aerodynamic load along the aerofoil due to a sharp-edged gust,
\be
\label{eq:DCp_g-SEG}
\Delta \bar{C}_{p_g}\left(s,\theta\right) = 4\alpha_0\,\cot{\f{\theta}{2}}\,\bar{\Psi}(s).
\ee 
The resulting lift response is
\be
\label{eq:L_g-SEG}
\bar{C}_{l_g} (s)=2\pi\alpha_0\,\bar{\Psi}(s),
\ee
which is the expected classical indicial lift of a rigid plate due to a sharp-edged gust. 
The distributed aerodynamic load \eqref{eq:DCp_g-SEG} is in fact the external force applied on the membrane through the term $\bH$ in \eqref{eq:MatrixEqUS-Lap} in the case of a sharp-edged gust. 
This applied force initiates a membrane deformation which brings about a change in the aerodynamic load through aeroelastic coupling.

\subsection{Unsteady lift response functions}\label{sec:LiftFunctions}

The membrane unsteady lift coefficient is derived by integration of the aerodynamic load \eqref{eq:Dcp_tot} along the membrane chord-line, which leads to
\be
\label{eq:CL_tot}
C_{l_m}(t) = C_{l_d}(t)+C_{l_f}(t) ,
\ee
where the normalised lift due to membrane deformation is
\be
\label{eq:Cl_d}
\f{C_{l_d}(t)}{2\pi\alpha_0} =  \int_0^t \Phi(t-\tau)\dot{f}(\tau)\md\tau +g(t) .
\ee
Here, $\Phi(t)$ is the time-domain Wagner function, and $f(t)$ and $g(t)$ are functions of the Fourier coefficients given by 
\ba
\label{eq:f_t}
f(t) &=& \f{1}{2}\mF_{1}(t) - \f{1}{2}\mF_{0}(t)  -\f{1}{4} \dot{\mF}_0(t) -\f{1}{4} \dot{\mF}_1(t) + \f{1}{4} \dot{\mF}_2(t) + \sum_{m=2}^{N/2}  \f{\dot{\mF}_{2m-1}(t)}{(2m-1)^2-1},\\
\label{eq:g_t}
g(t) &=& -\f{1}{4}\dot{\mF}_0(t) + \f{1}{4}\dot{\mF}_2(t) - \f{3}{16} \ddot{\mF}_1(t)  + \f{1}{8} \ddot{\mF}_3(t) +\f{1}{2} \sum_{m=3}^{N/2} \f{\ddot{\mF}_{2m-1}(t)}{(2m-1)^2-1} ,
\ea
where $\mF_n=F_n/\alpha_0$.
The term ${\alpha}_0$ is the unsteady angle of attack amplitude in the harmonic cases, or the steady angle of attack in the indicial cases; the reader may consult \S~\ref{sec:UnsteadyFlow} for details on the definition of ${\alpha}_0$ and the lift due to the unsteady flow, $C_{l_f}$, for each case of prescribed chord motion or gust encounter considered here.

The Laplace transform of $C_{l_d}(t)$,
\be
\label{eq:Cl_d-Lap}
\f{\bar{C}_{l_d}(s)}{2\pi\alpha_0 } = C(s)\bar{f}(s) +\bar{g}(s),
\ee
obtains the lift coefficient due to membrane deformation in the Laplace domain for indicial scenarios. 
Subsequent numerical Laplace inversion yields the membrane indicial lift response functions in the time domain.
For cases of harmonic oscillations, in which the membrane solution is obtained in the frequency domain, the lift coefficient due to membrane deformation is expressed in the frequency domain 
\be
\label{eq:Cl_d-Freq}
\f{\hat{C}_{l_d}(k)}{2\pi\alpha_0 } = C(k)\hat{f}(k) +\hat{g}(k),
\ee
where $\hat{f}(k)$ and $\hat{g}(k)$ are found by substituting $\mF_n(t)=\hat{\mF}_n(k)\, \me^{\mi kt}$ into \eqref{eq:f_t} and \eqref{eq:g_t}, respectively, and the auxiliary functions in the time domain become {$f(t)=\hat{f}(k)\, \me^{\mi kt}$} and {$g(t)=\hat{g}(k)\, \me^{\mi kt}$}.

Substitution of $C_{l_d}$ (\eqref{eq:Cl_d-Lap} or \eqref{eq:Cl_d-Freq} for the indicial or harmonic scenarios, respectively) and the case-specific $C_{l_f}$ into \eqref{eq:CL_tot} yields a closed-form expression for the total membrane lift coefficient, $C_{l_m}$, from which extensions to the classical unsteady aerodynamic functions are derived after a solution for the Fourier coefficients is obtained.

\subsubsection{Equivalent Theodorsen function}\label{sc:Formulation-EquivTheodorsen}

An equivalent Theodorsen function is constructed for a flexible membrane wing following the classical approach presented in \citet[p.~279]{Bisplinghoff_book1996}. An extension for Theodorsen's function is obtained by computing the membrane response to prescribed heave oscillations in the frequency domain:
\be
\label{eq:TheodorsenMem_def}
C_m(k) = \f{\hat{L}^C_h(k)}{\mi k h_0\, L_{s_\alpha}} ,
\ee
where $L_h^C$ is the circulatory lift due to heave oscillations, and $L_{s_\alpha}$ is the static (aeroelastic) membrane lift-curve slope.
The membrane circulatory lift is obtained by superposition between the circulatory lift due to membrane deformation (first term in \eqref{eq:Cl_d-Freq}) and the circulatory lift due to the unsteady flow (first term in \eqref{eq:CL_h-Freq}).
Substitution of the membrane circulatory lift expression into \eqref{eq:TheodorsenMem_def} yields a closed-form expression for the membrane Theodorsen function in the frequency domain,
\be
\label{eq:TheodorsenMem_expFn2}
C_m(k) = \f{2\pi}{C_{l_{s\alpha}}}\,C(k)\left[1+\hat{f}(k) \right].
\ee
The static membrane lift slope, 
\be
\label{eq:CLa_mem-static}
C_{l_{s\alpha}} = 2\pi \left[1+ \f{1}{2}\mF_{s_1} - \f{1}{2}\mF_{s_0} \right] ,
\ee
is established by direct integration of the static pressure load given by \cite{Nielsen1963}.
Here, $\mF_{s_n}$ are the static membrane Fourier coefficients, normalised by the static angle of attack, which depend solely on the tension coefficient. Note that as $k \rightarrow 0$ the Fourier coefficients in the unsteady membrane solution converge to the static solution, ${\hat{\mF}_n \xrightarrow[]{} \mF_{s_n}}$, where $\hat{\mF}_n = F_n/\alpha_0$ and $\alpha_0=\mi k h_0$. Application of this limit to \eqref{eq:TheodorsenMem_expFn2} after substituting the leading term in the asymptotic expansion of $\hat{f}(k)$ in \eqref{eq:a2_1-fk_lim_k0_1} recovers 
\be
\label{eq:TheodorsenMem_lim_k0}
C_m(k) = C(k) + \textit{O}\left(k\right) \quad \mbox{as}\quad k\ttz .
\ee
In other words, in the limit of small reduced frequencies the equivalent Theodorsen function converges to the standard Theodorsen function, as expected.

Finally, we note that the Fourier coefficients of the membrane solution converge to zero for a very stiff membrane as $C_T\rightarrow\infty$ at any reduced frequency other than the fluid-loaded resonance frequencies. Under these conditions, the auxiliary function $f(t)$ \eqref{eq:f_t} goes to zero, and the static membrane lift-curve slope converges on $2\pi$. Therefore, the equivalent Theodorsen function recovers the rigid plate function for $C_T\rightarrow\infty$, as expected. Appendix~\ref{a2_1:approx_k0} reports further details on this limit.

\subsubsection{Equivalent Wagner function}\label{sc:Formulation-EquivWagner}

We next derive the equivalent Wagner function for a flexible membrane wing following \citet[pp.~284-287]{Bisplinghoff_book1996}. 
Note that the equivalent Theodorsen function derived in \S~\ref{sc:Formulation-EquivTheodorsen} enables the computation of the equivalent Wagner function in the time domain through \citep[][p. 285]{Bisplinghoff_book1996}
\be
\label{eq:WagnerMem_def}
\Phi_m(t)  =  \f{L_{\alpha_0}^C}{\alpha_0\, L_{s_\alpha}} = 1+\f{2}{\pi}\int_0^\infty \f{\Im\left\{C_m(k)\right\}}{k}\cos{kt}\,\md k , \quad t>0,
\ee
where $L_{\alpha_0}^C$ is the circulatory lift of the membrane due to a step change in angle of attack, expressed in the time domain.
The above equation allows for the computation of the equivalent Wagner function from both the Laplace-domain solution (first equality) or the frequency-domain solution (second equality). 
However, whilst the frequency-domain solution is more efficient when computing the Theodorsen function, the application of this solution to \eqref{eq:WagnerMem_def} requires a broad range of frequencies to obtain an accurate initial (high-frequency) response of the membrane lift. Therefore, the equivalent Wagner function is presented using the Laplace-domain solution. 

Following the procedure presented in \S~\ref{sc:Formulation-EquivTheodorsen} for the equivalent Theodorsen function, a closed-form expression is derived for the equivalent Wagner function in the Laplace domain, 
\be
\label{eq:WagnerMem_expFn-LD}
\bar{\Phi}_m(s) = \f{2\pi}{C_{l_{s\alpha}}}\, \bar{\Phi}(s)\left[1+s\bar{f}(s)\right] ,
\ee
where $\bar{f}(s)$ is the Laplace transform of $f(t)$ from \eqref{eq:f_t}. In the time domain, the equivalent Wagner function becomes
\be
\label{eq:WagnerMem_expFn-TD}
\Phi_m(t) = \f{2\pi}{C_{l_{s\alpha}}}\left\{ \Phi(t)+\int_0^t \Phi(t-\tau) \dot{f}(\tau) \md\tau \right\}.
\ee
We note that  $\Phi_m(0) = {\pi}/{C_{l_{s\alpha}}}$ at time $t=0$, and the static membrane lift slope, $C_{l_{s\alpha}}$, is higher than the rigid plate lift slope due to the static membrane camber. Therefore, the initial value of the equivalent Wagner function is smaller than the classical value of $1/2$ predicted by the standard Wagner function for a rigid flat plate. 

At long times $(t\rightarrow\infty)$, we can use the fact that $\dot{f}(\tau)$ rapidly converges to zero as the membrane profile converges to the appropriate static profile. Thus,
\ba
\nonumber
\lim_{t\rightarrow\infty} \left\{\int_0^t \Phi(t-\tau) \dot{f}(\tau) \md\tau\right\} \cong&& \lim_{t\rightarrow\infty} \left\{\Phi(t)\int_0^t \dot{f}(\tau) \md\tau\right\} \\ 
\label{eq:WagnerMem_int_t_inf}
\cong&& \lim_{t\rightarrow\infty} \left\{\Phi(t)\, f(t)\right\} \cong \left(\f{C_{l_{s\alpha}}}{2\pi}-1\right) \Phi(t)
\ea
and the equivalent Wagner function converges to the standard Wagner function, ${\Phi_m(t\rightarrow\infty)\cong\Phi(t\rightarrow\infty)=1}$, as expected.
For a very stiff membrane $(C_T\rightarrow\infty)$, the static membrane lift slope becomes $2\pi$, and the equivalent Wagner function converges to the rigid-plate Wagner function during the entire time response.

\subsubsection{Equivalent Sears function}\label{sc:Formulation-EquivSears}

Extension of the canonical modified Sears function is next derived for a flexible membrane wing in the frequency domain, following the classical approach presented in \citet[pp. 286-287]{Bisplinghoff_book1996}. 
The equivalent Sears function is obtained by calculating the membrane response to oscillating gusts (\S~\ref{sc:Formulation-Sinus_gust}) and normalising the expression by the membrane static lift:
\be
\label{eq:SearsMem_def}
S_m(k) = \f{\hat{L}_{gs}(k)}{L_s},
\ee
where $\hat{L}_{gs}(k)$ is the (complex) amplitude of the lift due to harmonic gust with reduced frequency $k$.

Substitution of the membrane lift expression \eqref{eq:CL_tot} into \eqref{eq:SearsMem_def}, superposing the lift due to membrane deformation \eqref{eq:Cl_d-Freq} with the lift due to the sinusoidal gust \eqref{eq:L_g-Time-sinus} in the frequency domain, leads to a closed-form expression for the membrane equivalent Sears function,
\be
\label{eq:SearsMemExpr}
S_m (k) = \f{2\pi}{C_{l_{s\alpha}}} \left\{ S(k) + C(k)\hat{f}(k) + \hat{g}(k) \right\} .
\ee
The first term in \eqref{eq:SearsMemExpr} describes the rigid aerofoil lift response, the second term is the circulatory lift response due to membrane deformation, and the third term represents the non-circulatory lift response due to membrane deformation. 
Note that the unsteady membrane solution to an encounter with sinusoidal gusts converges to the static membrane solution as $k\rightarrow 0$, namely $\hat{\mF}_n \xrightarrow[k \rightarrow 0]{} \mF_{s_n}$, where $\hat{\mF}_n = \hat{F}_n/\alpha_0$.
Thus, for very low reduced frequencies the equivalent Sears function converges to
\ba
\nonumber
S_m(k) \cong&& \f{2\pi}{C_{l_{s\alpha}}} \left\{ S(k) + \left(\f{C_{l_{s\alpha}}}{2\pi} -1\right) \left[ 1 + k\left[\mi\ln{\f{k}{2}} -\f{\pi}{2}\right]\right] \right\}\\
\label{eq:SearsMem_lim_k0}
&&\mbox{} + \textit{O}\left( k\hat{\mF}_3 ; k\hat{\mF}_4 ; k^2\ln{k} ; k^2 \right)   \quad \mbox{as\ }\quad k\ttz .
\ea
We note that for the limiting case of $C_T\rightarrow\infty$ the static membrane lift slope approaches $2\pi$ and the lift due to the membrane deformation converges to zero for $k\neq\omega_{r_n}$. Therefore, the equivalent Sears function converges to the standard modified Sears function for ${C_T\rightarrow\infty}$, as expected. Appendix~\ref{a2_1:approx_k0} reports further details on these low-frequency limits.

\subsubsection{Equivalent K\"{u}ssner function}\label{sc:Formulation-EquivKussner}

The aeroelastic membrane K\"{u}ssner function in the time domain,
\be
\label{eq:KussnerMem_def}
\Psi_m(t)  =  \f{L_{\mathit{seg}}}{L_s} = 1+\f{2}{\pi}\int_0^\infty \f{\Im\left\{S_m(k)\right\}}{k}\cos{kt}\,\md k , \quad t>0,
\ee
follows from the equivalent Sears function determined in \S~\ref{sc:Formulation-EquivSears} using the procedure outlined in \S~\ref{sc:Formulation-EquivWagner} for the Wagner function \citep[e.g.,][p. 287]{Bisplinghoff_book1996}.
Here $L_{\mathit{seg}}$ is the time-domain membrane lift response to a sharp-edged gust. The above equation enables the computation of the equivalent K\"{u}ssner function from both the Laplace-domain solution (first equality) or the frequency-domain solution (second equality).

A closed-form expression for the equivalent K\"{u}ssner function is determined in the Laplace domain by superposing the rigid aerofoil indicial lift \eqref{eq:L_g-SEG} and the lift due to membrane deformation \eqref{eq:Cl_d-Lap}, with substitution of the resultant unsteady lift into the Laplace transform of \eqref{eq:KussnerMem_def}:
\be
\label{eq:KussnerMem_exp-LD}
\bar{\Psi}_m(s) = \f{2\pi}{C_{l_{s\alpha}}}\, \left\{\bar{g}(s) + \bar{\Psi}(s)+\bar{C}(s)\bar{f}(s)\right\} .
\ee
The equivalent K\"{u}ssner function in the time domain is clearly
\be
\label{eq:KussnerMem_exp-TD}
{\Psi}_m(t) = \f{2\pi}{C_{l_{s\alpha}}}\, \left\{g(t)+\Psi(t)+\int_0^t \Phi(t-\tau) \dot{f}(\tau) \md\tau\right\} .
\ee
We note that the initial value of the equivalent K\"{u}ssner function is ${\Psi_m(0) = 0}$. For $t\rightarrow\infty$ the equivalent K\"{u}ssner function asymptotically converges to unity, and for a very stiff membrane of $C_T\rightarrow\infty$ the rigid-plate K\"{u}ssner function is recovered.

\section{Results and discussion}\label{sec:results}

The membrane response to prescribed chord motion or an incoming gust is derived in both the time domain (via inverse Laplace transform) and the frequency domain, the latter of which is used to study the steady-state response to harmonic motions or sinusoidal gusts.
Results are shown for four canonical cases: harmonic heave oscillations, step change in angle of attack, sinusoidal gust and sharp-edged gust, for which extensions of the four respective classical unsteady lift functions are presented for a flexible membrane wing.
We begin by studying the lift and dynamic response of a nominal membrane of $\mu=1$ and $C_T=2.5$, followed by analysis of the role of each of the membrane parameters.

\subsection{Prescribed motion}\label{sec:results-pres}

The membrane response to prescribed chord motion is derived for two canonical problems: harmonic heave oscillations, from which Theodorsen's function is derived, and a step change in angle of attack (also known as Wagner's problem). 
In both cases the membrane is free to deform around the chord-line, which adheres to the prescribed motion.
Extensions of the classical Theodorsen and Wagner functions are presented for flexible membrane wings, along with a discussion on the membrane dynamic response to these unsteady flow conditions and the role of the membrane parameters $\left(\mu, C_T\right)$ in its aerodynamic performance.

\subsubsection{Harmonic heave oscillations}\label{sec:results-pres-heave}

\begin{figure}
	\begin{center}
	\begin{subfigure}[]{0.46\textwidth}
			\includegraphics[width=\textwidth]{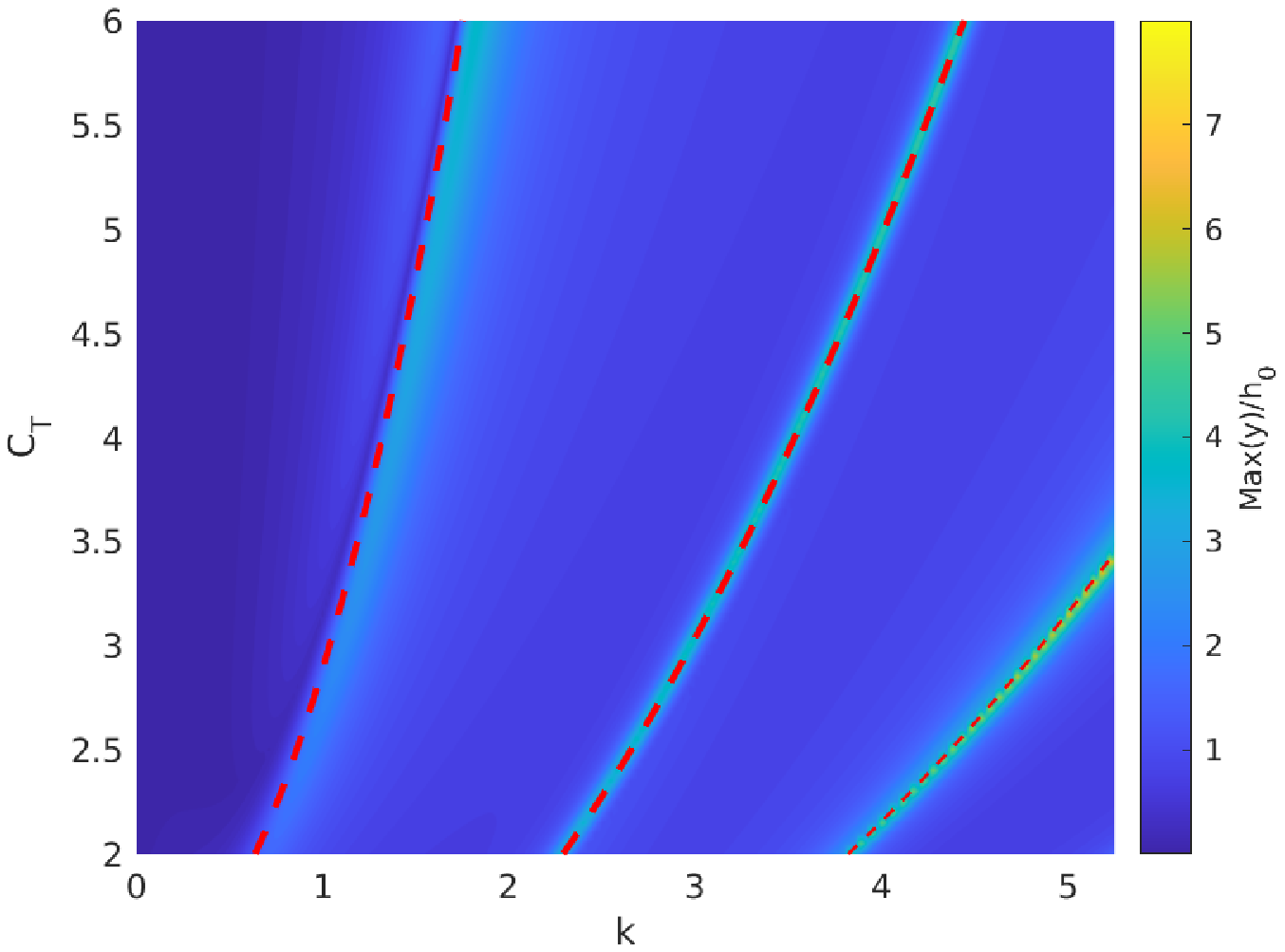}
			\caption{}
			\label{fig:HeaveSSMem_CT_vs_k-mu1_max_amp_map}
		\end{subfigure}
		\quad
		\begin{subfigure}[]{0.46\textwidth}
			\includegraphics[width=\textwidth]{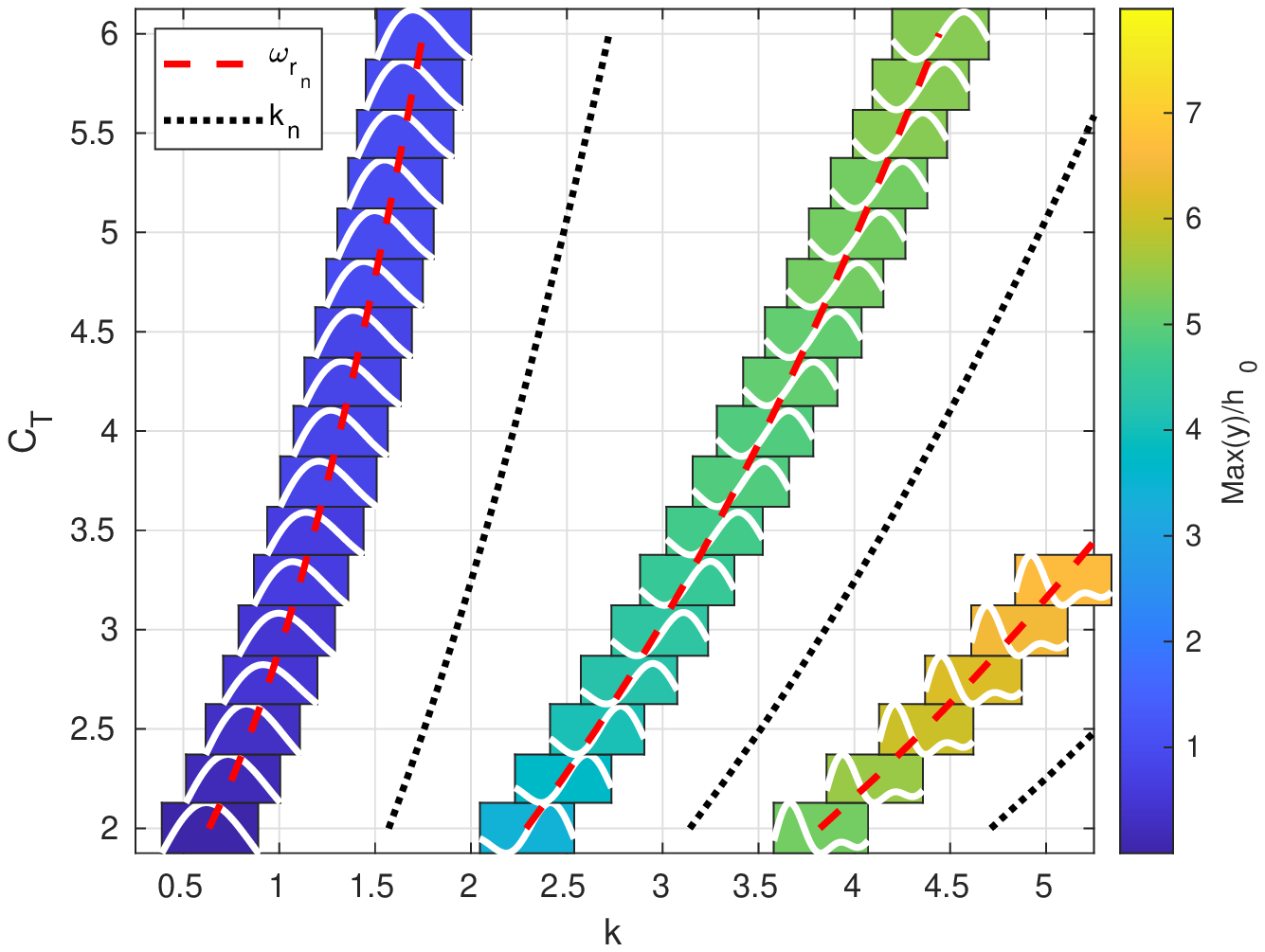}
			\caption{}
			\label{fig:HeaveSSMem_CT_vs_k-mu1}
		\end{subfigure}
		\\
		\begin{subfigure}[]{0.46\textwidth}
			\includegraphics[width=\textwidth]{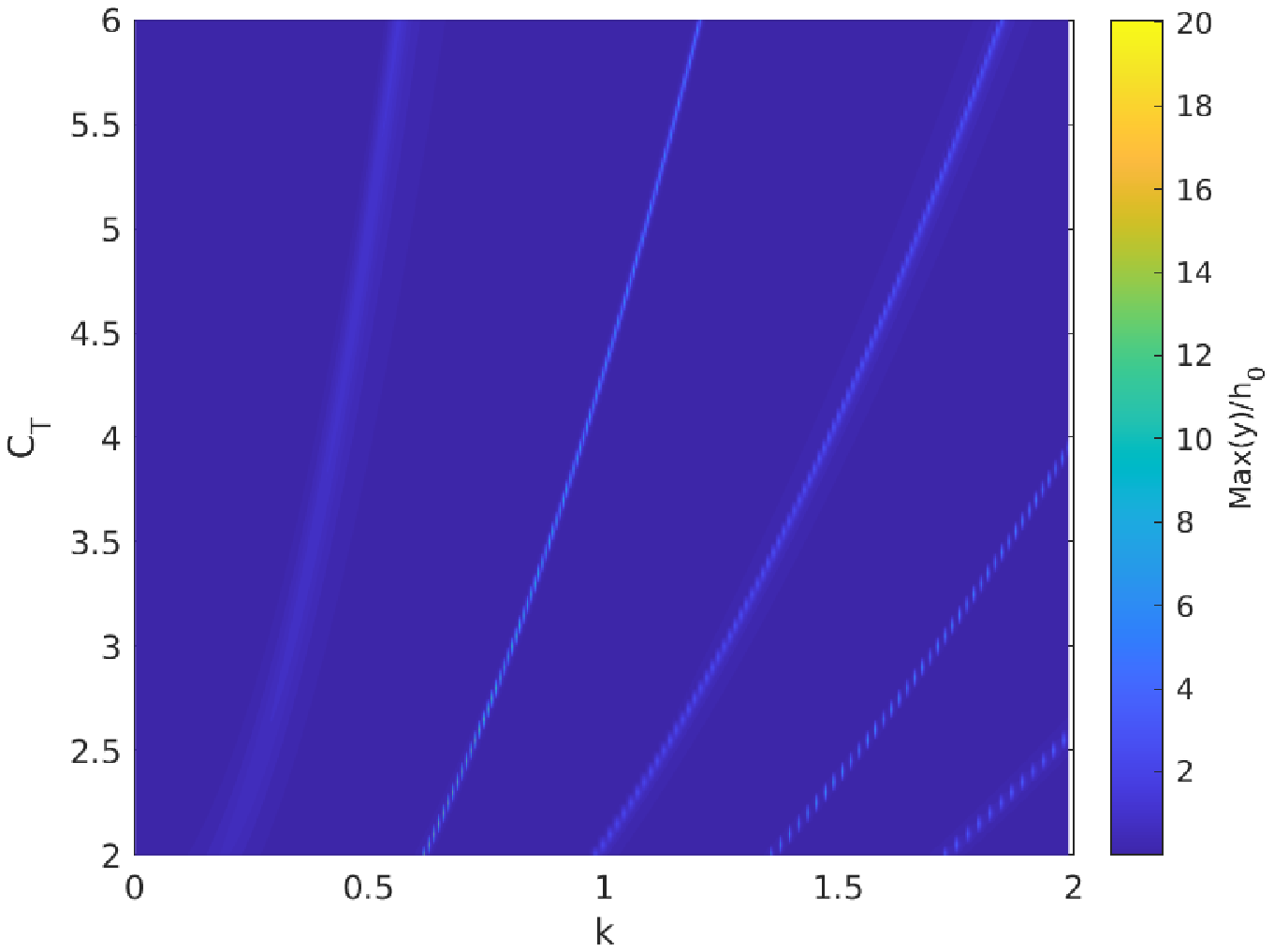}
			\caption{}
			\label{fig:HeaveSSMem_CT_vs_k-mu18_max_amp_map}
		\end{subfigure}
		\quad
		\begin{subfigure}[]{0.46\textwidth}
			\includegraphics[width=\textwidth]{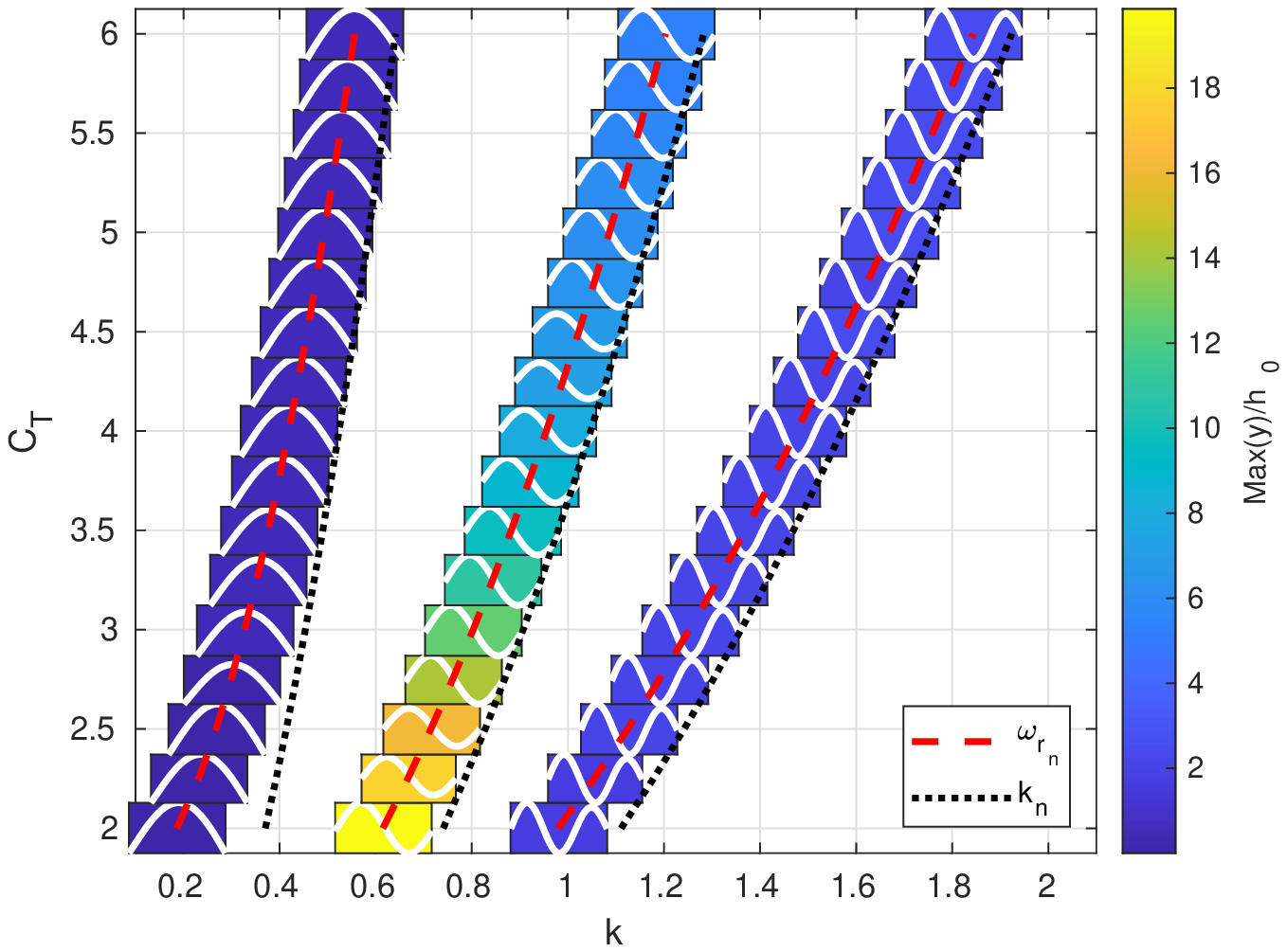}
			\caption{}
			\label{fig:HeaveSSMem_CT_vs_k-mu18}
		\end{subfigure}
		\caption{Membrane amplitude in response to heave oscillations: (a), (c) maximum amplitude maps obtained for various values of $C_T$ and $k$ for $\mu=1$ and $\mu=18$, respectively. (b), (d) present the membrane amplitude profiles obtained at the fluid-loaded resonance frequencies for $\mu=1$ and $\mu=18$, respectively. Background colour depicts maximum amplitude of membrane deformation, normalised by the heave motion amplitude, $h_0$. Red dashed lines describe the first, second and third resonance frequencies of the fluid-loaded membrane, and black dotted lines present membrane in vacuo natural frequencies. A large value of the mass ratio, $\mu=18$, is chosen to assess the membrane response near the onset of flutter, which \cite{Tiomkin2017} predict to occur for $\mu\ge18.8$ when $C_T=2$. Aerodynamic damping leads to finite membrane amplitudes at resonance.}
	\label{fig:HeaveSSMem_CT_vs_k}
	\end{center}
\end{figure}

To assess the membrane wing response to prescribed oscillations in heave, we compute first the membrane amplitude at various reduced frequencies of oscillation, $k$, for various tension coefficients and two mass ratios (figure~\ref{fig:HeaveSSMem_CT_vs_k}). In addition, the resonance frequencies of the fluid-loaded membrane system ($\omega_{r_1}, \omega_{r_2},$ etc.) are computed from the homogeneous system of \eqref{eq:MatrixEqUS-Freq}, following the method of \cite{Kornecki1976}. The parametric dependence of the fluid-loaded resonance frequencies on $C_T$ and $k$ is illustrated with dashed red lines in figure~\ref{fig:HeaveSSMem_CT_vs_k}. 
The left column of figure~\ref{fig:HeaveSSMem_CT_vs_k} presents maps of the resulting maximum membrane amplitude, obtained for $\mu=1$ (upper row) and $\mu=18$ (lower row) for varying tension coefficient and reduced frequency.
As expected, significant amplitudes of oscillation occur for frequencies near the resonance frequencies of the fluid-loaded membrane. In the heavy membrane case, $\mu=18$, where the mass ratio is encroaching upon the flutter instability threshold, predicted by \cite{Tiomkin2017} at $\mu\ge18.8$ for $C_T=2$, the peaks in the maximal amplitude map are more concentrated, with a significantly increased amplitude along the second fluid-loaded resonance frequency. Note that these narrow peaks in figure~\ref{fig:HeaveSSMem_CT_vs_k-mu18_max_amp_map} prevent the addition of the resonance frequencies to this plot, as these lines cover the peaks entirely; the relevant fluid-loaded resonance frequencies are plotted in figure~\ref{fig:HeaveSSMem_CT_vs_k-mu18} for reference.
We further note that the presence of aerodynamic damping leads to finite amplitudes of the membrane at resonance in this linear analysis.
The amplitude peaks along the fluid-loaded resonance frequencies reach large values that are beyond the validity range of the current study (especially in the heavy membrane case). However, away from these very narrow peaks the results across the rest of the frequency regime satisfy the ansatz of linear dynamics assumed by the present work.

The right column of figure~\ref{fig:HeaveSSMem_CT_vs_k} shows the membrane amplitude profiles obtained along the fluid-loaded resonance frequencies, with background colour used to indicate the maximum amplitude value. Dotted black lines indicate the in vacuo natural frequencies, $k_1, k_2,$ etc., where $k_n=n\pi\sqrt{C_T/8\mu}$, which are compared against the fluid-loaded resonance frequencies. 
Significant differences between the in vacuo and the fluid-loaded membrane resonance frequencies are obtained for the lowest mass ratio, $\mu=1$, in figure~\ref{fig:HeaveSSMem_CT_vs_k-mu1}. 
This difference in frequencies evokes oscillations with membrane amplitude profiles that are noticeably different from the membrane in vacuo modes.
In addition, the membrane amplitude increases with reduced frequency, following the behaviour of the excitation term in \eqref{eq:DCp_h-Freq} whose amplitude increases monotonically with $k$.
As the mass ratio increases in figure~\ref{fig:HeaveSSMem_CT_vs_k-mu18}, the gap between the resonance and the structural frequencies diminishes, and second-mode oscillations become dominant; this is the first dynamically unstable membrane mode \citep[][]{Nielsen1963,Tiomkin2017}. In this case the membrane amplitude increases as the tension decreases (although the resonance frequency also decreases), which is typical of the membrane-wing behaviour on the verge of instability \citep{Tiomkin2017,Mavroyiakoumou2020,Mavroyiakoumou2021}.

The differences between the in vacuo natural frequencies and the fluid-loaded resonance frequencies are mainly due to the added mass of the surrounding fluid, which must be taken into account when computing the total inertia of the coupled system. This effect may be quantified in non-dimensional terms as an added mass ratio, $\mu_\mathit{add}$, which can be computed by assuming $\omega_{r_1}=\pi\sqrt{C_T/8(\mu+\mu_\mathit{add})}$.
For a rigid plate the added mass is commonly taken as $\mu_\mathit{add}=\pi/4$ \citep[][pp. 385-387]{KatzPlotkin2001}. For a membrane wing, \cite{Alon2019} found a constant added mass value of $\mu_\mathit{add}=0.5$, and \cite{Minami1998} determined that $\mu_{\mathit{add}} = 0.68$ for a membrane oscillating in quiescent air. 
\cite{Minami1998} used standing modes to describe the membrane deformation, without considering the tension along the membrane. 
However, \cite{Yadkyn2003} showed that the added mass of flexible plates is strongly affected by the mode of vibration.
Figures~\ref{fig:HeaveSSMem_CT_vs_k-mu1} and \ref{fig:HeaveSSMem_CT_vs_k-mu18} reveal in the current investigation that the membrane parameters and the reduced frequency of the harmonic heave motion control the amplitude profile of the oscillating membrane. Therefore, the added mass ratio in fact depends on both the membrane mass ratio and tension coefficient when considering the coupled problem of the membrane passive deformation in response to unsteady flow.

\begin{figure}
	\begin{center}
	    \begin{subfigure}[h!]{0.46\textwidth}
			\includegraphics[width=\textwidth]{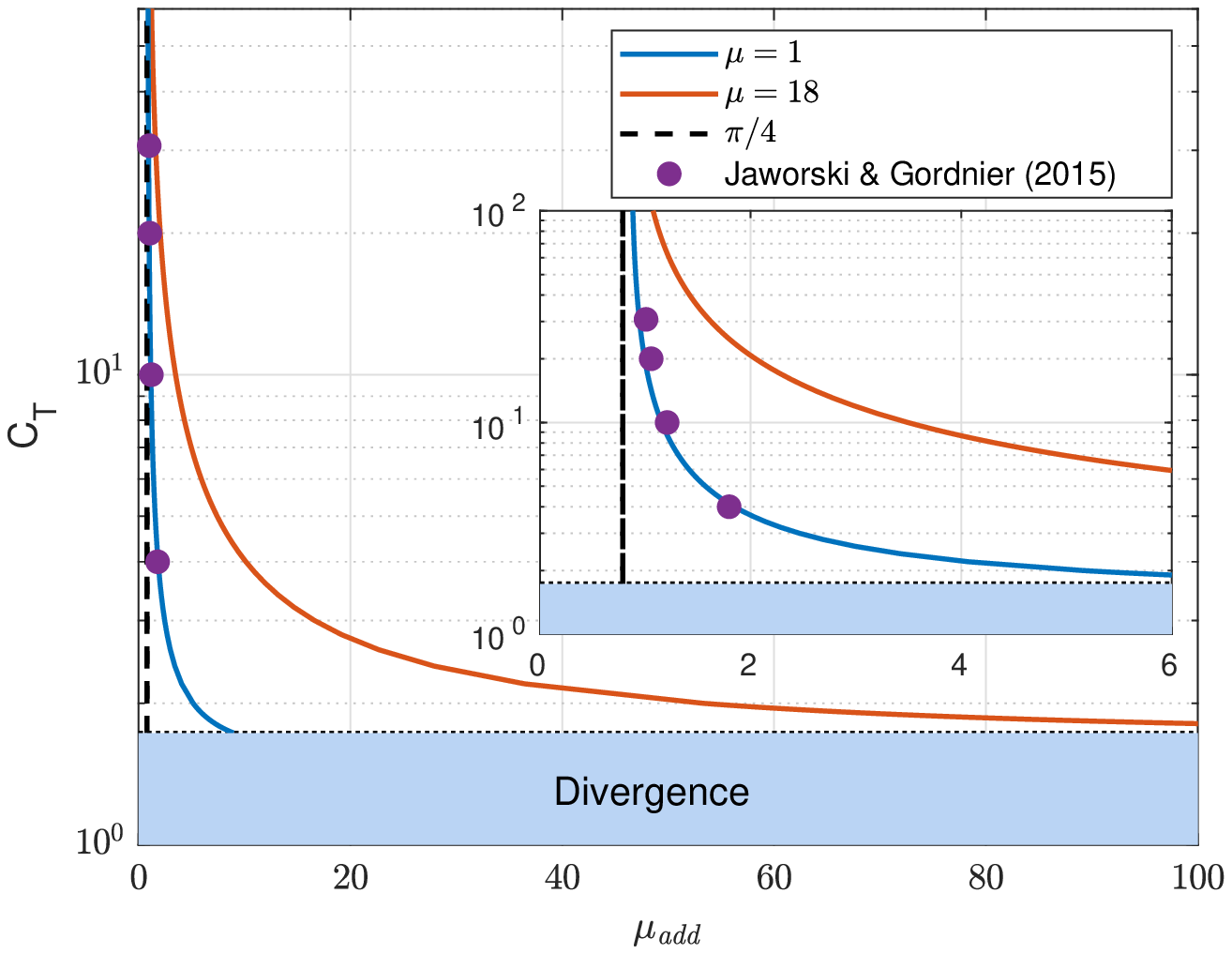}
			\caption{}
			\label{fig:MemHeave_mu1_a_18-Mu_add}
		\end{subfigure}
	    \quad
		\begin{subfigure}[h!]{0.46\textwidth}
		    \includegraphics[width=\textwidth]{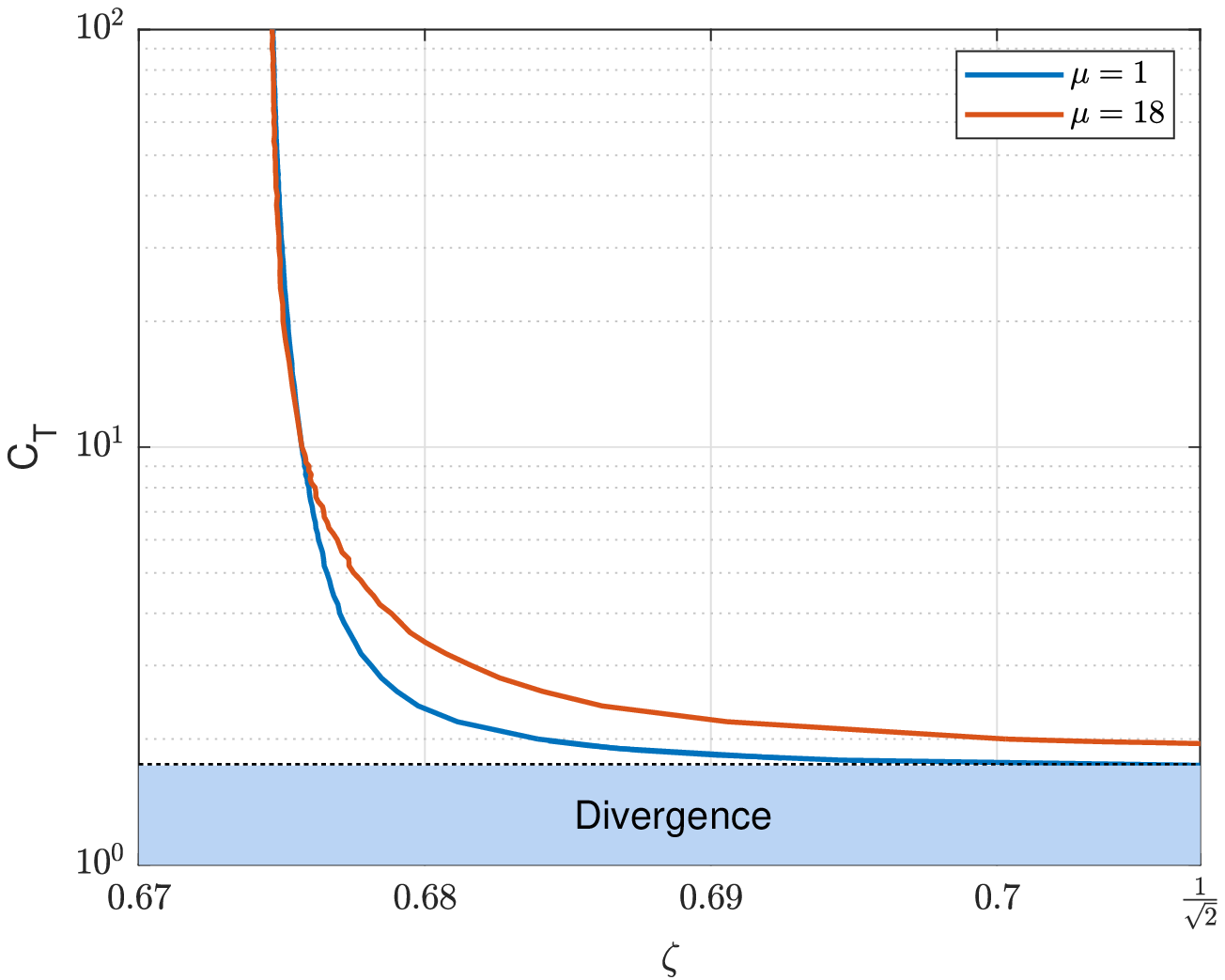}
		    \caption{}
		    \label{fig:MemHeave_mu1_a_18-zeta}
	    \end{subfigure}
		\caption{The dependence of non-dimensional added mass and aerodynamic damping on membrane tension coefficient for membrane mass ratios of $\mu=1, 18$: (a) added mass ratio, $\mu_\mathit{add}$, compared against the rigid plate added mass value of $\pi/4$, and the added mass obtained by \cite{Jaworski2015} for $\mu=1.2065$; (b) damping coefficient, $\zeta$, computed from the frequency ratio at the lift amplitude peak that sustains $\left(k/\omega_{r_1}\right)_\mathit{peak}=\sqrt{1-2\zeta^2}$. The membrane added mass approaches the rigid plate value as $C_T\to\infty$, and is in close agreement with \cite{Jaworski2015}. The aerodynamic damping is effectively constant for all of the examined values of $C_T$, except at the lowest values near the divergence instability threshold, $C_T\cong1.73$, where the damping coefficient approaches $\zeta=1/\sqrt{2}$. Resonance peaks do not occur for $\zeta>1/\sqrt{2}$ \citep[][p.~274]{Rao_book2007}.}
	\label{fig:MemHeave_mu1_a_18}
	\end{center}
\end{figure}

Figure~\ref{fig:MemHeave_mu1_a_18-Mu_add} presents the membrane added mass ratio values for $\mu=1$ and $\mu=18$ as a function of the tension coefficient over $1.73\le C_T\le100$, which are compared against the known rigid plate added mass ratio of $\pi/4$ and the results of \cite{Jaworski2015} obtained for $\mu=1.2065$. It is evident that the added mass ratio increases with the membrane mass ratio and decreases with increase in tension coefficient, where an asymptotic approach to the rigid plate solution as $C_T\to\infty$ is noted.
In addition, a good agreement is obtained with the results computed by \citet[Table 1]{Jaworski2015} assuming quasi-steady aerodynamics. \cite{Jaworski2015} argued that circulatory effects are negligible in the computation of the resonance frequency of the system, and the close agreement in figure~\ref{fig:MemHeave_mu1_a_18-Mu_add} substantiates this claim.

An aerodynamic damping coefficient, $\zeta$, may also be computed for the fluid-loaded membrane using the frequency ratio at the peak of the unsteady lift amplitude value, $\left(k/\omega_{r_1}\right)_\mathit{peak}=\sqrt{1-2\zeta^2}$ \citep[][pp.~271--274]{Rao_book2007}.
Figure~\ref{fig:MemHeave_mu1_a_18-zeta} plots the damping coefficient for two values of mass ratio as a function of the tension coefficient and shows that it is practically constant for $C_T\ge3$. The damping coefficient approaches the limit of $\zeta=1/\sqrt{2}$ as the tension coefficient is further reduced, which is near the divergence instability threshold of $C_T\cong1.73$ \citep{Tiomkin2017}. This limit describes the aerodynamic damping value beyond which no resonance peak is obtained, as would be expected for any harmonically forced linear system \citep[][p. 274]{Rao_book2007}. In general, all of the examined cases possess substantial aerodynamic damping, which explains the finite amplitudes obtained at the fluid-loaded resonance frequency conditions in the present linear analysis.

\begin{figure}
	\begin{center}
	    \begin{subfigure}[]{0.46\textwidth}
			\includegraphics[width=\textwidth]{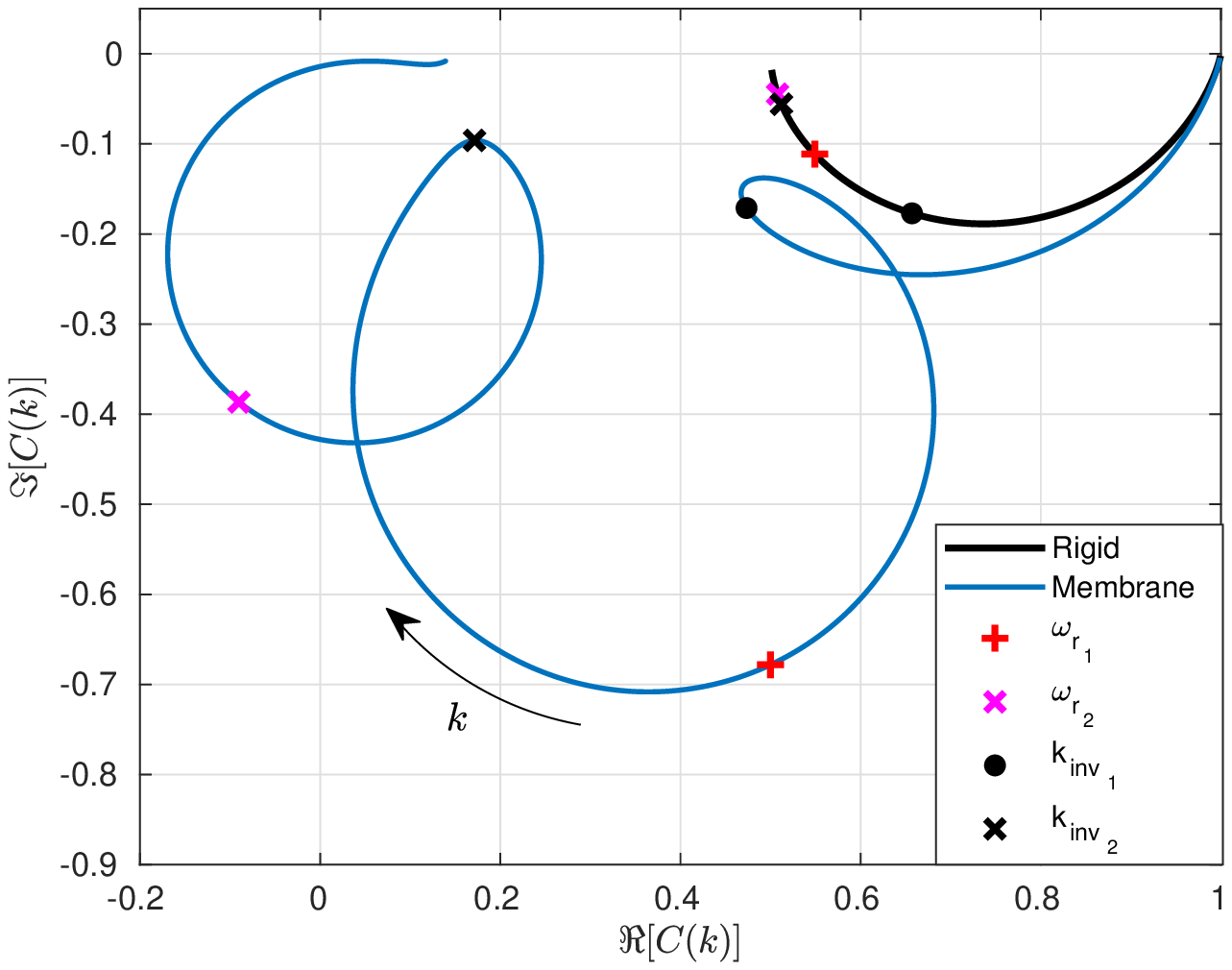}
			\caption{}
			\label{fig:HeaveMem_mu1CT2_5-Theodorsen}
		\end{subfigure}
		\quad
		\begin{subfigure}[]{0.46\textwidth}
		    \includegraphics[width=\textwidth]{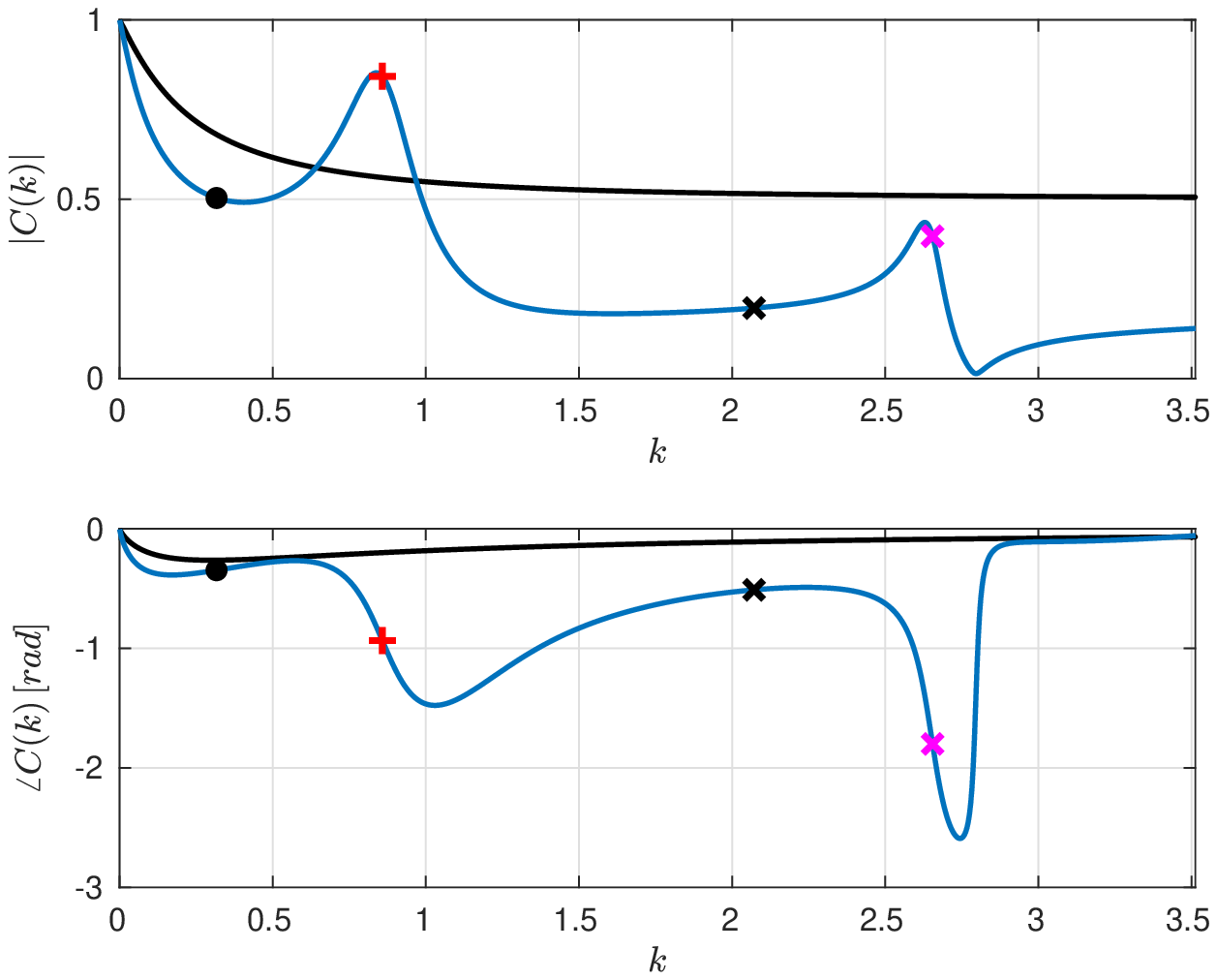}
		    \caption{}
		    \label{fig:HeaveMem_mu1CT2_5-Theodorsen_abs_phase}
	    \end{subfigure}
	    \caption{Membrane lift response to heave oscillations, obtained for $C_T=2.5, \mu=1$ in terms of the membrane equivalent Theodorsen function: (a) Argand diagram; (b) modulus and phase. Frequencies of inflection points are denoted with black circles ($k_{{inv}_1}$) and crosses ($k_{{inv}_1}$), and resonance frequencies are denoted with red pluses ($\omega_{{r}_1}$) and magenta crosses ($\omega_{{r}_2}$).}
	\label{fig:HeaveMem_mu1CT2_5-Lift}
	\end{center}
\end{figure}

The membrane lift response to heave oscillations is next evaluated by comparing the membrane equivalent Theodorsen function \eqref{eq:TheodorsenMem_expFn2} to the standard Theodorsen function of a rigid flat plate.
Figure~\ref{fig:HeaveMem_mu1CT2_5-Lift} illustrates this comparison for the nominal membrane of $\mu=1$ and $C_T=2.5$ as an Argand diagram (figure~\ref{fig:HeaveMem_mu1CT2_5-Theodorsen}) and in terms of modulus and phase (figure~\ref{fig:HeaveMem_mu1CT2_5-Theodorsen_abs_phase}).
This representation of a complex-valued function as a two-dimensional plot is used to describe the unsteady lift amplitude and the phase lag relative to the heaving motion of the aerofoil. When $\Im{[C_m(k)]}<0$ in the Argand diagram, the lift response lags the rigid motion (negative phase), whereas the lift precedes the heaving motion (positive phase) when ${\Im{[C_m(k)]}>0}$.
For low reduced frequencies, the membrane lift response follows the general behaviour of the rigid plate response, with reduced amplitude and increased phase lag. 
As the reduced frequency of heave oscillations increases,
at some point (typically for $k$ smaller than the first resonance frequency) the membrane equivalent Theodorsen function changes its direction abruptly, where the unsteady lift response amplitude increases, rather than converging to zero as it would for rigid aerofoils. We mark this inflection point by reduced frequency $k_{{inv}_1}$. With a further increase in $k$ beyond $k_{{inv}_1}$, a circular path is obtained in the complex plain plot until the next inflection point is reached at $k=k_{inv_2}$, and so on. In figure~\ref{fig:HeaveMem_mu1CT2_5-Theodorsen}, we present results for reduced frequencies up to the second in vacuo natural frequency, $k_2\cong3.5$, for the sake of clarity. Each of these circular arcs contains one of the system's resonance frequencies for which a local maximum is observed in the lift response amplitude (figure~\ref{fig:HeaveMem_mu1CT2_5-Theodorsen_abs_phase}).
A region of special interest is revealed around the first resonance frequency, for $0.65\le k\le 0.96$, where the membrane aerofoil demonstrates a substantial increase in lift magnitude over a rigid aerofoil. Interestingly, oscillations with lower or higher reduced frequency (in the examined range of $k\le 3.5$) result in substantial deficit in unsteady lift amplitude.
Viewed in a practical context, these results for the lift amplitude may be used to extract the maximum unsteady wing load in the design process. Therefore, cases where the flexible membrane presents higher maximum loads than a rigid aerofoil could be hazardous when using predictions of the standard Theodorsen function, for example.

\begin{figure}
	\begin{center}
	    \begin{subfigure}[]{0.46\textwidth}
			\includegraphics[width=\textwidth]{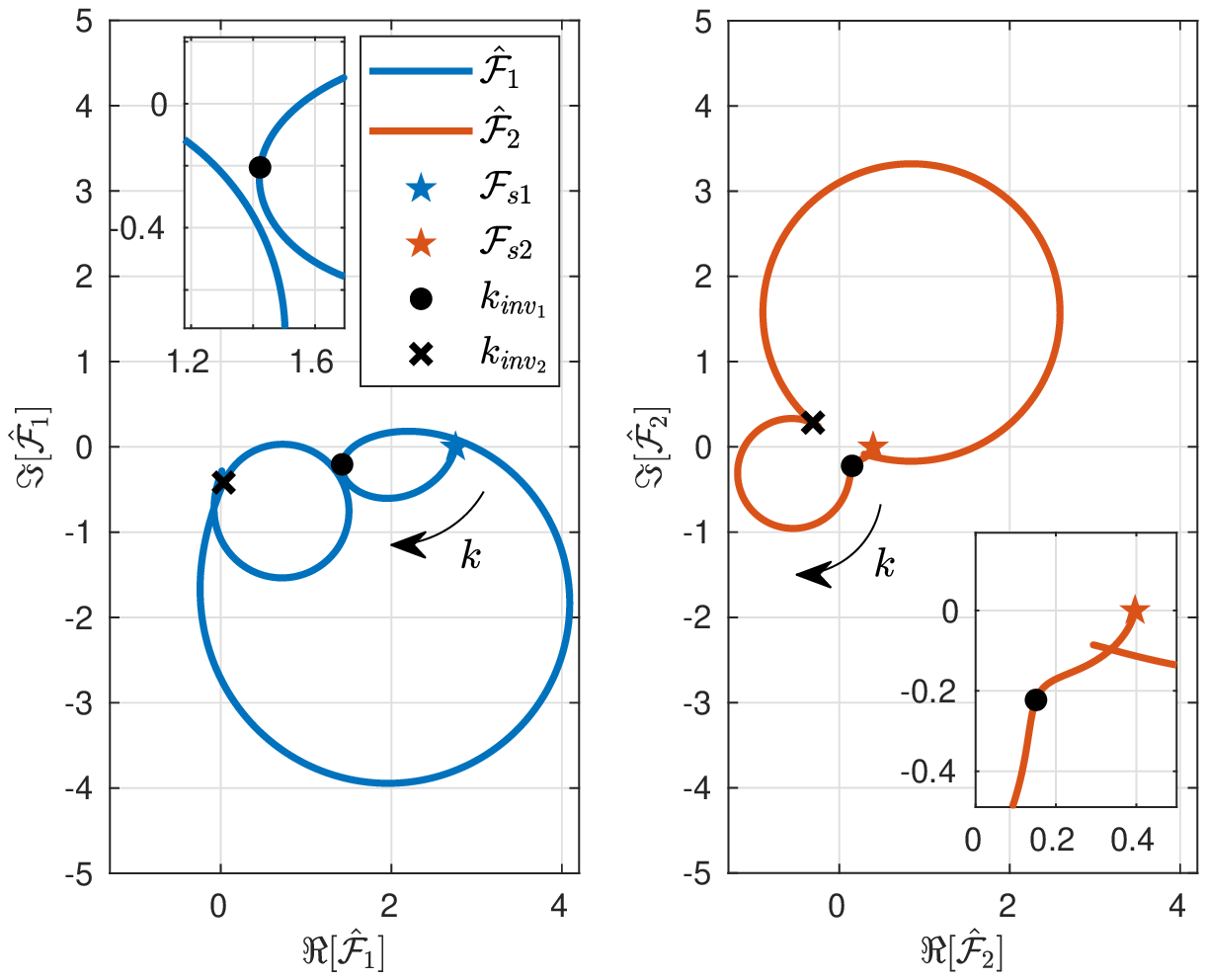}
			\caption{}
			\label{fig:HeaveMem_mu1CT2_5-Fn_ik_cmplx}
		\end{subfigure}
		\quad
		\begin{subfigure}[]{0.46\textwidth}
		    \includegraphics[width=\textwidth]{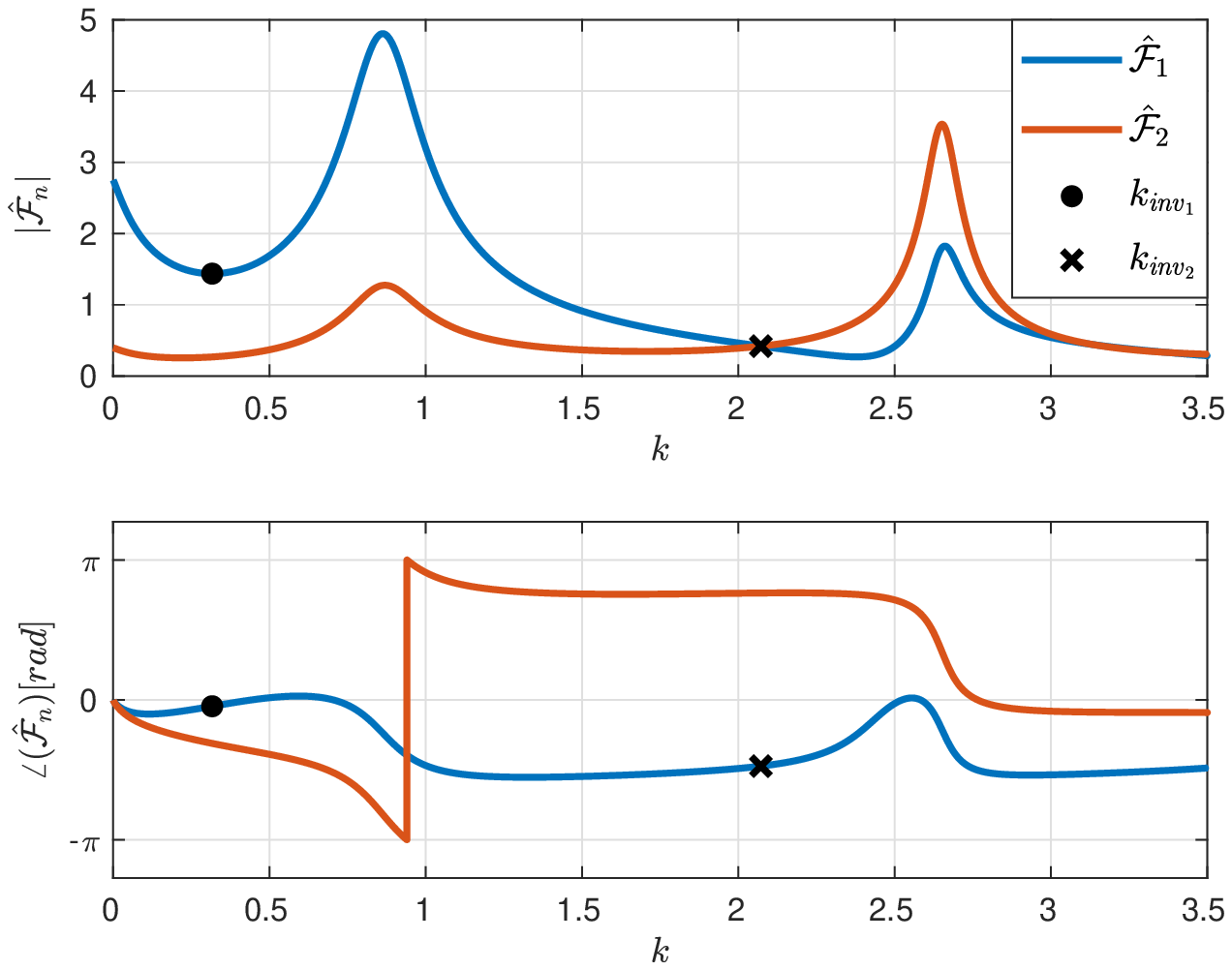}
		    \caption{}
		    \label{fig:HeaveMem_mu1CT2_5-Fn_ik_abs_phase}
	    \end{subfigure}
	    \caption{The first two (most dominant) complex-valued Fourier coefficients, normalised by $\alpha_0=\mi k h_0$, for a nominal membrane of $C_T=2.5, \mu=1$ undergoing harmonic heave oscillations of reduced frequency $k$: (a) Argand diagram; (b) modulus and phase.
	    Static solutions are denoted by pentagram markers and recovered by the unsteady results as $k\rightarrow 0$. First inflection point is marked with black circles and identified by the first local minimum of $|\hat{\mathcal{F}}_1|$. Second inflection point is denoted by black crosses, identified by an intersection between $|\hat{\mathcal{F}}_1|$ and $|\hat{\mathcal{F}}_2|$. The insets in (a) introduce a zoom-in on the first inflection point region for illuminating the trend of both functions as $k$ passes $k_{\mathit{inv}_1}$.}
	\label{fig:HeaveMem_mu1CT2_5-Fn_ik}
	\end{center}
\end{figure}

To further examine the origin of the circular paths in the membrane Theodorsen function, we recall that the equivalent Theodorsen function \eqref{eq:TheodorsenMem_expFn2} is in fact a product of the standard Theodorsen function and a function of the membrane Fourier coefficients. Figure~\ref{fig:HeaveMem_mu1CT2_5-Fn_ik} presents the evolution of the first two Fourier coefficients (normalised by $\alpha_0=\mi k h_0$) with varying reduced frequency, obtained for the nominal membrane case. These normalised Fourier coefficients are the most dominant coefficients in the Fourier series used to describe the membrane slope \eqref{eq:SailSlope-Fourier}, and are plotted as an Argand diagram (figure~\ref{fig:HeaveMem_mu1CT2_5-Fn_ik_cmplx}) and in terms of its modulus and phase (figure~\ref{fig:HeaveMem_mu1CT2_5-Fn_ik_abs_phase}). 
Note that the unsteady solution recovers the static aeroelastic membrane results for $k\rightarrow 0$, as expected, where the static results are marked by pentagrams in the Argand diagram.
As the reduced frequency is increased from the static limit, the amplitudes of both of the normalised Fourier coefficients decrease at first, yielding a smaller amplitude of the oscillating membrane shape, and then increase as the reduced frequency approaches the first fluid-loaded resonance frequency. The first inflection point in the equivalent Theodorsen function corresponds to the first local minimum of $|\hat{\mathcal{F}}_1|$, which is followed by a circular path in the complex plane plots of all coefficients (figure~\ref{fig:HeaveMem_mu1CT2_5-Fn_ik_cmplx}). 
This entire frequency regime, in which the first circle appears in the Fourier coefficients, is dominated by the membrane's first~mode, as is evident by the dominance of the first Fourier coefficient in figure~\ref{fig:HeaveMem_mu1CT2_5-Fn_ik_abs_phase}.
This dominance is maximum near the first resonance frequency and diminishes as $k$ increases, which continues until the second resonance frequency is approached and the second mode coefficient becomes dominant. The frequency at which $|\hat{\mathcal{F}}_2|$ crosses $|\hat{\mathcal{F}}_1|$ is in fact $k_{{inv}_2}$, which marks the transition from the first circle to the second circle in the equivalent Theodorsen function in figure~\ref{fig:HeaveMem_mu1CT2_5-Theodorsen}. This behaviour, in which the inflection points are identified by a switch of dominance between the membrane modes, also continues to higher modes as the frequency is further increased, but is not shown here for the sake of clarity and brevity. 
Thus, it can be concluded that the circular arcs in the equivalent Theodorsen function are due to the membrane dynamic response, where each circle is related to a different dominant mode, and the inflection points between circles occur at the intersection between the modulus functions of two consecutive normalised Fourier coefficients.

\begin{figure}
	\begin{center}
	   \begin{subfigure}[]{0.46\textwidth}
			\includegraphics[width=\textwidth]{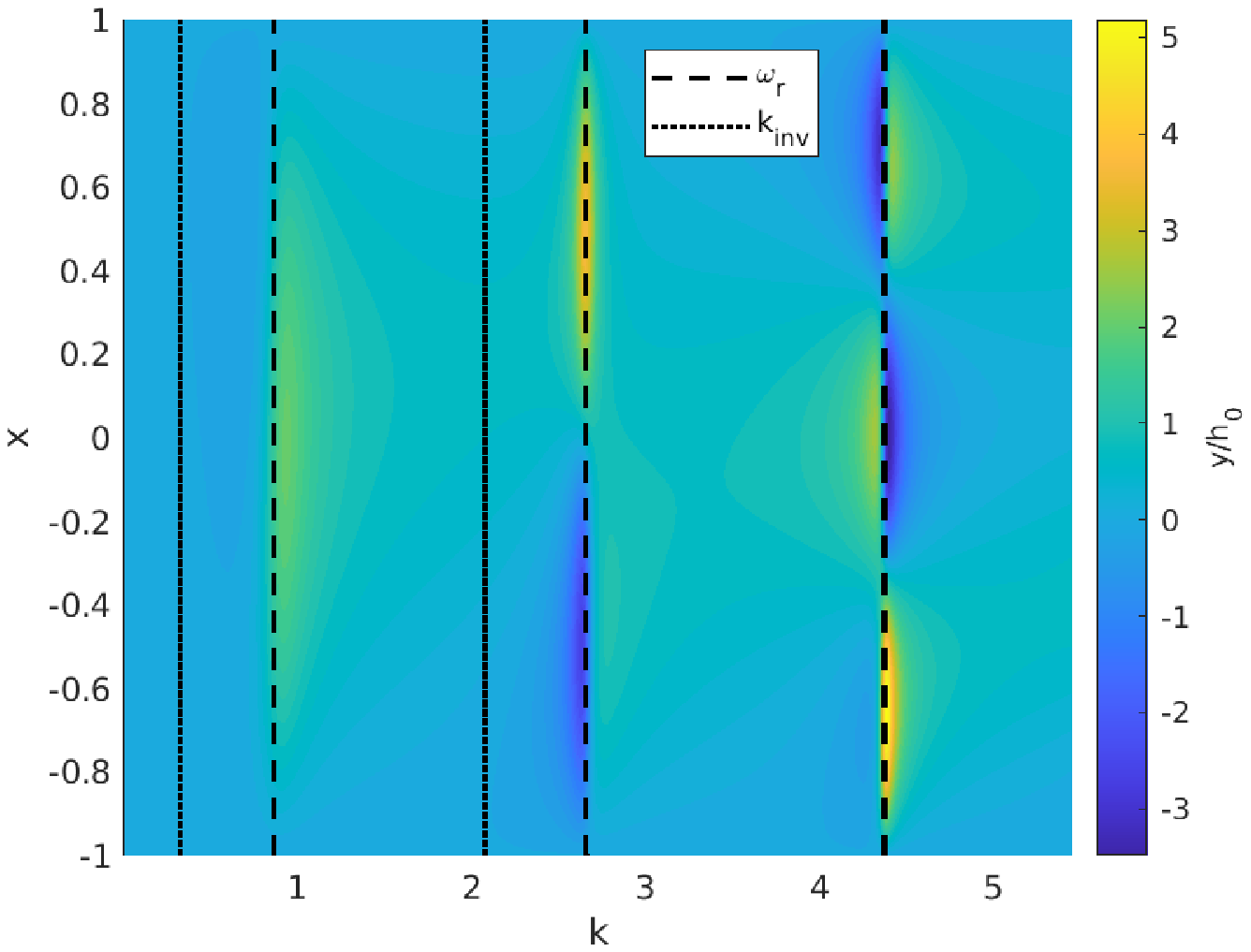}
			\caption{}
			\label{fig:HeaveMem_mu1CT2_5-MemContourf}
		\end{subfigure}
		\\
	    \begin{subfigure}[]{0.46\textwidth}
			\includegraphics[width=\textwidth]{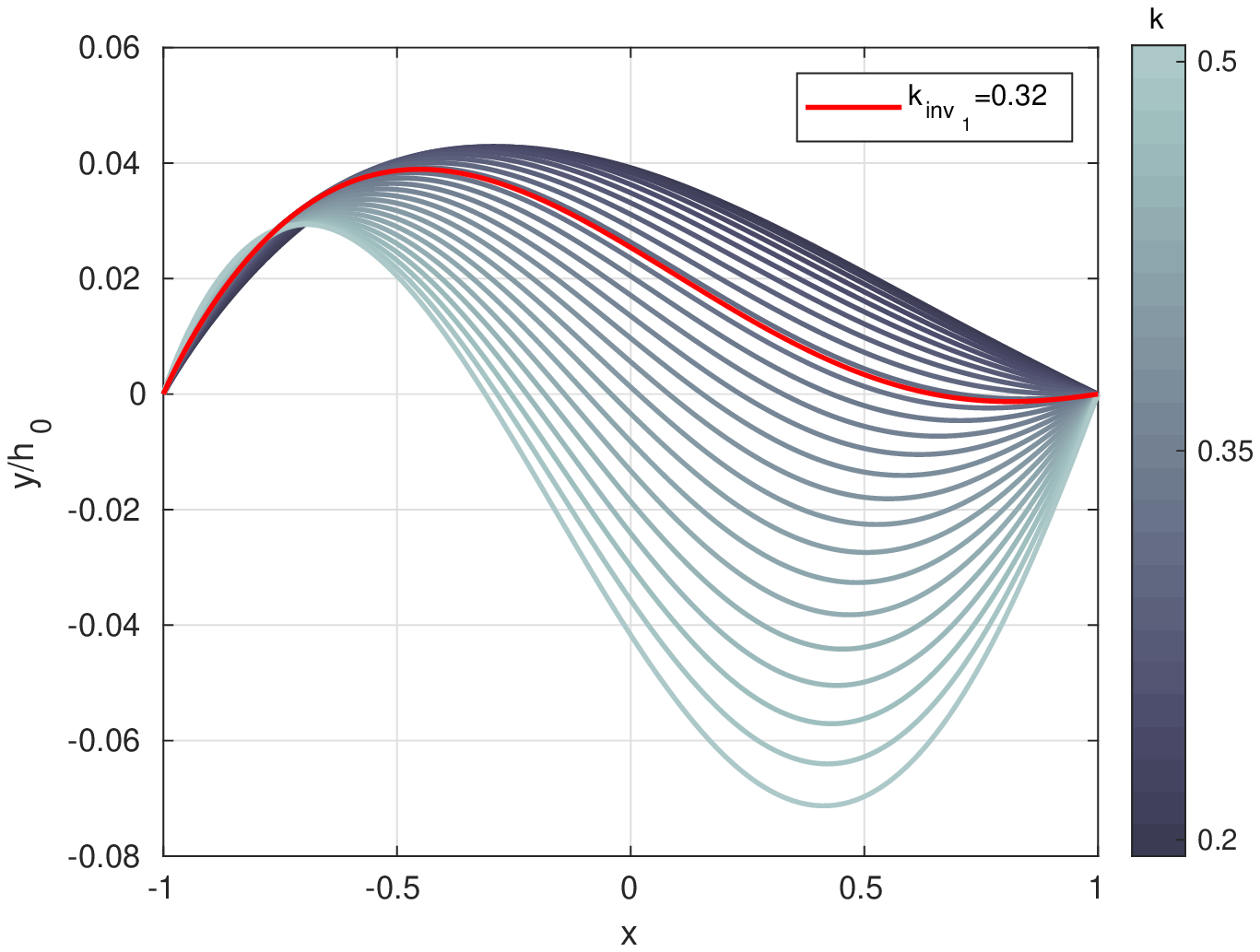}
			\caption{}
			\label{fig:HeaveMem_mu1CT2_5-Mem_kinv1}
		\end{subfigure}
	    \quad
		\begin{subfigure}[]{0.46\textwidth}
		    \includegraphics[width=\textwidth]{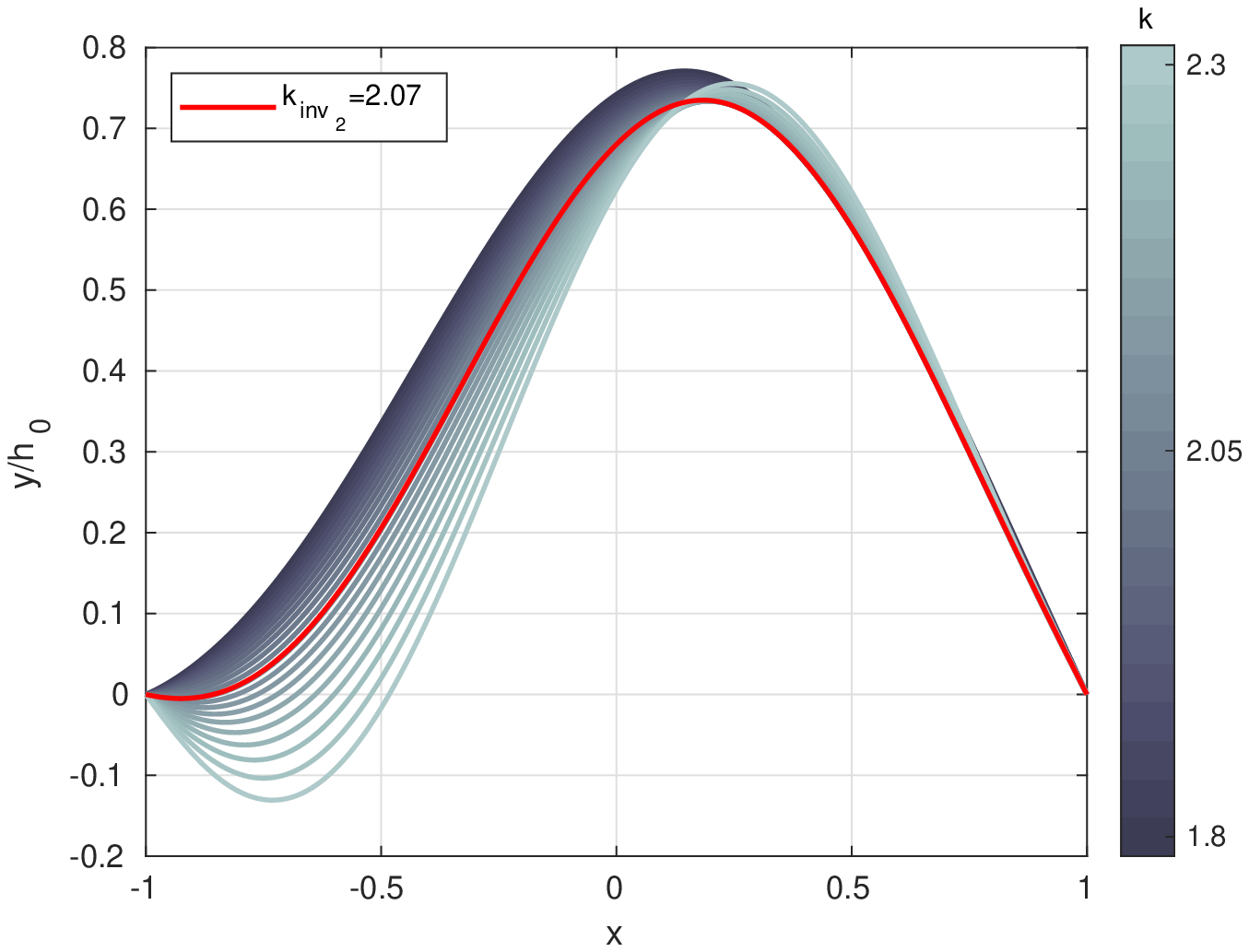}
		    \caption{}
		    \label{fig:HeaveMem_mu1CT2_5-Mem_kinv2}
	    \end{subfigure}
	    \caption{Membrane dynamic response to harmonic heave oscillations, obtained for a nominal membrane of $C_T=2.5, \mu=1$: (a) amplitude profiles computed for varying values of reduced frequency. Resonance frequencies are denoted with dashed black lines, and inflection point frequencies are marked with dotted black lines. (b) and (c) present the membrane amplitude profiles obtained for reduced frequencies around the first and second inflection points, respectively.}
	\label{fig:HeaveMem_mu1CT2_5-Mem}
	\end{center}
\end{figure}

Figure~\ref{fig:HeaveMem_mu1CT2_5-Mem} presents the membrane amplitude profiles computed for a nominal membrane undergoing heave oscillations of varying reduced frequency, where the profile represents the amplitude of oscillation at every point along the membrane chord.
A contour plot of the amplitude profiles is presented in figure~\ref{fig:HeaveMem_mu1CT2_5-MemContourf} for varying reduced frequency, $k$, with black dashed and dotted lines denoting the resonance and inflection point frequencies, respectively. 
Small deformations relative to the heave amplitude are obtained along the entire frequency range, except in the vicinity of the system's (fluid-loaded) resonance frequencies, for which large-amplitude profiles are observed with a shape similar to the membrane structural modes.
This observation suggests a close coupling between the unsteady lift amplitude and the membrane amplitude in response to harmonic heave oscillations. Furthermore, a favorable lift is clearly achievable only for odd resonance frequencies ($\omega_{r_1}, \omega_{r_3}, ... $), which correspond to shapes that are symmetric around the mid-chord point, rather than anti-symmetric in the even-mode cases, as illustrated by comparing the lift response in figure~\ref{fig:HeaveMem_mu1CT2_5-Theodorsen_abs_phase} with the membrane amplitude profiles in figure~\ref{fig:HeaveMem_mu1CT2_5-MemContourf}.

Figures~\ref{fig:HeaveMem_mu1CT2_5-Mem_kinv1} and \ref{fig:HeaveMem_mu1CT2_5-Mem_kinv2} provides a more detailed view of the membrane amplitude profiles obtained for reduced frequencies near $k_{{inv}_1}$ and $k_{{inv}_2}$, respectively.
In accordance with the normalised Fourier coefficients (figure~\ref{fig:HeaveMem_mu1CT2_5-Fn_ik}), we note that small-amplitude profiles are obtained for small values of reduced frequency (figure~\ref{fig:HeaveMem_mu1CT2_5-Mem_kinv1}). These membrane shapes are convex, with a maximum camber point at the fore section of the aerofoil, in accordance with the static membrane solution \citep{Nielsen1963}. As the reduced frequency increases, the amplitude of the profile decreases until for $k_{{inv}_1}$ an inflection point appears in the membrane profile and drastic changes in the profile shape are evoked with further increase in reduced frequency.
These deformations indicate the excitation of the membrane structural modes as the reduced frequency approaches the system's first resonance frequency.
For larger reduced frequencies near the second inflection point, $k_{{inv}_2}$, figure~\ref{fig:HeaveMem_mu1CT2_5-Mem_kinv2} shows that significant membrane oscillations are evoked, in which the inflection point ($k_{{inv}_2}$) represents the shift in the membrane amplitude profile, from the fully convex shape obtained for the first resonance frequency to the second mode shape obtained for the second resonance frequency. This shift is identified by an inflection point that appears near the leading edge of the membrane profile for $k\ge k_{{inv}_2}$, after which the second mode of the membrane becomes most dominant. 
This result is in accordance with the behaviour of the Fourier coefficients, presented in figure~\ref{fig:HeaveMem_mu1CT2_5-Fn_ik}, affirming the conclusion that the inflection points in the complex plane plot of the equivalent Theodorsen function mark the shift in dominance between consecutive membrane modes.

\begin{figure}
	\begin{center}
	    \begin{subfigure}[h!]{0.46\textwidth}
			\includegraphics[width=\textwidth]{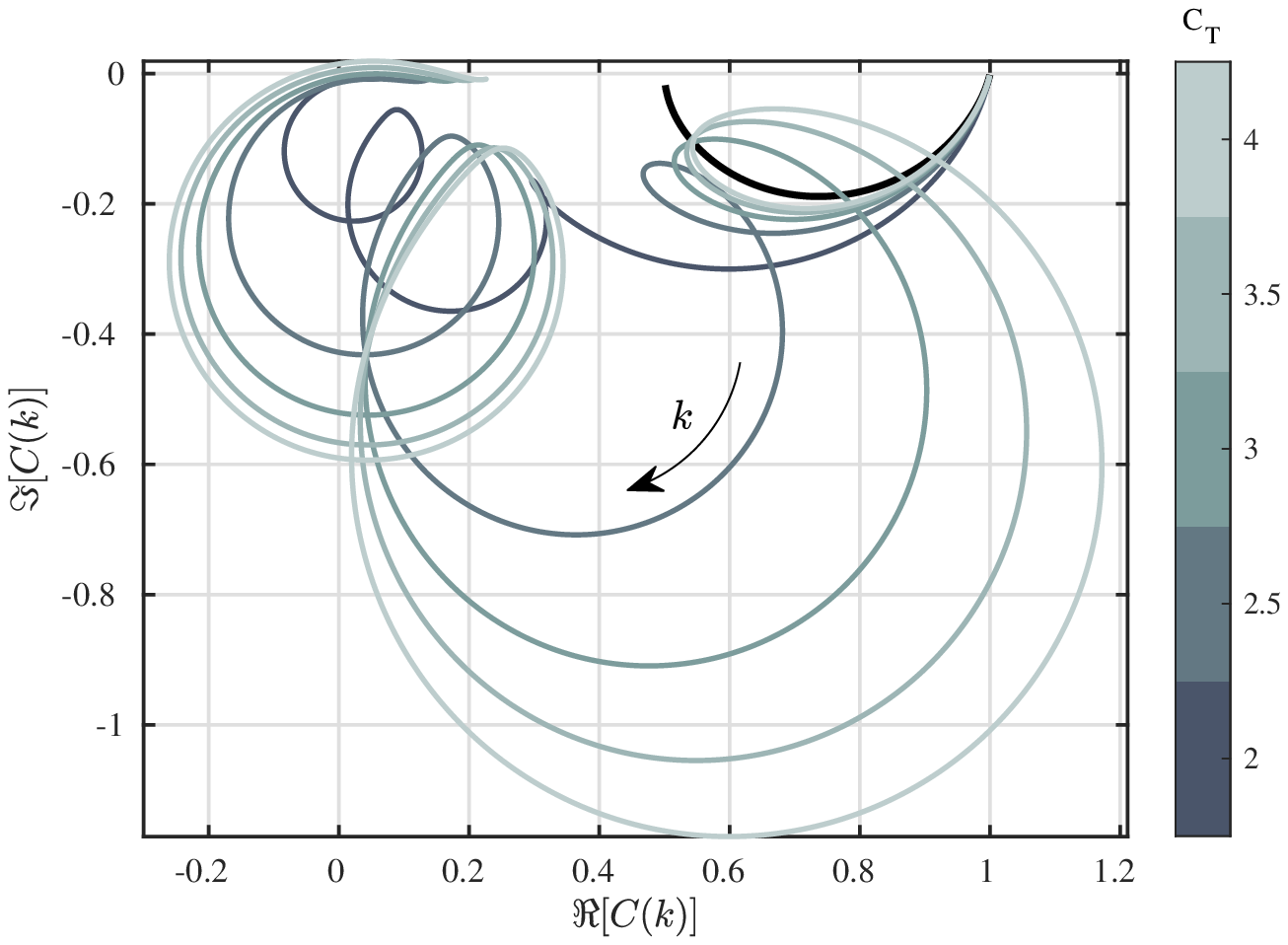}
			\caption{}
			\label{fig:TheodorsenMemLight-CTeffect_complx}
		\end{subfigure}
	    \quad
		\begin{subfigure}[h!]{0.46\textwidth}
		    \includegraphics[width=\textwidth]{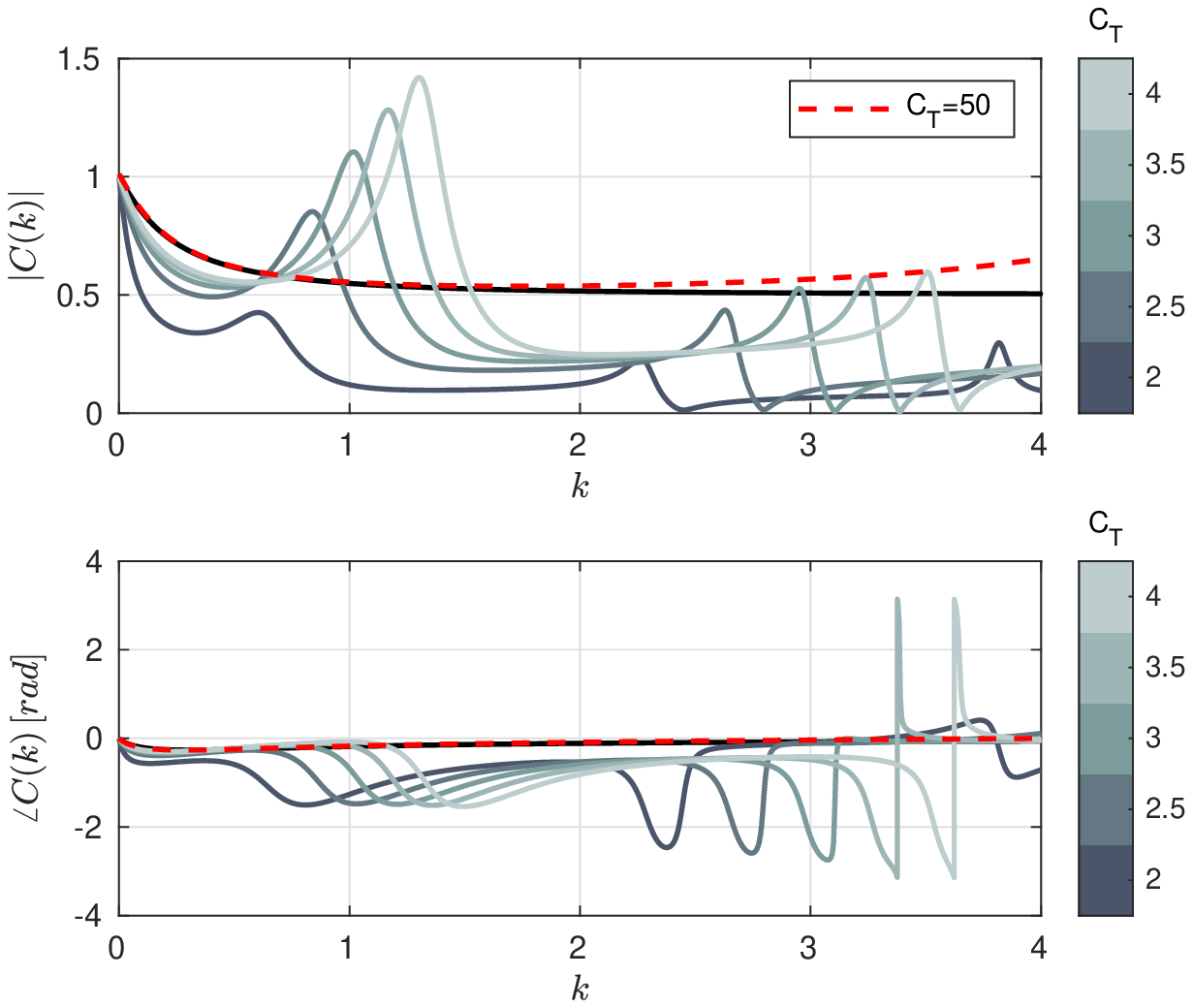}
		    \caption{}
		    \label{fig:TheodorsenMemLight-CTeffect_amp_phase}
	    \end{subfigure}
		\caption{Effect of membrane tension coefficient on the membrane equivalent Theodorsen function for $\mu=1$: (a) Argand diagram; (b) modulus and phase. The rigid-plate Theodorsen function is presented with a black line for comparison with the membrane equivalent Theodorsen function. The solution for a very large tension coefficient of $C_T=50$ is presented with a red dashed line in (b) and indicates asymptotic convergence of the unsteady solution to the rigid plate solution as $C_T \rightarrow\infty$.}
	\label{fig:TheodorsenMemLight-CTeffect}
	\end{center}
\end{figure}

\begin{figure}
	\begin{center}
	    \begin{subfigure}[h!]{0.46\textwidth}
			\includegraphics[width=\textwidth]{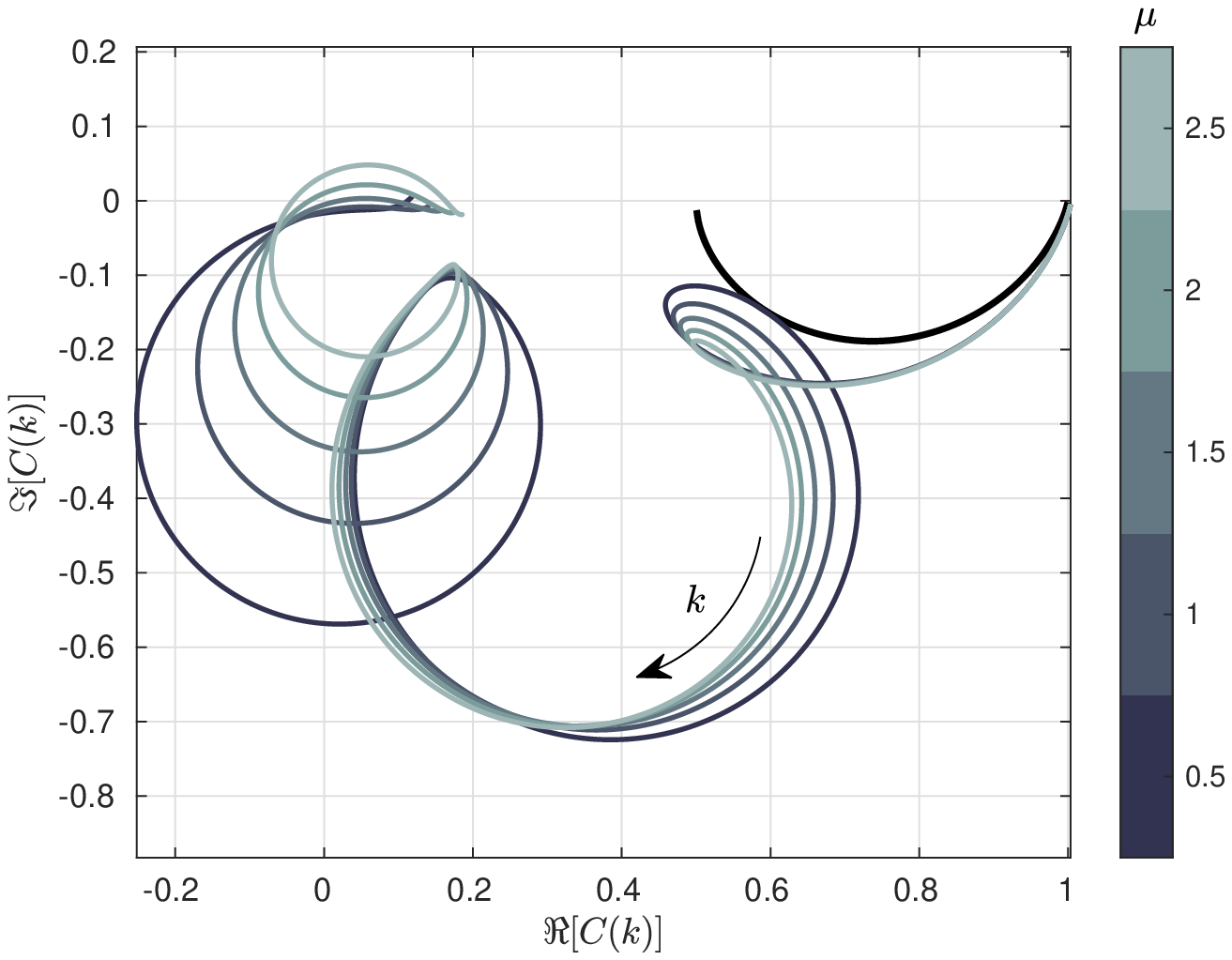}
			\caption{}
			\label{fig:TheodorsenMemLight-Mueffect_complx}
		\end{subfigure}
	    \quad
		\begin{subfigure}[h!]{0.46\textwidth}
		    \includegraphics[width=\textwidth]{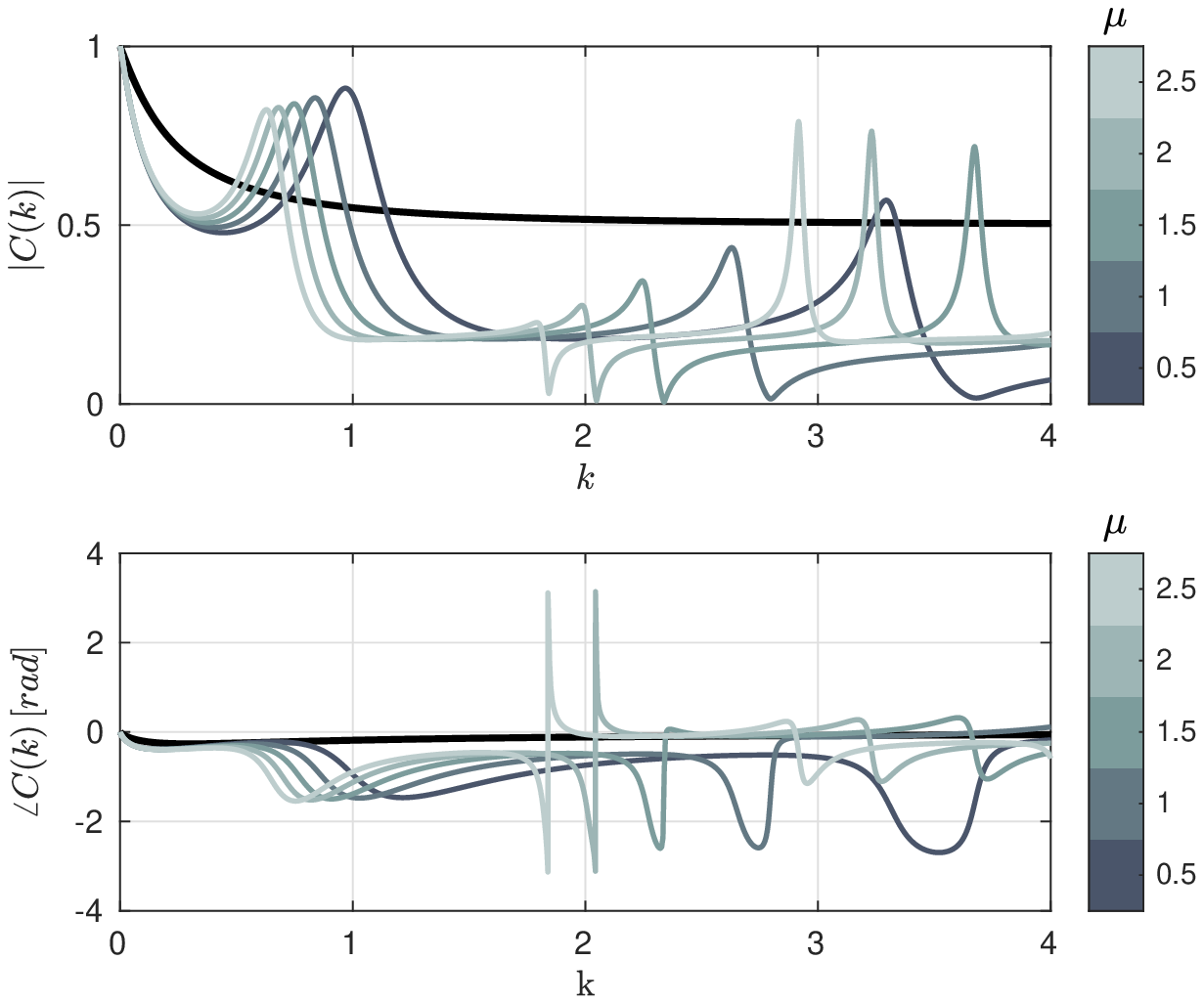}
		    \caption{}
		    \label{fig:TheodorsenMemLight-Mueffect_amp_phase}
	    \end{subfigure}
		\caption{Effect of membrane mass ratio on the membrane equivalent Theodorsen function for $C_T=2.5$: (a) Argand diagram; (b)modulus and phase (b). The rigid-plate Theodorsen function is presented with a black line for comparison with the membrane equivalent Theodorsen function.}
	\label{fig:TheodorsenMemLight-Mueffect}
	\end{center}
\end{figure}

Figures~\ref{fig:TheodorsenMemLight-CTeffect} and \ref{fig:TheodorsenMemLight-Mueffect} illustrate the separate effects of the tension coefficient and the membrane mass ratio on the membrane Theodorsen function. Tension coefficients between $2$ and $4$, and mass ratios between $0.5$ and $2.5$ are chosen  to represent realistic membrane wings \citep[e.g.,][]{Rojratsirikul2010,Tiomkin2021}, while still remaining in the membrane stable regime, as the membrane loses stability via divergence for {$C_T<1.73$} and loses stability via flutter only in the case of heavy membranes of {$\mu>18.8$} \citep[see][for details]{Tiomkin2017}.
Results are presented for reduced frequencies up to the second in vacuo frequency $(k=k_2)$ for the sake of clarity.
The membrane stiffness is strongly influenced by the tension coefficient, and figure~\ref{fig:TheodorsenMemLight-CTeffect_amp_phase} shows that the membrane lift response to low-frequency oscillations approaches the rigid plate response with increasing tension coefficient, as expected.
This result is further validated by examining an extreme case of $C_T=50$, presented in figure~\ref{fig:TheodorsenMemLight-CTeffect_amp_phase} with a dashed red line.
For this high tension coefficient the membrane is practically rigid, and indeed the resulting equivalent Theodorsen function follows closely the standard Theodorsen function for a wide range of frequencies up to about $k=2$, where differences in the amplitudes arise due to parametric proximity to the system's first resonance frequency. Because the system's resonance frequencies increase with $C_T$, the range of frequencies for which the equivalent Theodorsen function follows the standard Theodorsen function increases with $C_T$ as well, and the first resonance circle occurs at a larger reduced frequency (i.e., $k_{{inv}_1}$ increases).
In addition, the circle diameter increases with $C_T$ for $\mu=1$, indicating that an increase in tension coefficient leads locally to an increased amplitude of the unsteady lift response near the resonance frequency, as the membrane oscillation amplitude also increases (figure~\ref{fig:HeaveSSMem_CT_vs_k-mu1}).
Thus, in a practical sense, the hazardous region where a substantial increase in unsteady lift amplitude is obtained is clearly controlled by the tension coefficient, suggesting the possibility of optimizing the flapping wing performance by controlling the tension along the membrane.

Analysis of the mass ratio effect on the membrane response to prescribed heave oscillations (figure~\ref{fig:TheodorsenMemLight-Mueffect}) shows that at small reduced frequencies the mass ratio has practically no effect on the lift response, as expected.
For higher reduced frequencies the first resonance circle appears earlier (at lower $k$) as the mass ratio is increased, in accordance with the decrease in resonance frequency, while the amplitude of the lift response is practically unaffected. 
The hazardous region is also controlled by the mass ratio, as the peak in the lift amplitude follows the movement of the resonance frequency. However, in practical applications, this variable is often harder to control than the tension coefficient.

\begin{figure}
	\begin{center}
	    \begin{subfigure}[h!]{0.46\textwidth}
			\includegraphics[width=\textwidth]{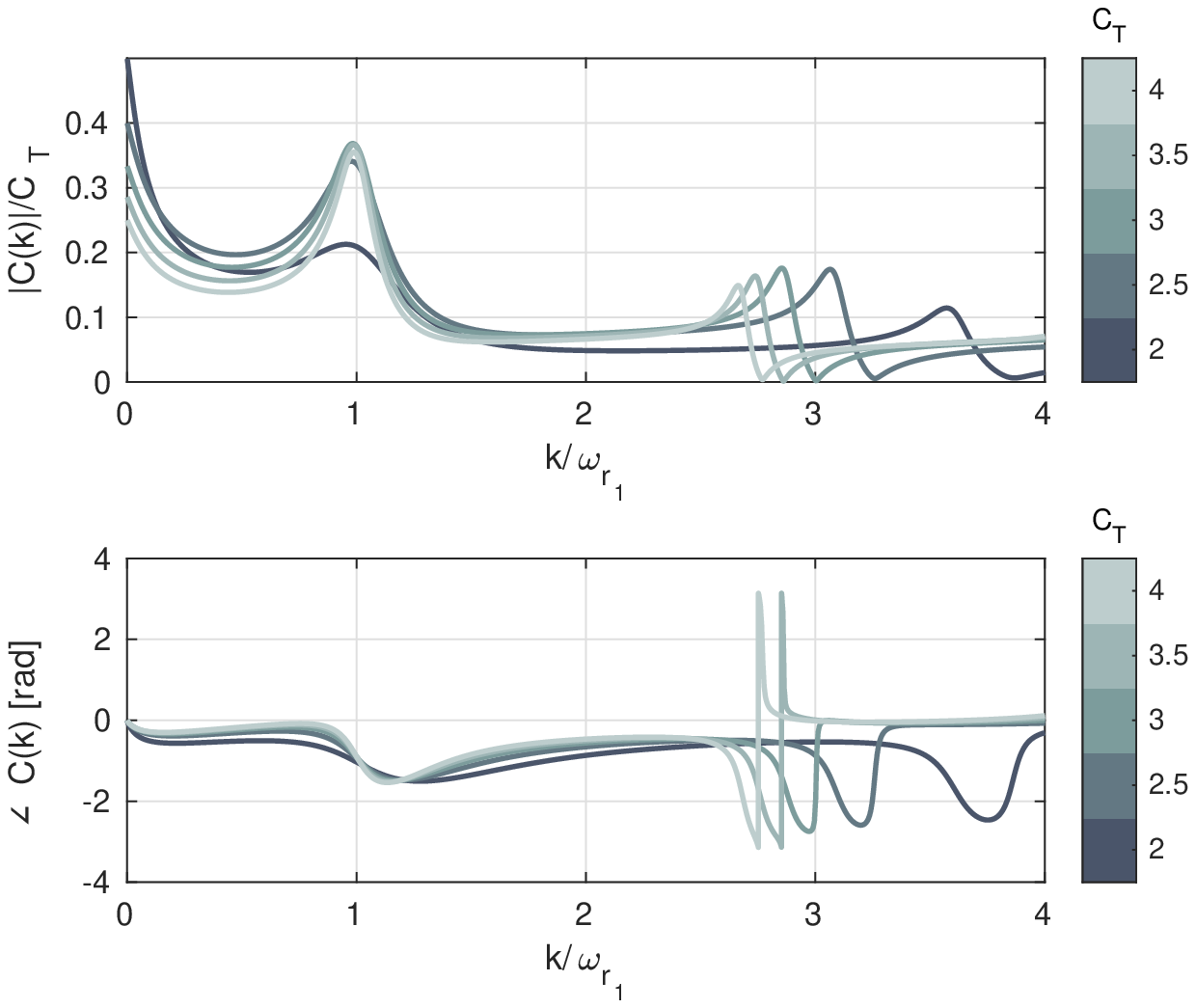}
			\caption{}
			\label{fig:TheodorsenMem_vs_k_wr1-CTeffect}
		\end{subfigure}
	    \quad
		\begin{subfigure}[h!]{0.46\textwidth}
		    \includegraphics[width=\textwidth]{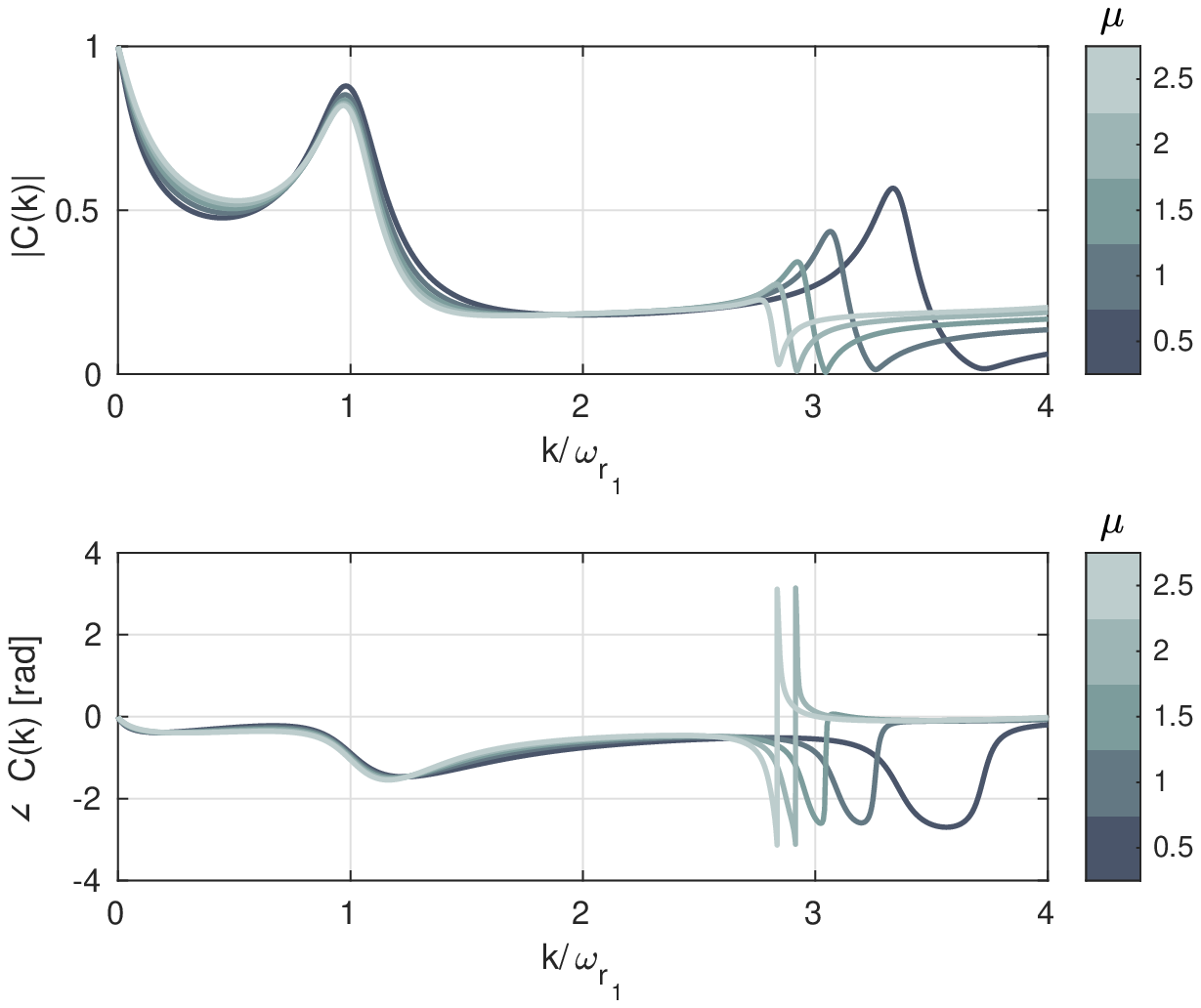}
		    \caption{}
		    \label{fig:TheodorsenMem_vs_k_wr1-Mueffect}
	    \end{subfigure}
		\caption{Effect of membrane tension coefficient (a) and mass ratio (b) on the membrane-equivalent Theodorsen function, obtained for $\mu=1$ and $C_T=2.5$, respectively. The modulus and phase of the equivalent Theodorsen functions are plotted against the reduced frequency of the prescribed motion, normalized by the first fluid-loaded resonance frequency. Aerodynamic damping leads to a finite peak in the modulus of the equivalent Theodorsen function, which scales on $C_T$, and to a peak frequency that is smaller than the fluid-loaded resonance frequency.}
	\label{fig:TheodorsenMem_vs_k_wr1}
	\end{center}
\end{figure}

Some insight into the scaling of the membrane lift response to harmonic heave oscillations is gleaned from figures~\ref{fig:TheodorsenMem_vs_k_wr1-CTeffect} and \ref{fig:TheodorsenMem_vs_k_wr1-Mueffect}, which present the effects of the tension coefficient and the membrane mass ratio on the membrane Theodorsen function, respectively, as a function of the normalised reduced frequency, $k/\omega_{r_1}$. 
For varying mass ratio, all of the examined cases collapse to a single curve for reduced frequency ratios up to $k/\omega_{r_1}\cong2.6$, beyond which the second fluid-loaded resonance peak is approached.
Variation in the tension coefficient shows that the membrane Theodorsen function modulus peak at the first fluid-loaded resonance frequency is linearly proportional to the tension coefficient for all membranes of $C_T\ge 2.5$; note that the modulus is normalised by $C_T$ in figure~\ref{fig:TheodorsenMem_vs_k_wr1-CTeffect}. This dependence of the amplitude peak on $C_T$ suggests that the aerodynamic damping that controls the peak amplitude is effectively constant for $C_T\ge 2.5$, as was shown in figure~\ref{fig:MemHeave_mu1_a_18-zeta} for $\mu=1$. For lower values of the tension coefficient, a larger aerodynamic damping is obtained ($\zeta\to1/\sqrt{2}$), and the amplitude peak reduces significantly.

\subsubsection{Step angle of attack}\label{sec:results-pres-AoA}

\begin{figure}
	\begin{center}
	    \begin{subfigure}[]{0.46\textwidth}
			\includegraphics[width=\textwidth]{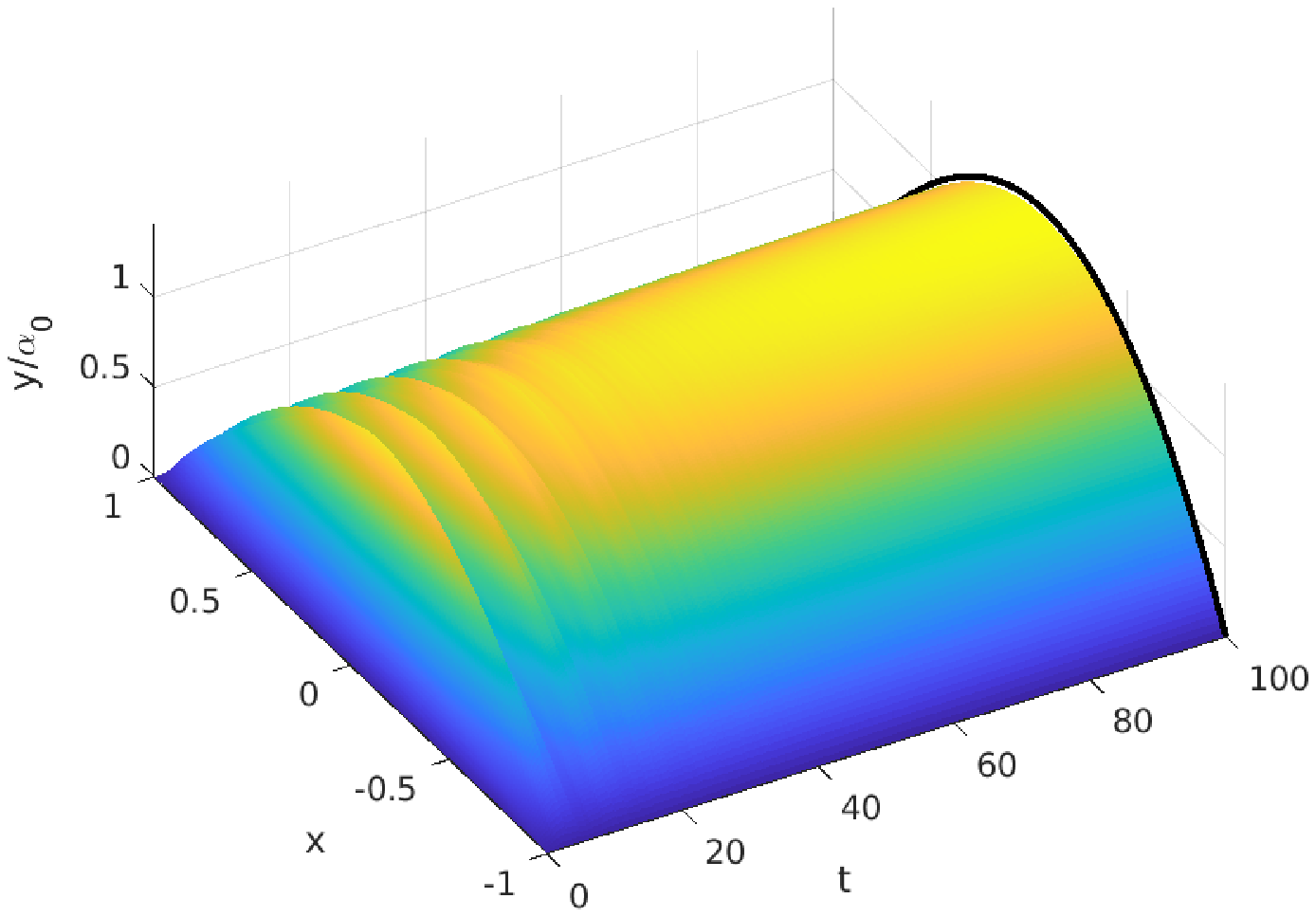}
			\caption{}
			\label{fig:MemAoAstepRes_mu1_CT2_5-Mem}
		\end{subfigure}
		\quad
		\begin{subfigure}[]{0.46\textwidth}
		    \includegraphics[width=\textwidth]{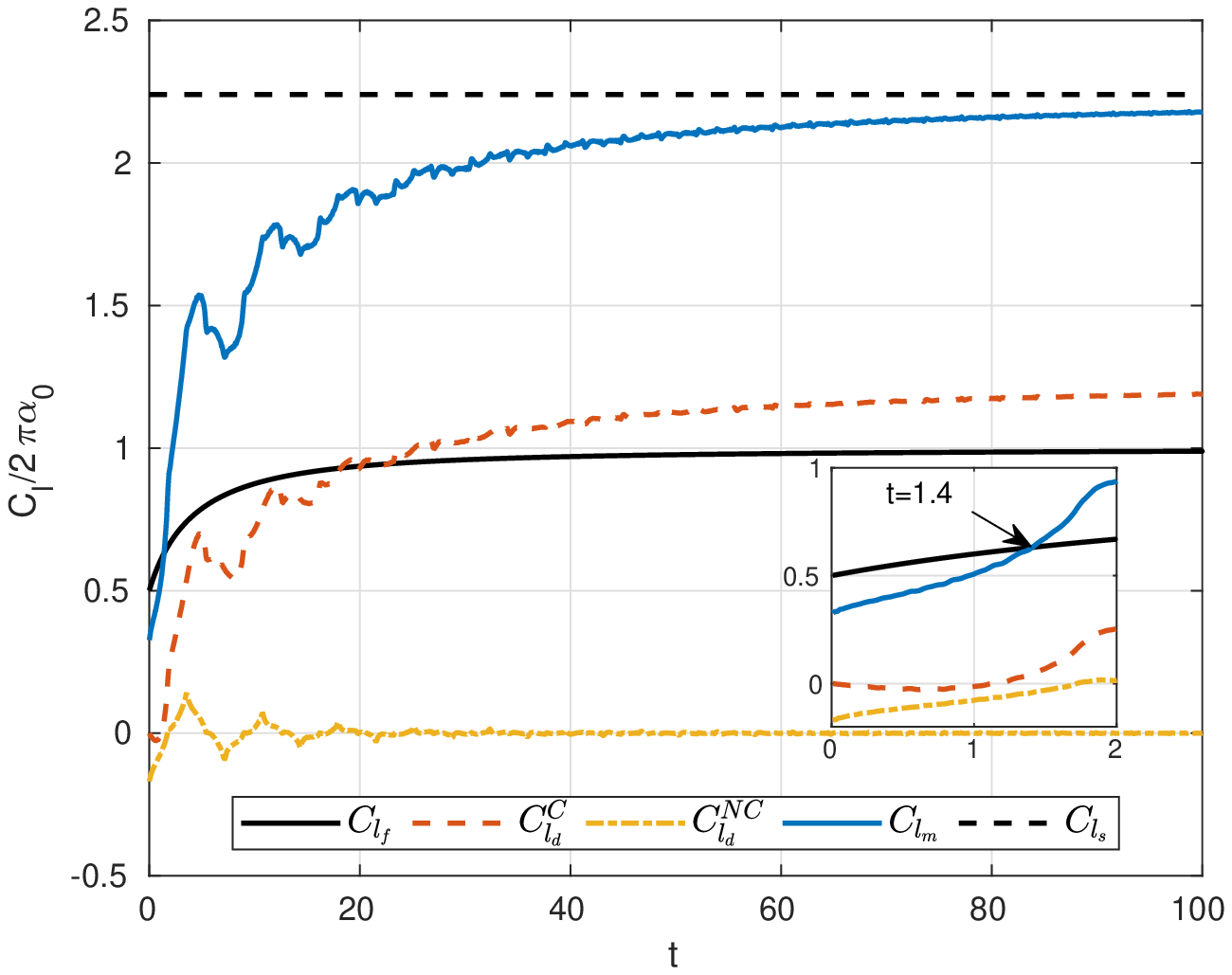}
		    \caption{}
		    \label{fig:MemAoAstepRes_mu1_CT2_5-CL}
	    \end{subfigure}
		\caption{Membrane dynamic response (a) and lift response (b) to a step in angle of attack, obtained for a nominal membrane of $C_T=2.5$ and $\mu=1$. Black line in (a) and dashed black line in (b) denote the static solution \citep[][]{Nielsen1963}. The membrane unsteady lift coefficient, $C_{l_m}$, is computed by superposition between the rigid plate indicial lift, $C_{l_f}$, and the lift due to membrane deformation, $C_{l_d}$, which is composed of the circulatory and non-circulatory terms, $C_{l_d}^C$ and $C_{l_d}^\mathit{NC}$, respectively.}
	    \label{fig:MemAoAstepRes_mu1_CT2_5}
	\end{center}
\end{figure}

We next examine the response of the nominal membrane to a step in angle of attack, in terms of its dynamic and aerodynamic response (figure~\ref{fig:MemAoAstepRes_mu1_CT2_5}). The membrane dynamic response begins with an initially taut profile, followed by membrane oscillations as a result of the abrupt change in angle of attack. These oscillations decrease in amplitude with time until a steady-state profile is obtained that is identical to the respective static solution (figure~\ref{fig:MemAoAstepRes_mu1_CT2_5-Mem}). The resulting lift coefficient history (figure~\ref{fig:MemAoAstepRes_mu1_CT2_5-CL}) presents a similar trend to the membrane deformation history and suggests a close coupling between the two. 
In addition, when comparing the nominal membrane lift history in response to a step in angle of attack, $C_{l_m}$, with the rigid plate response, $C_{l_f}$, we see that the membrane wing achieves a larger lift across almost the entire response, converging to a value more than double that of the rigid-plate lift, due to membrane camber.
Interestingly, during the initial transient stage of the response (for $t<1.4$) the membrane lift is lower than the rigid plate lift. 
Namely, the membrane deformation due to the abrupt change in angle of attack produces negative lift, as evident by the plot of $C_{l_d}^C$ and $C_{l_d}^{\mathit{NC}}$ in figure~\ref{fig:MemAoAstepRes_mu1_CT2_5-CL}, which describe the circulatory and non-circulatory terms in $C_{l_d}$ \eqref{eq:Cl_d}.
As the membrane is initially still and taut, the membrane surface accelerates in response to the sudden change in flow conditions, leading to a negative apparent mass lift at $t=0$ and zero circulatory lift. 
The membrane inflates with time and its acceleration reduces, which yields a decrease in the non-circulatory lift magnitude.
The initial acceleration subsequently yields a negative circulatory lift, but it also causes an increase in the membrane velocity, which increases the circulatory lift.
This trend continues until at time $t=1.4$ the circulatory lift is able to compensate the lift deficit due to $C_{l_d}^{\mathit{NC}}$ and the membrane lift exceeds the rigid plate lift. For longer times the membrane deformation yields a higher circulatory lift that contributes to the further increase in the total membrane lift, as it proceeds to converge to the static membrane lift coefficient, $C_{l_s}$.

\begin{figure}
	\begin{center}
	    \begin{subfigure}[]{0.46\textwidth}
			\includegraphics[width=\textwidth]{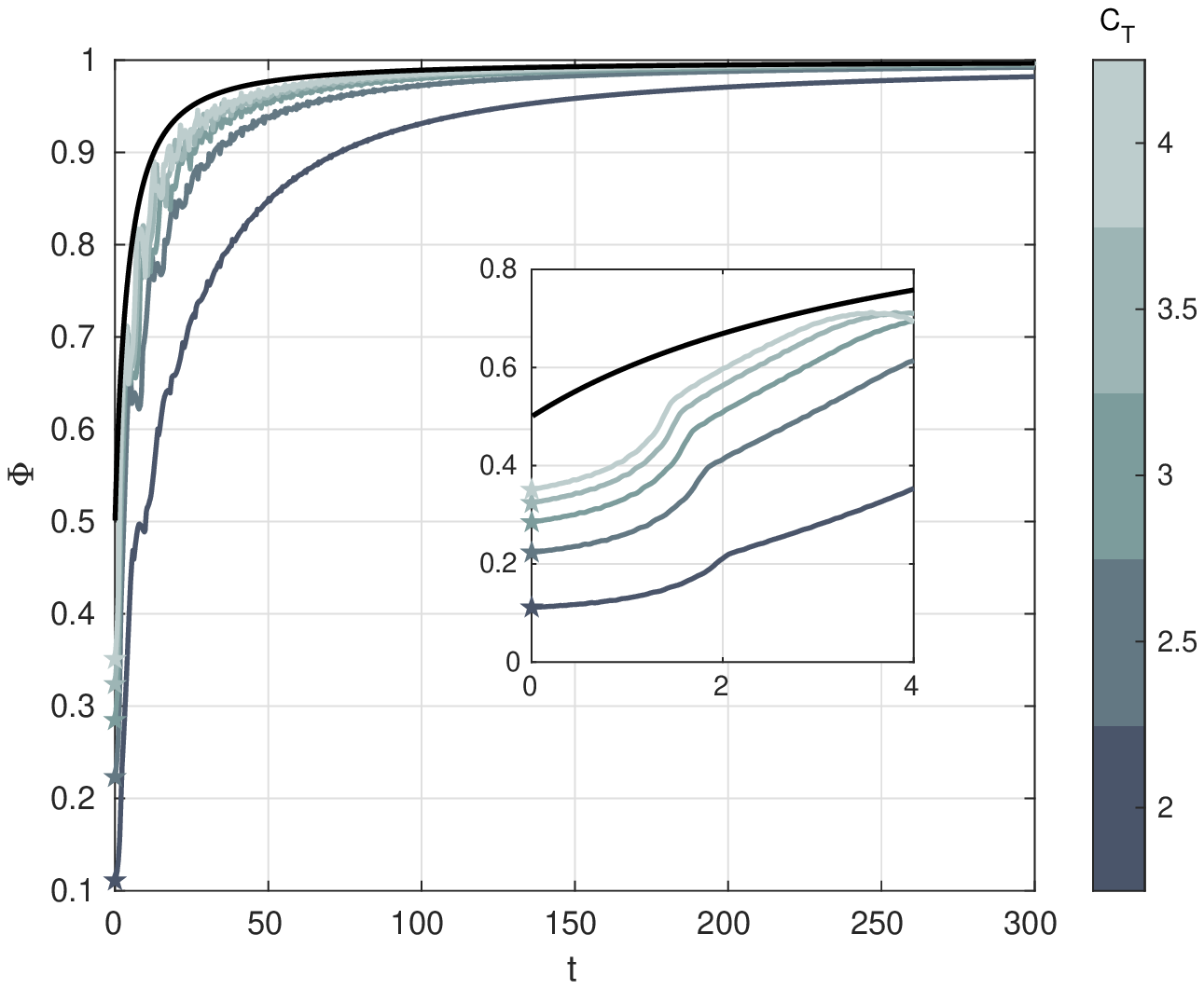}
			\caption{$\mu=1$}
			\label{fig:WagnerMem-Mu1}
		\end{subfigure}
		\quad
		\begin{subfigure}[]{0.46\textwidth}
		    \includegraphics[width=\textwidth]{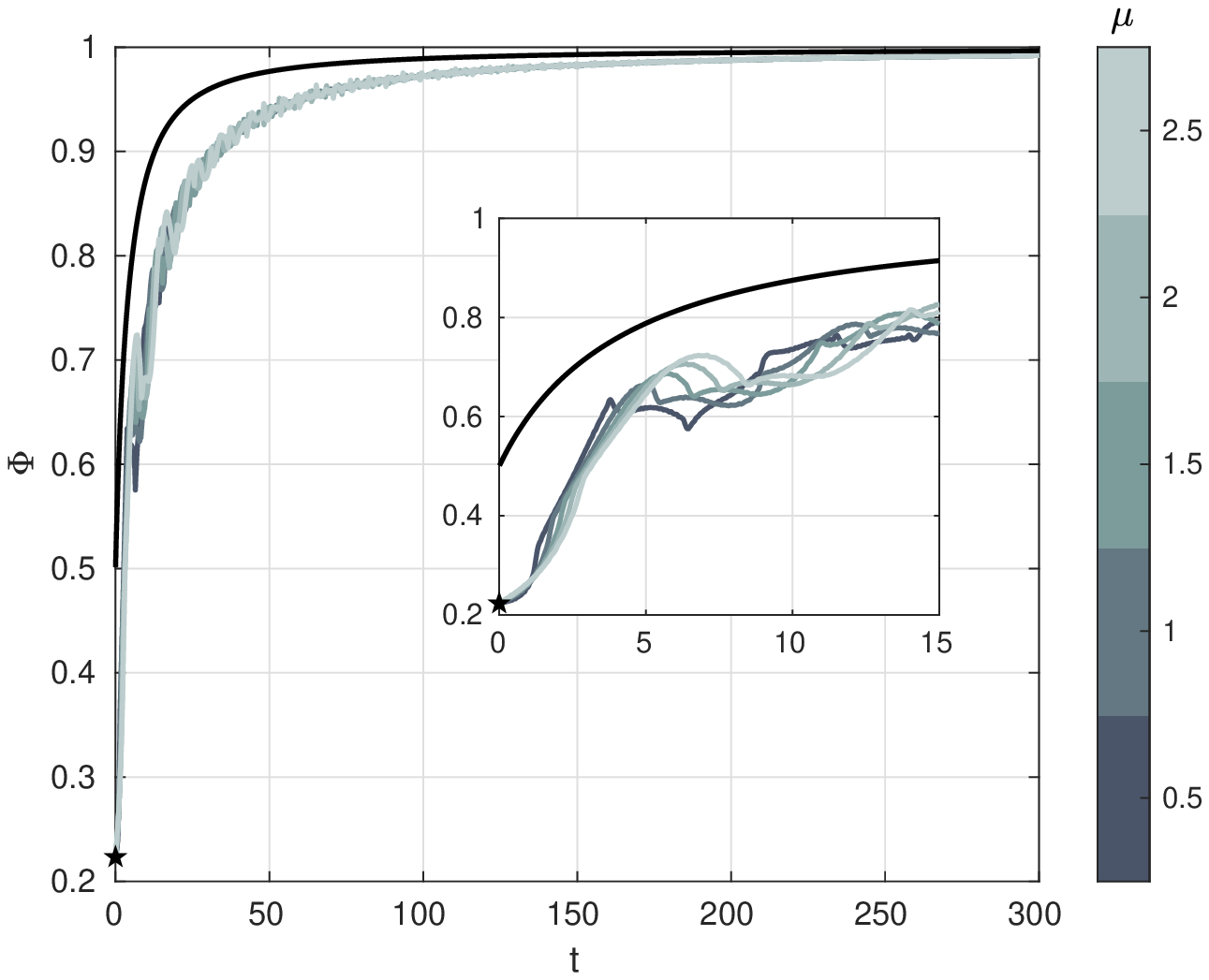}
		    \caption{$C_T=2.5$}
		    \label{fig:WagnerMem-CT2_5}
	    \end{subfigure}
	    \caption{Effect of the tension coefficient (a) and mass ratio (b) on the membrane equivalent Wagner function, as compared to the standard Wagner function (black line). Results are obtained via Laplace-domain solution and are verified against the expected initial values marked with pentagrams.}
	\label{fig:WagnerMem}
	\end{center}
\end{figure}

Figure~\ref{fig:WagnerMem} illustrates the separate effects of the membrane tension coefficient and mass ratio on the membrane lift response to a step in angle of attack in terms of the equivalent Wagner function, and compares it to the standard Wagner function for rigid aerofoils.
The equivalent Wagner functions were computed in the Laplace domain \eqref{eq:WagnerMem_expFn-LD} and transformed to the time domain via numerical Laplace inversion.
In general, for all of the examined cases, the membrane lift response is slower than the rigid plate response. 
However, we recall that the static membrane lift-curve slope is determined by the tension coefficient \eqref{eq:CLa_mem-static}, and can be significantly larger than the rigid-plate lift slope due to aeroelastic camber.
Thus, for example, while for $C_T=2,\mu=1$, at time $t=100$ the equivalent Wagner function reaches only $93.8\%$ of its steady-state solution, compared to $99\%$ for a rigid flat plate, the steady-state lift in this case is substantially higher for the membrane wing ($C_{l_{s\alpha}}\cong28$), yielding a lift that is more than $4$ times larger than the rigid plate lift.
As the tension along the membrane is increased, the membrane Wagner function approaches the classical solution for a rigid plate, as expected (figure~\ref{fig:WagnerMem-Mu1}). 
In addition, the initial value of the equivalent Wagner function is significantly lower than its rigid value, as it recovers the result predicted in \S~\ref{sc:Formulation-EquivWagner}, $\Phi_m(0)=\pi/C_{l_{s\alpha}}$, marked by pentagrams in figure~\ref{fig:WagnerMem}. 

The effect of the mass ratio on the equivalent Wagner function is much less pronounced than the tension coefficient effect (figure~\ref{fig:WagnerMem-CT2_5}) and is discernible only for short time periods (i.e., high frequencies) when inertial effects are important. For long time durations, the lift responses are equivalent for any practical use.

\subsection{Gust response}\label{sec:results-gust}

The membrane response to encounters with transverse gusts is now studied for two canonical cases: (i) a sinusoidal gust, and (ii) a sharp-edged gust. The sinusoidal and sharp-edged gusts produce unsteady lift responses described by the equivalent Sears and K\"{u}ssner functions, respectively. These extensions of the classical Sears and K\"{u}ssner functions are presented for flexible membrane wings, along with discussion on the membrane dynamic response to these unsteady flow conditions and the role of the membrane parameters $\left(\mu, C_T\right)$ in its aerodynamic performance. We note that while the sharp-edged gust has no physical meaning by itself, it is a very useful tool when predicting the aerofoil's response to an arbitrary (small amplitude) transverse gust by appeal to convolution theory \citep[][p. 288]{Bisplinghoff_book1996}.

\subsubsection{Sinusoidal gust}\label{sec:results-gust-sin}

The membrane lift and dynamic response to an encounter with a sinusoidal gust is controlled by the membrane tension coefficient, mass ratio, and the gust reduced frequency.
The response of a nominal membrane wing to sinusoidal gusts of varying reduced frequency is analyzed first, followed by a separate analysis of the effect of each of the membrane parameters on the resulting unsteady lift and dynamic response of the membrane.

\begin{figure}
	\begin{center}
		\begin{subfigure}[]{0.46\textwidth}
			\includegraphics[width=\textwidth]{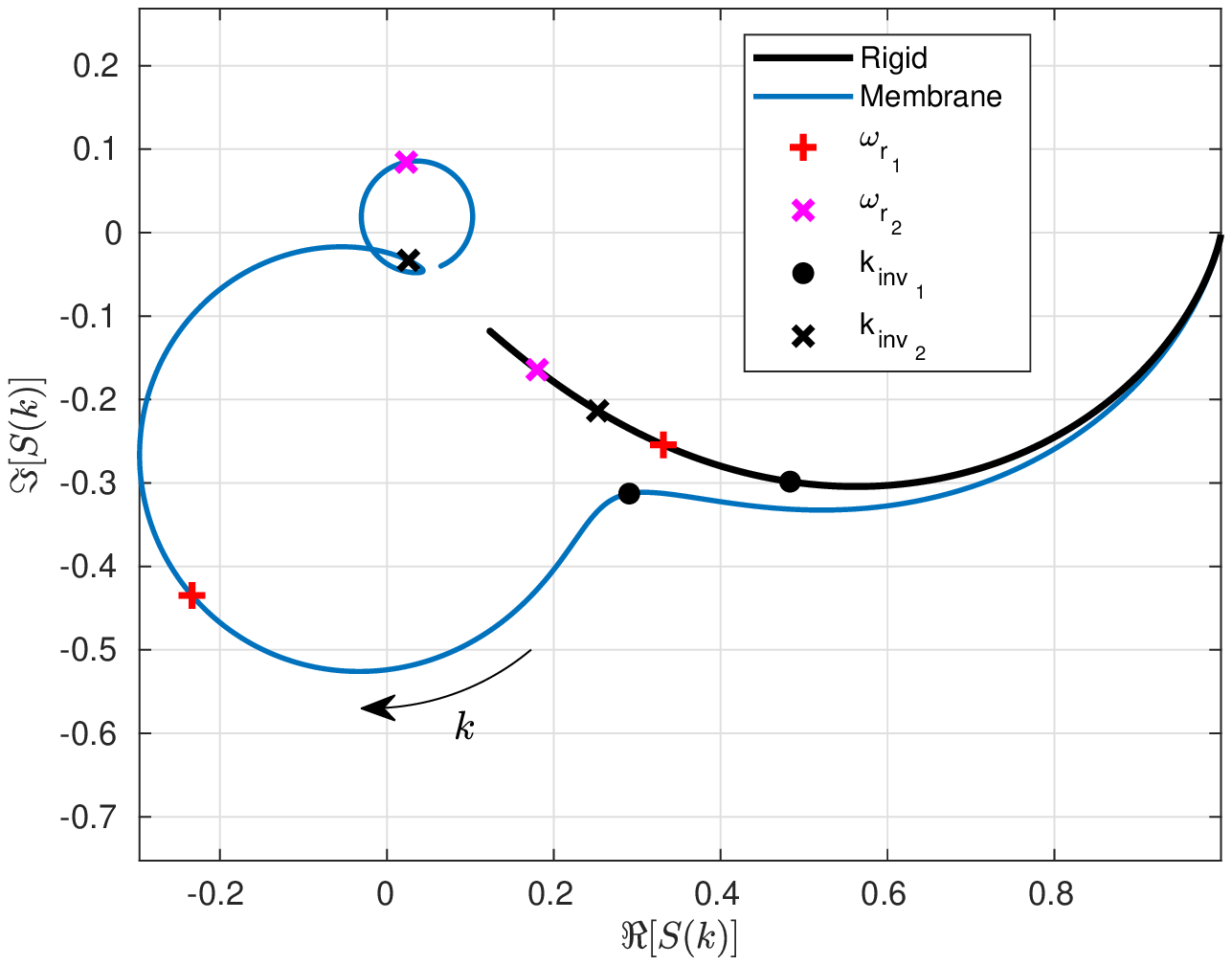}
			\caption{}
			\label{fig:SearsCompMu1CT2_5-complx}
		\end{subfigure}
		\quad
		\begin{subfigure}[]{0.46\textwidth}
			\includegraphics[width=\textwidth]{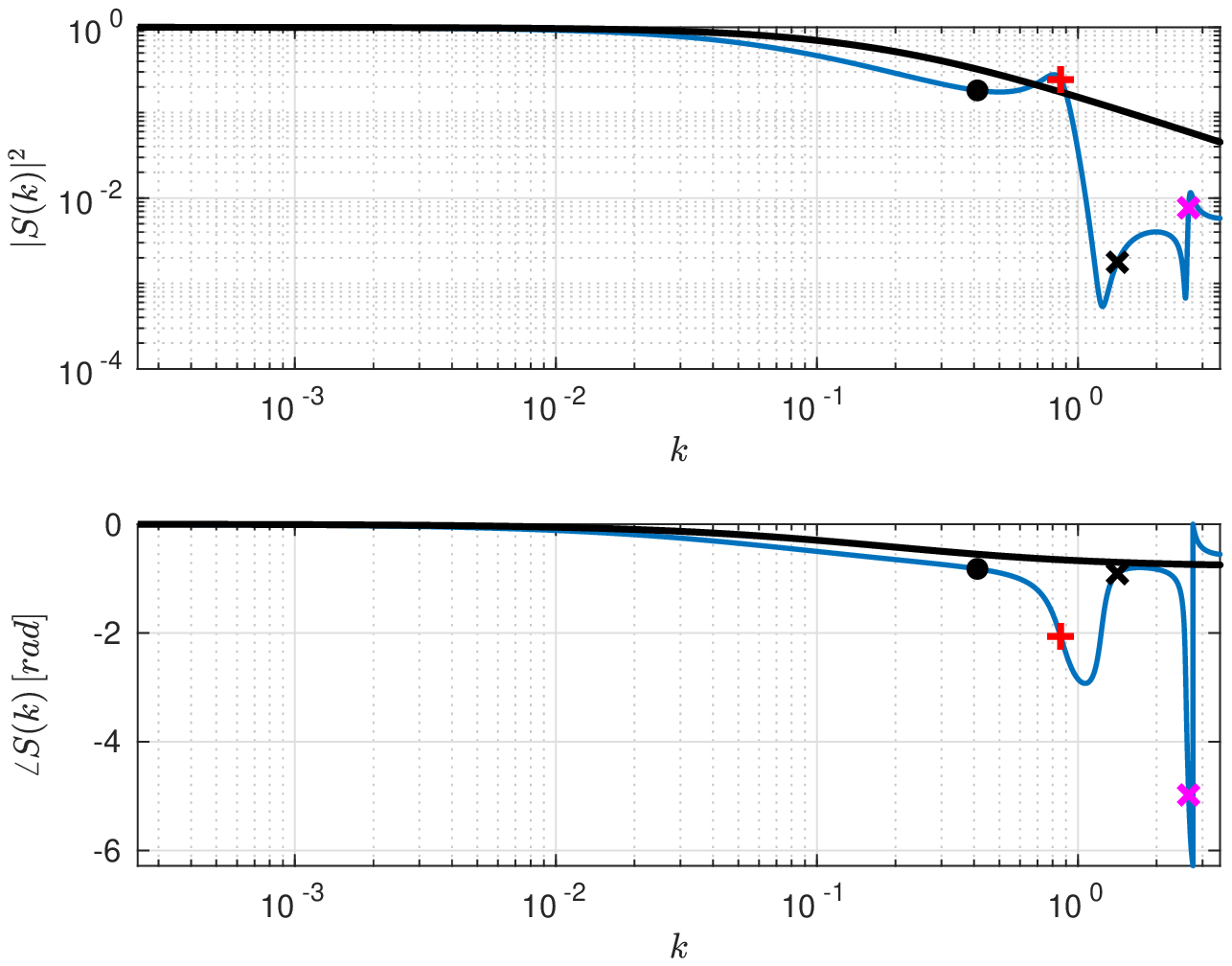}
			\caption{}
			\label{fig:SearsCompMu1CT2_5-amp2_phase}
		\end{subfigure}
		\caption{Membrane lift response to sinusoidal gusts of various frequencies, obtained for a nominal membrane of $C_T=2.5, \mu=1$ in terms of the membrane equivalent Sears function: (a) Argand diagram, and (b) squared modulus and phase. Frequencies of the inflection points are denoted with black circles ($k_{{inv}_1}$) and crosses ($k_{{inv}_2}$), and resonance frequencies are denoted with red pluses ($\omega_{{r}_1}$) and magenta cross signs ($\omega_{{r}_2}$).}
		\label{fig:SearsCompMu1CT2_5}
	\end{center}
\end{figure}

Figure~\ref{fig:SearsCompMu1CT2_5} presents the lift response of the nominal membrane to sinusoidal gusts in terms of the equivalent Sears function \eqref{eq:SearsMemExpr}.
This equivalent Sears function is compared against the classical modified Sears function for a rigid flat plate in figure~\ref{fig:SearsCompMu1CT2_5-complx} using an Argand diagram and in figure~\ref{fig:SearsCompMu1CT2_5-amp2_phase} in terms of the squared magnitude and the phase.
The choice of a squared amplitude plot rather than a modulus plot follows the convention originated by \cite{Drischler1956} for harmonic gusts.
For low reduced frequencies of $k<k_{{inv}_1}\cong0.41$, the membrane equivalent Sears function closely follows the classical modified Sears function, with a slightly decreased amplitude and increased phase lag. However, at the point of inflection ($k=k_{{inv}_1}$) the lift amplitude begins to increase significantly with reduced frequency, creating a circular path in the complex plane, in a manner similar to the  equivalent Theodorsen function (cf. figure~\ref{fig:HeaveMem_mu1CT2_5-Lift}). The lift amplitude increases beyond the rigid plate response in the vicinity of the first fluid-loaded resonance frequency ($k=\omega_{r_1}$), which is followed by a sharp decrease in amplitude at higher reduced frequencies. This behaviour suggests the existence of a reduced frequency region for which gust mitigation is achievable using flexible membrane wings, while also revealing a range of frequencies (near the first resonance frequency) for which membrane flexibility could have adverse results.

\begin{figure}
	\begin{center}
	    \begin{subfigure}[]{0.46\textwidth}
			\includegraphics[width=\textwidth]{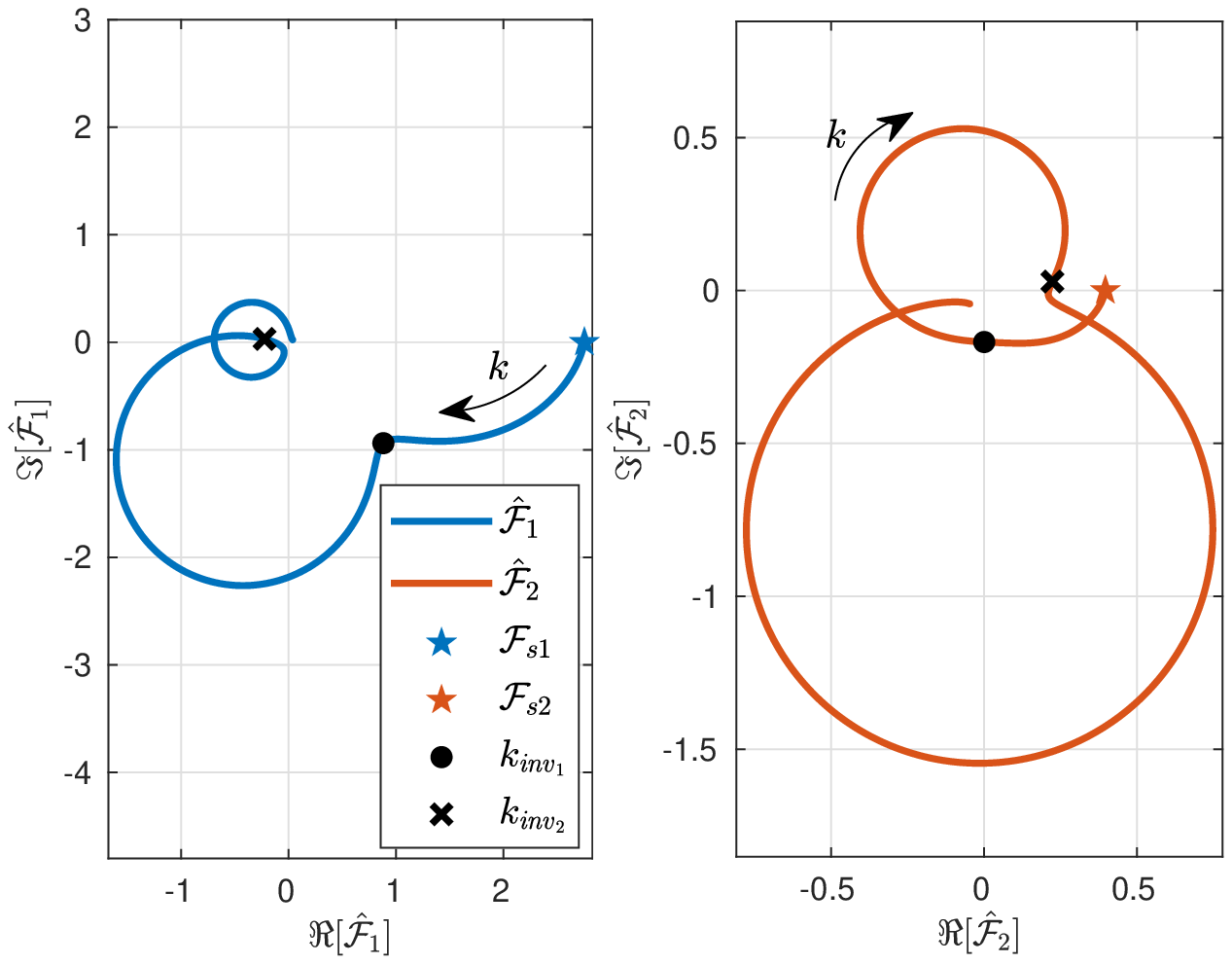}
			\caption{}
			\label{fig:SinGustMem_mu1CT2_5-Fn_cmplx}
		\end{subfigure}
		\quad
		\begin{subfigure}[]{0.46\textwidth}
		    \includegraphics[width=\textwidth]{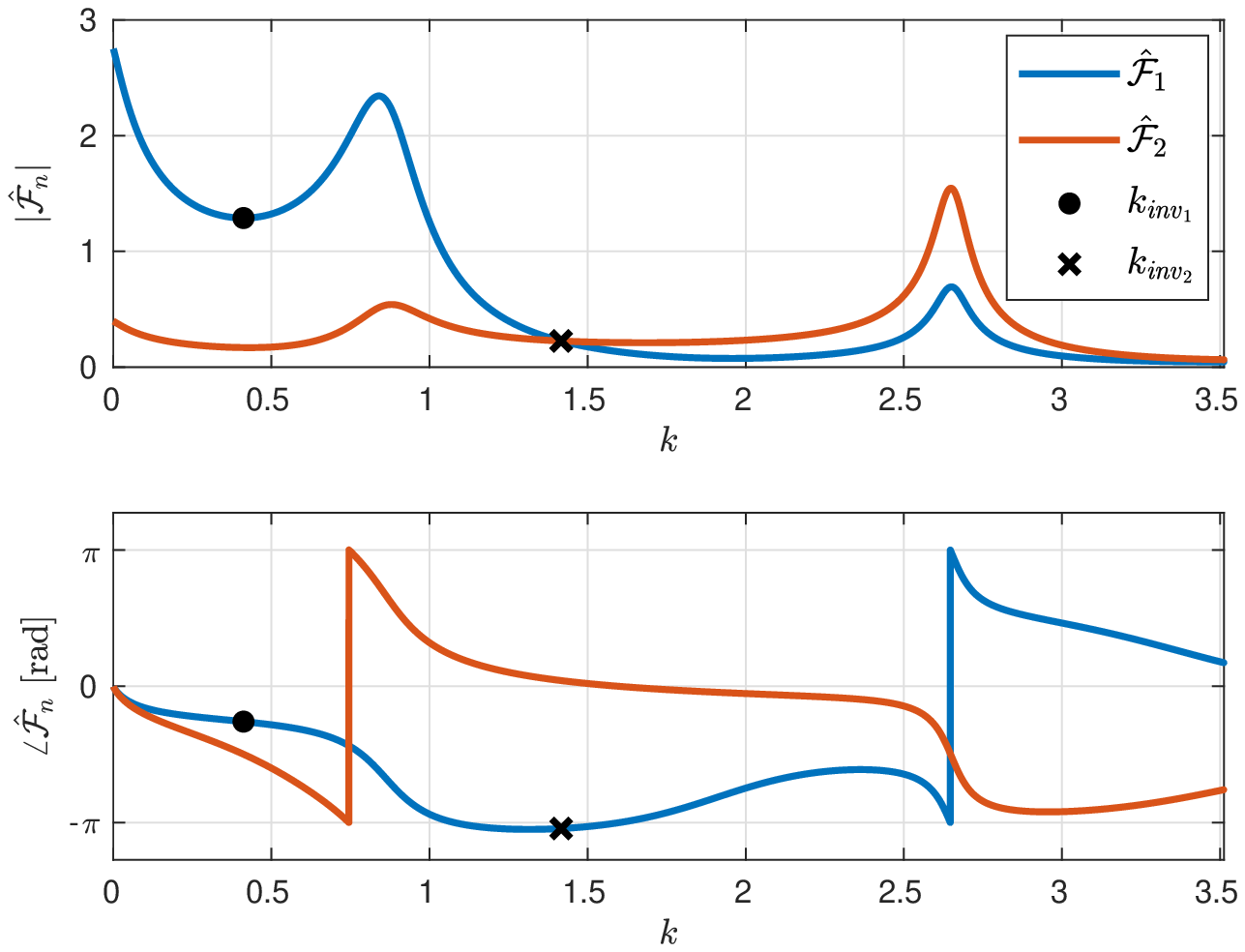}
		    \caption{}
		    \label{fig:SinGustMem_mu1CT2_5-Fn_abs_phase}
	    \end{subfigure}
	    \caption{The first two (most dominant) complex-valued normalized Fourier coefficients obtained for a nominal membrane of $C_T=2.5, \mu=1$ that encounters a sinusoidal gust of reduced frequency $k$: (a) Argand diagram; (b) modulus and phase.
	    Static solutions are denoted by pentagram markers and recovered by the unsteady results as $k\rightarrow 0$. First inflection point is marked with black circles and identified by the first local minimum of $|\hat{\mathcal{F}}_1|$. Second inflection point is denoted by black crosses, identified by an intersection between $|\hat{\mathcal{F}}_1|$ and $|\hat{\mathcal{F}}_2|$.}
	\label{fig:SinGustMem_mu1CT2_5-Fn}
	\end{center}
\end{figure}

To further study the origin of the inflection points in the complex plane plot of the equivalent Sears function, we recall that the equivalent Sears function \eqref{eq:SearsMemExpr} depends on the standard modified Sears function, the standard Theodorsen function, and the Fourier coefficients used to describe the membrane deformation. 
Figure~\ref{fig:SinGustMem_mu1CT2_5-Fn} presents the behaviour of the first two normalized Fourier coefficients obtained for the nominal membrane in response to sinusoidal gusts of varying reduced frequency. These Fourier coefficients are the most dominant coefficients in the membrane dynamic response for the range of reduced frequencies examined.
For $k\rightarrow 0$, the Fourier coefficients converge to the appropriate static solution, marked by pentagrams in figure~\ref{fig:SinGustMem_mu1CT2_5-Fn_cmplx}.
As the reduced frequency increases, the changes in both Fourier coefficients resemble the behaviours of the normalised coefficients in the heaving membrane case (cf. figure~\ref{fig:HeaveMem_mu1CT2_5-Fn_ik}), while the amplification at the resonance frequency is less pronounced for the sinusoidal gust response.
The first and second inflection points in the equivalent Sears function are identified here, similarly to the heaving membrane case, by locating the first local minimum in $|\hat{\mathcal{F}}_1|$, and the crossing between $|\hat{\mathcal{F}}_1|$ and $|\hat{\mathcal{F}}_2|$ as the dominance transfers from the first membrane mode to the second mode around $k_{{inv}_2}$. Namely, the circles in the complex plane plot of the equivalent Sears function are due to the membrane dynamic response, just like in the equivalent Theodorsen function, where each circle corresponds to a different dominant mode in the membrane oscillations. However, both the location of these circles in the frequency domain and the magnitude of the lift amplification differ significantly from the heaving membrane case, as the gust encounter introduces different flow mechanisms due to shedding of the unsteady flow conditions along the aerofoil.

\begin{figure}
	\begin{center}
	   \begin{subfigure}[]{0.46\textwidth}
			\includegraphics[width=\textwidth]{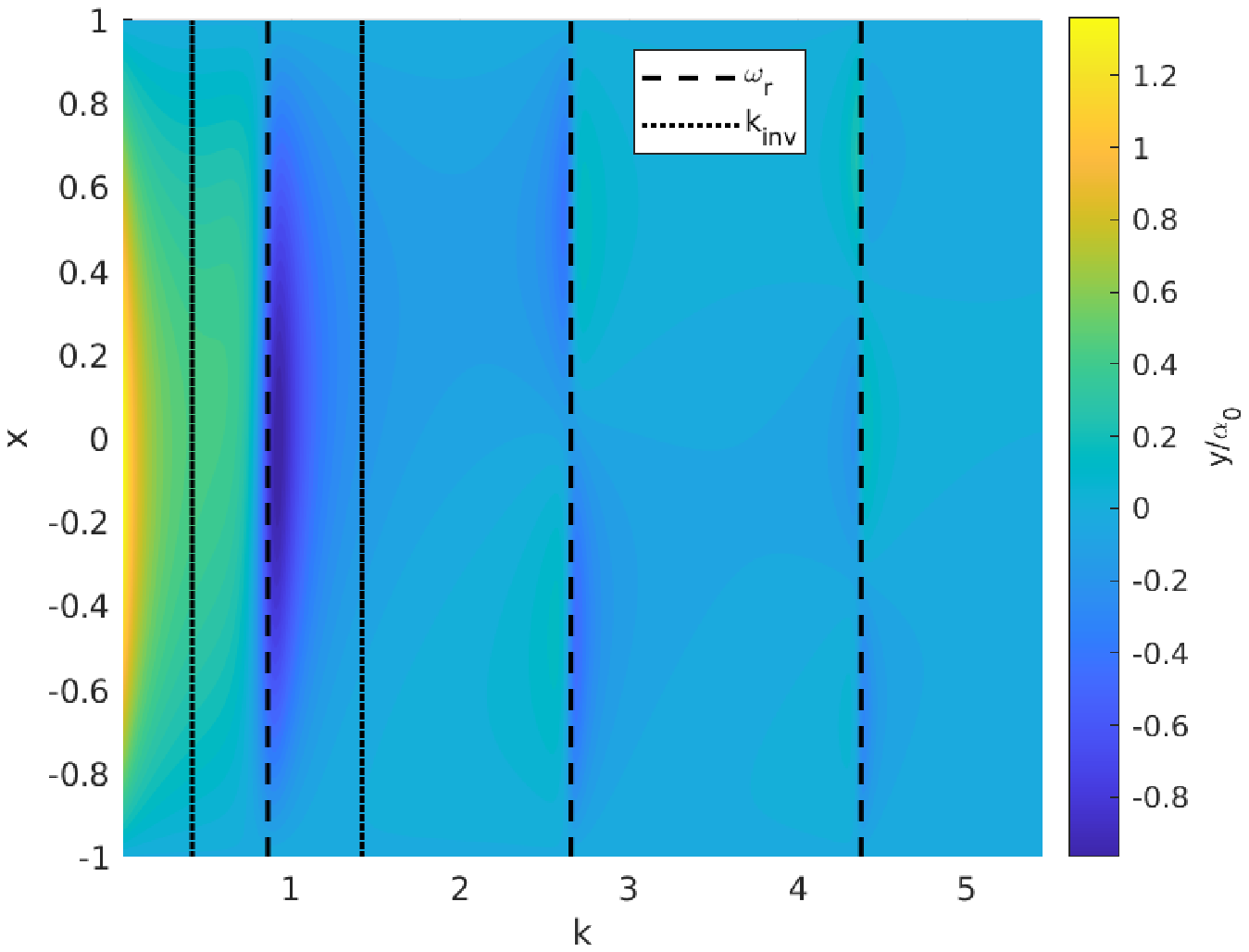}
			\caption{}
			\label{fig:MemSinGust_mu1CT2_5-MemContourf}
		\end{subfigure}
		\\
	    \begin{subfigure}[]{0.46\textwidth}
			\includegraphics[width=\textwidth]{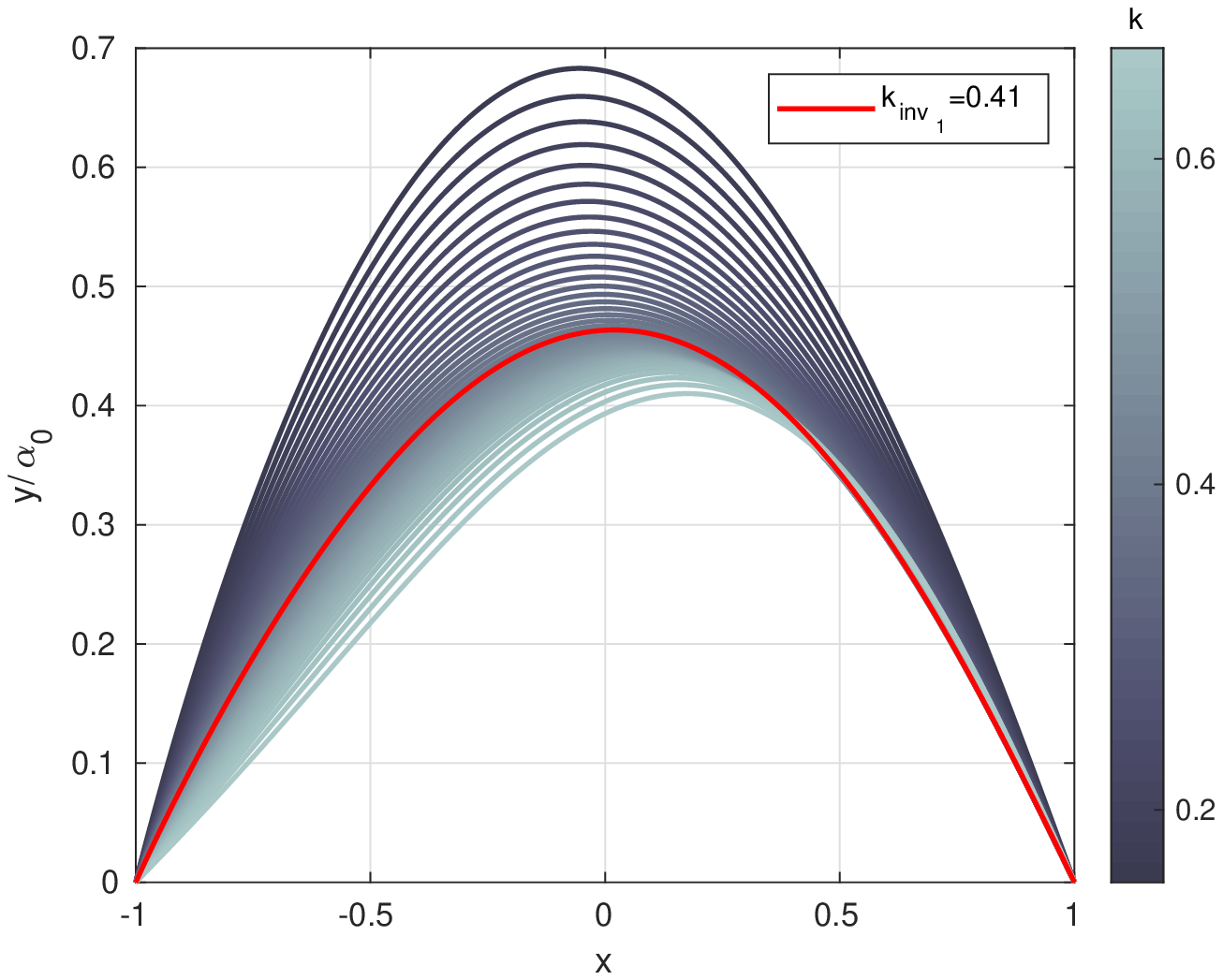}
			\caption{}
			\label{fig:MemSinGust_mu1CT2_5-Mem_kinv1}
		\end{subfigure}
	    \quad
		\begin{subfigure}[]{0.46\textwidth}
		    \includegraphics[width=\textwidth]{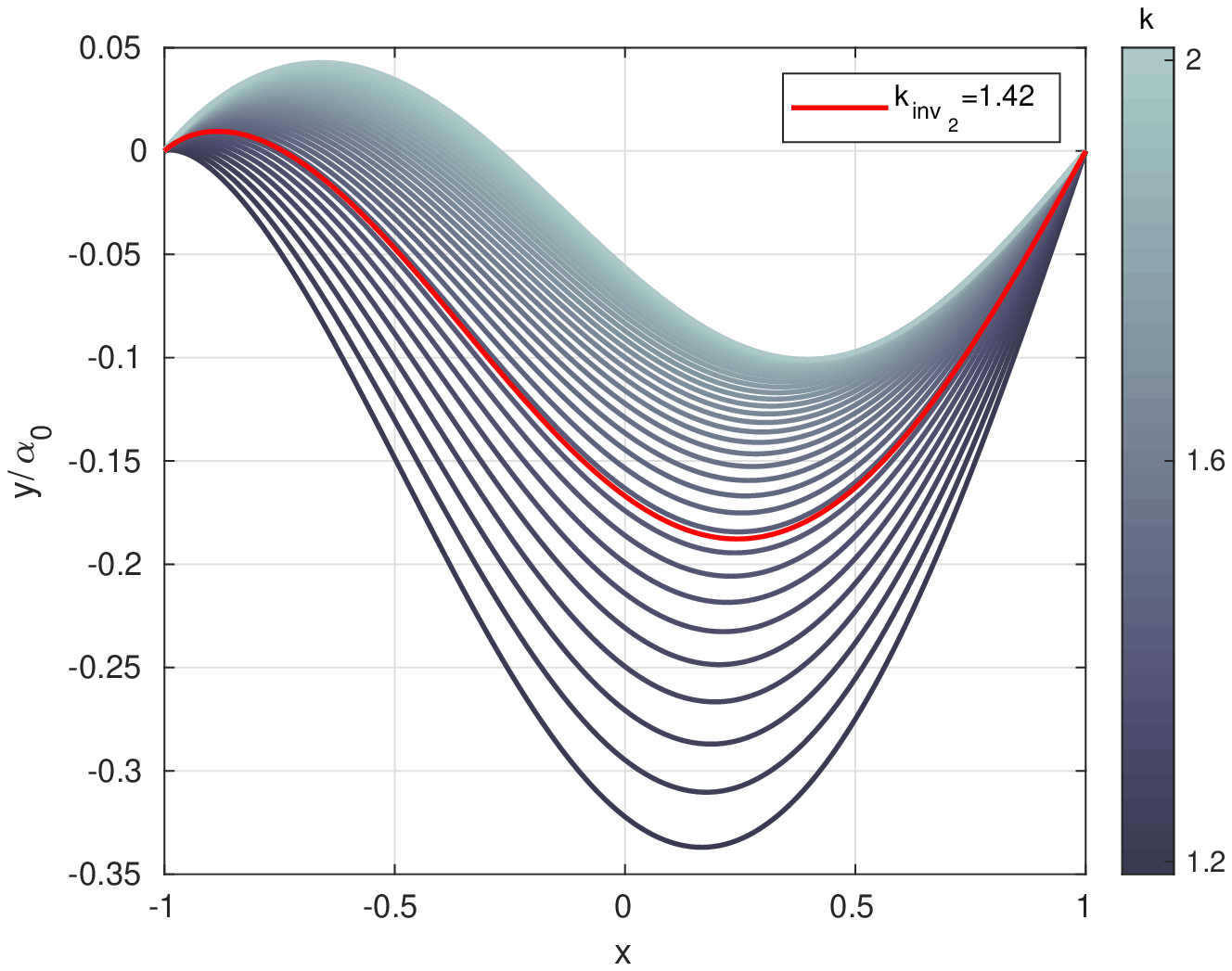}
		    \caption{}
		    \label{fig:MemSinGust_mu1CT2_5-Mem_kinv2}
	    \end{subfigure}
	    \caption{Membrane dynamic response to an encounter with a sinusoidal gust, obtained for a nominal membrane of ${C_T=2.5}, {\mu=1}$. (a) amplitude profile computed for varying values of reduced frequency. Resonance frequencies, $\omega_r$, are denoted with dashed black lines and frequencies of inflection points, $k_{inv}$, are denoted by dotted black lines. Membrane amplitude profiles near the (b) first and (c) second inflection points.}
	\label{fig:MemSinGust_mu1CT2_5-Mem}
	\end{center}
\end{figure}

Figure~\ref{fig:MemSinGust_mu1CT2_5-MemContourf} presents the membrane amplitude profiles computed during steady state oscillations of a nominal membrane that encounters sinusoidal gusts of various reduced frequencies. For $k\to0$ a convex amplitude profile is obtained, in accordance with the static membrane solution. Then, as the reduced frequency is increased, the membrane amplitude profile is flattened until $k$ approaches the first resonance frequency, for which a large maximum amplitude is obtained. Figure~\ref{fig:MemSinGust_mu1CT2_5-Mem_kinv1} illustrates the membrane amplitude profiles computed for sinusoidal gusts at reduced frequencies near the first inflection point frequency, $k_{{inv}_1}$. For reduced frequencies smaller than $k_{{inv}_1}$ a convex shape with a maximum camber point at the fore part of the aerofoil is obtained. As the reduced frequency increases to $k_{{inv}_1}$ the maximum amplitude of the membrane decreases, and the maximum camber point slowly approaches the mid-chord location. A further increase in the reduced frequency beyond $k_{{inv}_1}$ yields a sudden shift of the maximum camber point downstream, as the membrane profile bears a close resemblance to the first unstable eigenshape of the membrane in cases of divergence instability, as reported by \cite{Sygulski2007} and \cite{Tiomkin2017}. This shift in the membrane amplitude profile signals the excitation of the membrane structural modes as the first fluid-loaded resonance frequency is approached.
As we further increase the reduced frequency to the vicinity of the second inflection point frequency, $k_{{inv}_2}$, in figure~\ref{fig:MemSinGust_mu1CT2_5-Mem_kinv2} we see a clear change in the membrane amplitude profile from a shape that is dominated by the first structural mode to a shape in which the second structural mode is most dominant, as supported by the Fourier coefficients in figure~\ref{fig:SinGustMem_mu1CT2_5-Fn}.
Thus, any inflection point in the complex plane plot of the equivalent Sears function is related to a shift in dominance between two consecutive membrane mode shapes. As the gust frequency is increased, higher membrane modes become dominant.
We further note that for the frequency regime in which membrane oscillations amplify the lift response (around $k=\omega_{r_1}$), large amplitude deformations are obtained with a convex amplitude profile. For higher reduced frequencies, the membrane amplitude profile is no longer convex due to the appearance of additional nodal points along the profile, and smaller maximum camber is obtained. Membrane oscillations with these amplitude profiles attenuate the aerofoil's lift response, presenting a reduced lift amplitude relative to the rigid plate lift (figure~\ref{fig:SearsCompMu1CT2_5-amp2_phase}).

\begin{figure}
	\begin{center}
		\begin{subfigure}[]{0.46\textwidth}
			\includegraphics[width=\textwidth]{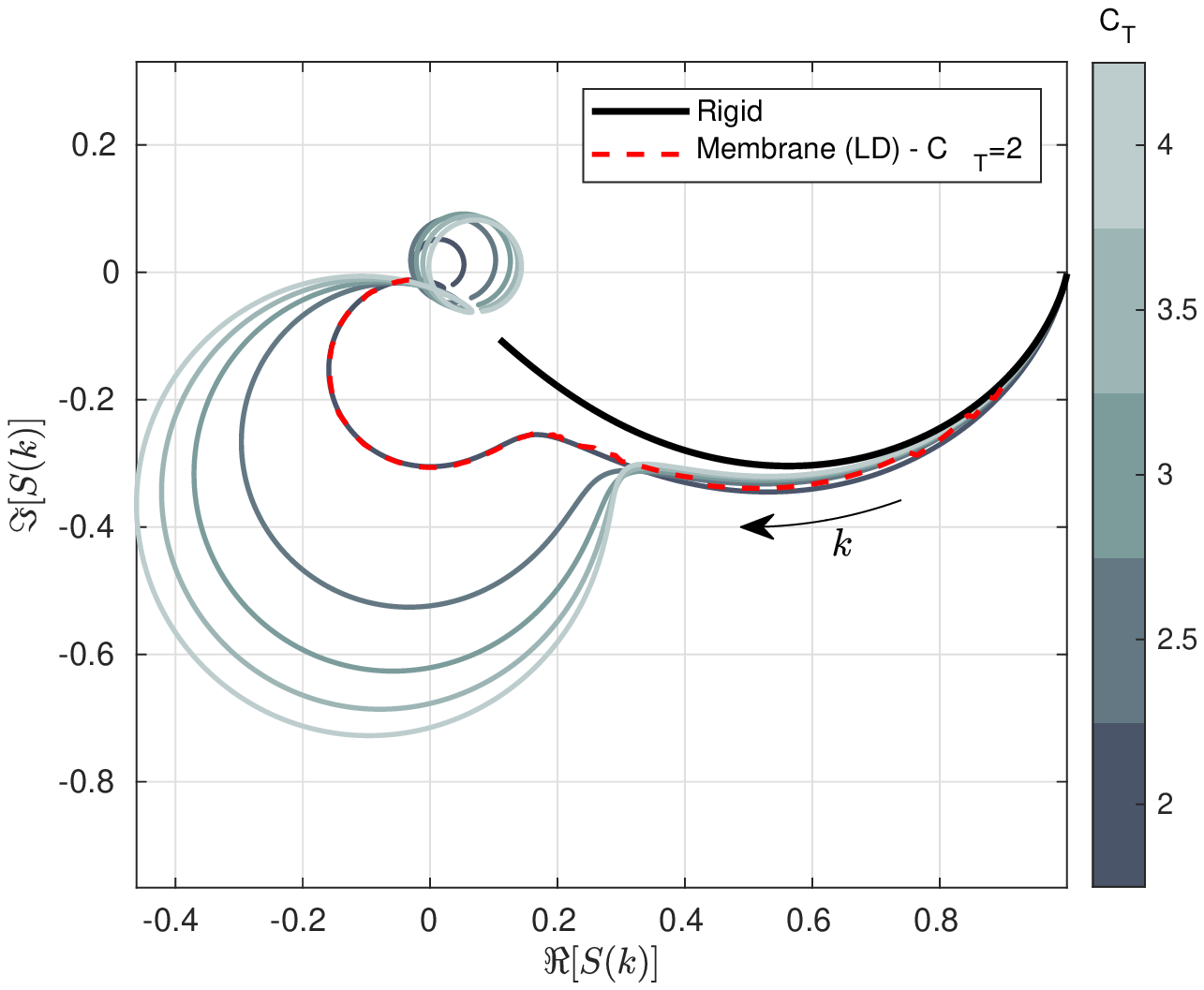}
			\caption{}
			\label{fig:SearsCompMu1CTvar-complx}
		\end{subfigure}
		\quad
		\begin{subfigure}[]{0.46\textwidth}
			\includegraphics[width=\textwidth]{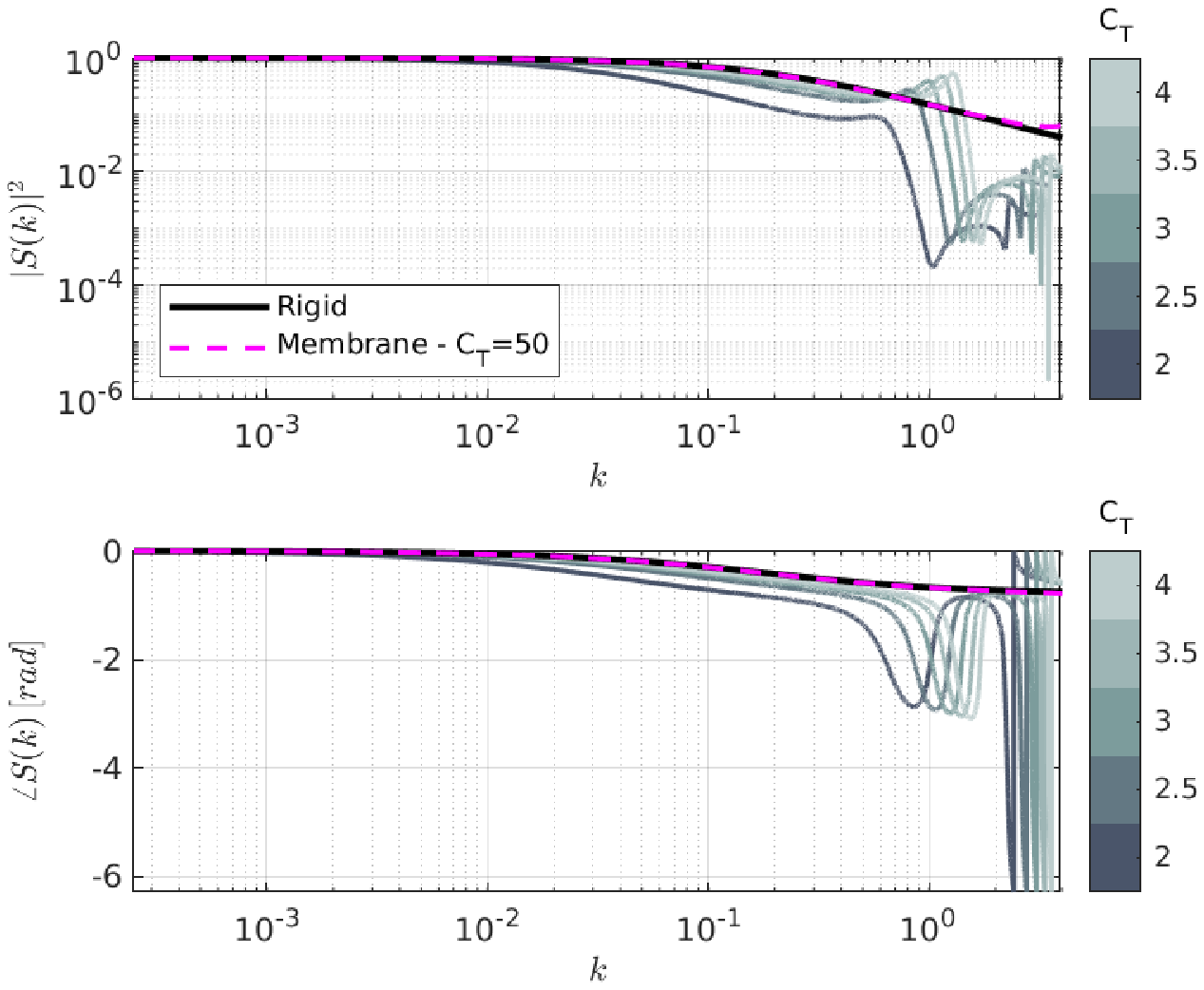}
			\caption{}
			\label{fig:SearsCompMu1CTvar-amp2_phase}
		\end{subfigure}
		\caption{Effect of tension coefficient on the membrane equivalent Sears function, for $\mu=1$: (a) Argand diagram; (b) squared amplitude and phase, as compared to the rigid-plate modified Sears function (black line). 
		An additional solution for a very large tension coefficient of $C_T=50$ is presented with a magenta dashed line in (b), to validate the unsteady solution by convergence to the rigid plate solution.
		All results are obtained with the frequency-domain solution, presented for frequencies up to $k=k_2$ for clarity. Direct comparison with the Laplace-domain solution is made for $C_T=2$ (red dashed line in (a)), where the strong agreement indicates the equivalence of the methods.
		}
		\label{fig:SearsCompMu1CTvar}
	\end{center}
\end{figure}

Figure~\ref{fig:SearsCompMu1CTvar} illustrates the effect of the tension coefficient on the equivalent Sears function, which is presented as a curve in the complex plane and in terms of its squared amplitude and phase, as compared to the classical modified Sears function. These results are obtained using the frequency-domain solution, which are verified by comparison with results of the Laplace domain solution (see dashed red line in figure~\ref{fig:SearsCompMu1CTvar-complx}).
The equivalent Sears function approaches the classical rigid aerofoil function uniformly at low reduced frequencies as the tension coefficient, $C_T$, increases.
An extreme case of $C_T=50$, in which the membrane is practically rigid, is presented in figure~\ref{fig:SearsCompMu1CTvar-amp2_phase} to confirm the approach of the analytical solution to that of the rigid aerofoil at large $C_T$. The equivalent Sears function of this extreme case follows closely the rigid-plate modified Sears function for a wide range of reduced frequencies up to about $k=2$ and validates our solution for large tension coefficients.
As the tension coefficient is increased from $2$ to $4$ in figure~\ref{fig:SearsCompMu1CTvar}, the first inflection point is delayed to a higher reduced frequency while also producing a larger unsteady lift amplification near the first resonance frequency.
The beneficial region, where significant gust mitigation is obtained, is also shifted to higher values of $k$ with an increase in the tension coefficient.
This unique characteristic of the membrane wing presents a special opportunity to calibrate the pre-tension on the membrane to successfully alleviate gusts of specific frequency regime, which is known to be one of the crucial challenges in SUAV design \citep[][]{Mohamed2014}.

\begin{figure}
	\begin{center}
	    \begin{subfigure}[]{0.46\textwidth}
			\includegraphics[width=\textwidth]{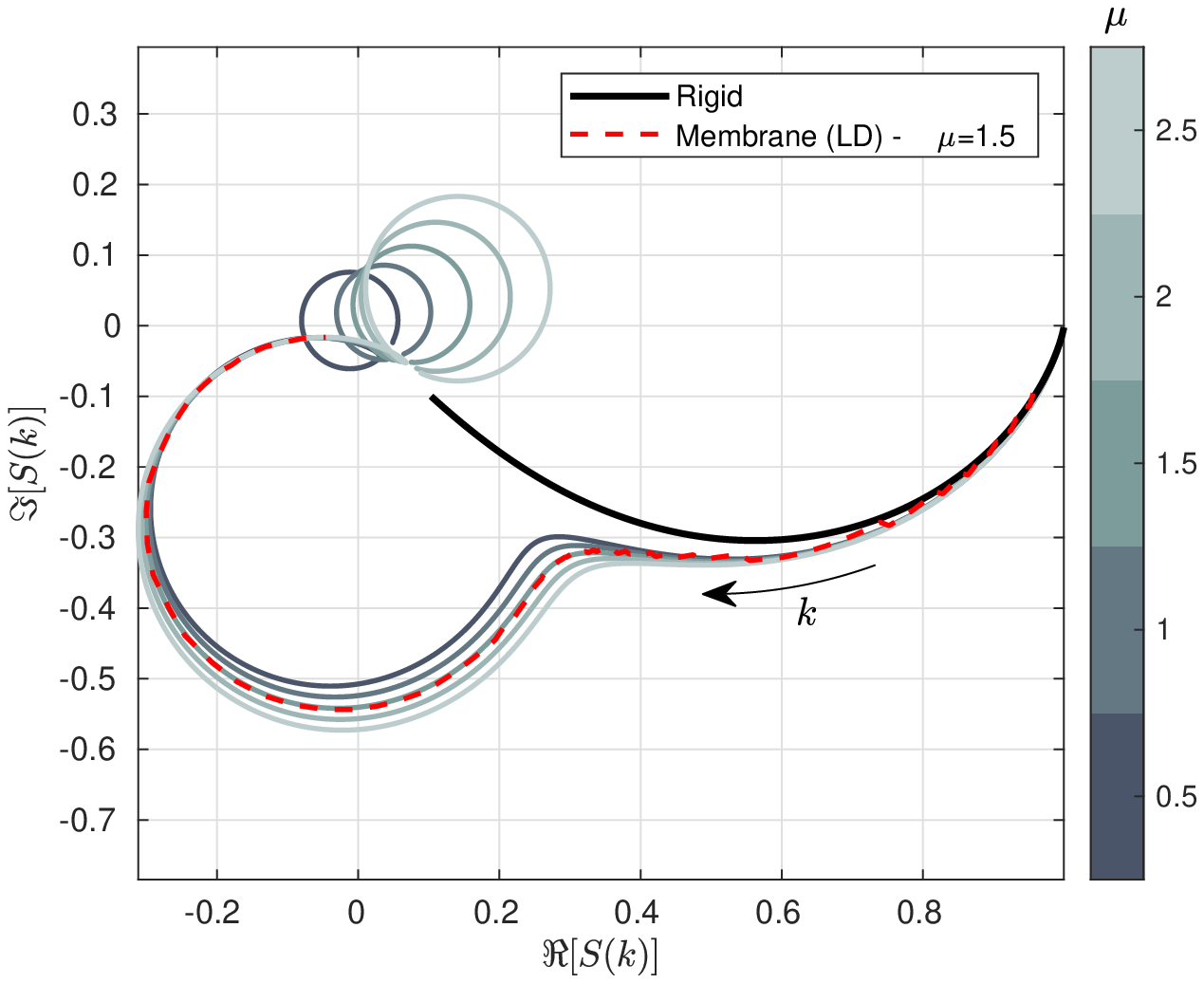}
			\caption{}
			\label{fig:SearsCompCT2Muvar-complx}
		\end{subfigure}
		\quad
		\begin{subfigure}[]{0.46\textwidth}
			\includegraphics[width=\textwidth]{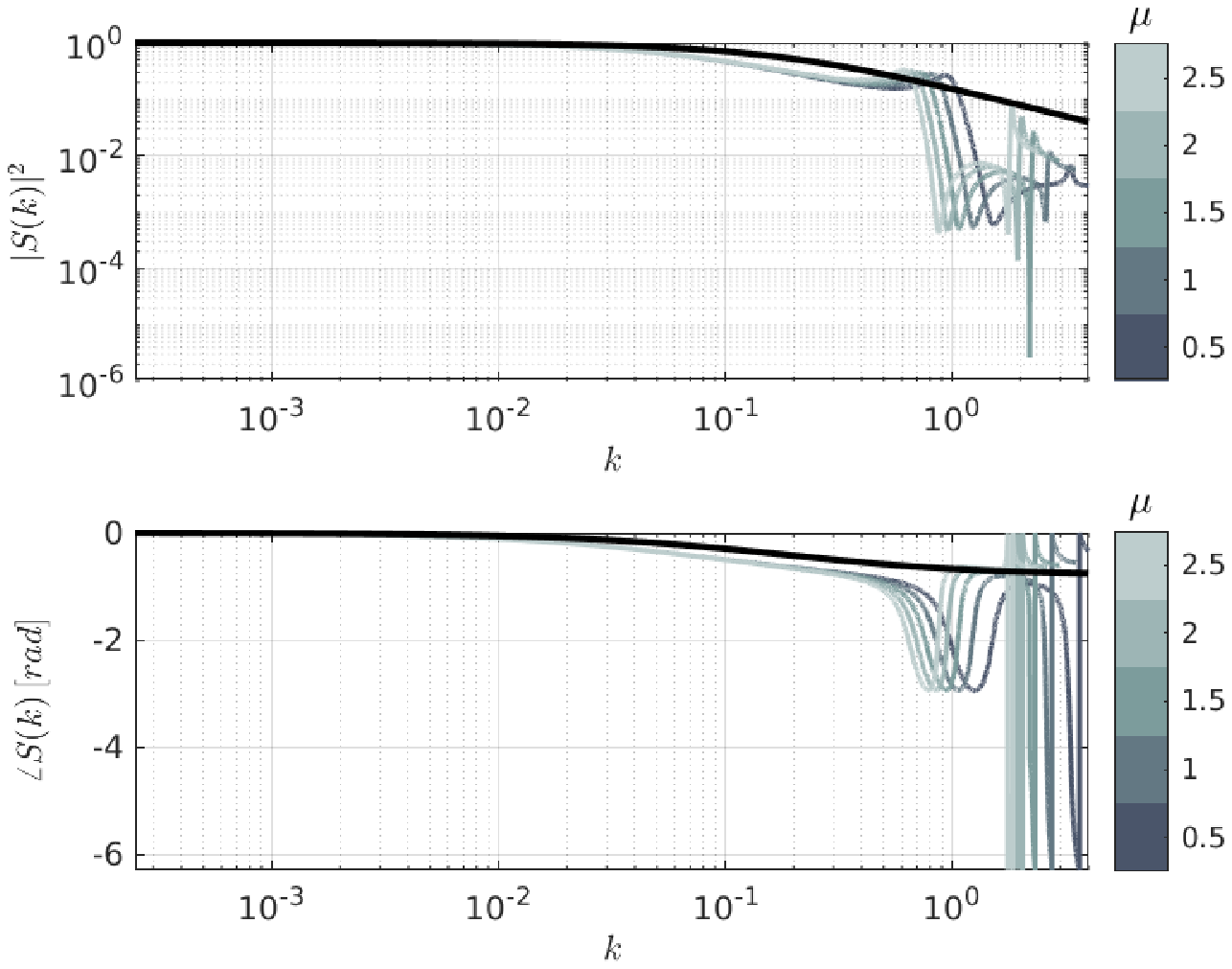}
			\caption{}
			\label{fig:SearsCompCT2Muvar-amp2_phase}
		\end{subfigure}
		\caption{Effect of membrane mass ratio on the membrane equivalent Sears function plotted, $C_T=2.5$: (a) Argand diagram; (b) squared amplitude and phase, as compared to the rigid-plate modified Sears function (black line). 
		All results are obtained with the frequency-domain solution, presented for frequencies up to $k=k_2$ for clarity. Results from the Laplace-domain solution are shown in (a) with a red dashed line, presenting good agreement with the frequency-domain solution.}
		\label{fig:SearsCompCT2Muvar}
	\end{center}
\end{figure}

Looking into the effect of the mass ratio on the membrane lift response to a sinusoidal gust, figure~\ref{fig:SearsCompCT2Muvar} presents the equivalent Sears functions which are computed for varying values of the mass ratio, $\mu$, and are compared to the classical modified Sears function.
For very low frequencies, up to the point of inflection $k_{{inv}_1}$ (whose value decreases with increase in $\mu$), the lift response is practically unaffected by changes in the mass ratio, as predicted by the asymptotic analysis \eqref{eq:SearsMem_lim_k0}. Thus, the response can be considered quasi-steady for this frequency regime.
As the gust frequency is increased beyond the inflection point, a circular path appears at the complex plane plot with a radius (and amplitude) that slightly increases with $\mu$. 
The membrane oscillates with the gust frequency so that an increase in $\mu$ leads to increase in the membrane inertia, which is proportional to $\mu k^2$, resulting with a larger amplitude of oscillation for a given gust frequency. However, as the system's resonance frequencies decrease with increase in $\mu$ (c.f., figure~\ref{fig:HeaveSSMem_CT_vs_k}), the lift amplification at the first resonance frequency increased only slightly with an increase in $\mu$.
We note that the in vacuo natural frequencies of the membrane are proportional to $\mu^{-\f{1}{2}}$, indicating that for these frequencies the mass ratio has no effect on the membrane amplitude. Thus, the small effect of the mass-ratio on the lift amplitude, illustrated in figure~\ref{fig:SearsCompMu1CTvar}, is attributed to the added mass of the fluid-loaded membrane.
For frequencies beyond the first resonance frequency, the gust mitigation region is controlled by the membrane mass ratio in a manner that is similar to the effect of the tension coefficient (cf., figure~\ref{fig:SearsCompMu1CTvar-amp2_phase}), where a decrease in $\mu$ delays the region to higher values of $k$. However, in contrast to the tension coefficient, this variable is typically fixed in membrane wing applications, and thus is expected to be less useful for aerodynamic design purposes.

\subsubsection{Sharp-edged gust}\label{sec:results-gust-SEG}

\begin{figure}
	\begin{center}
		\begin{subfigure}[]{0.46\textwidth}
			\includegraphics[width=\textwidth]{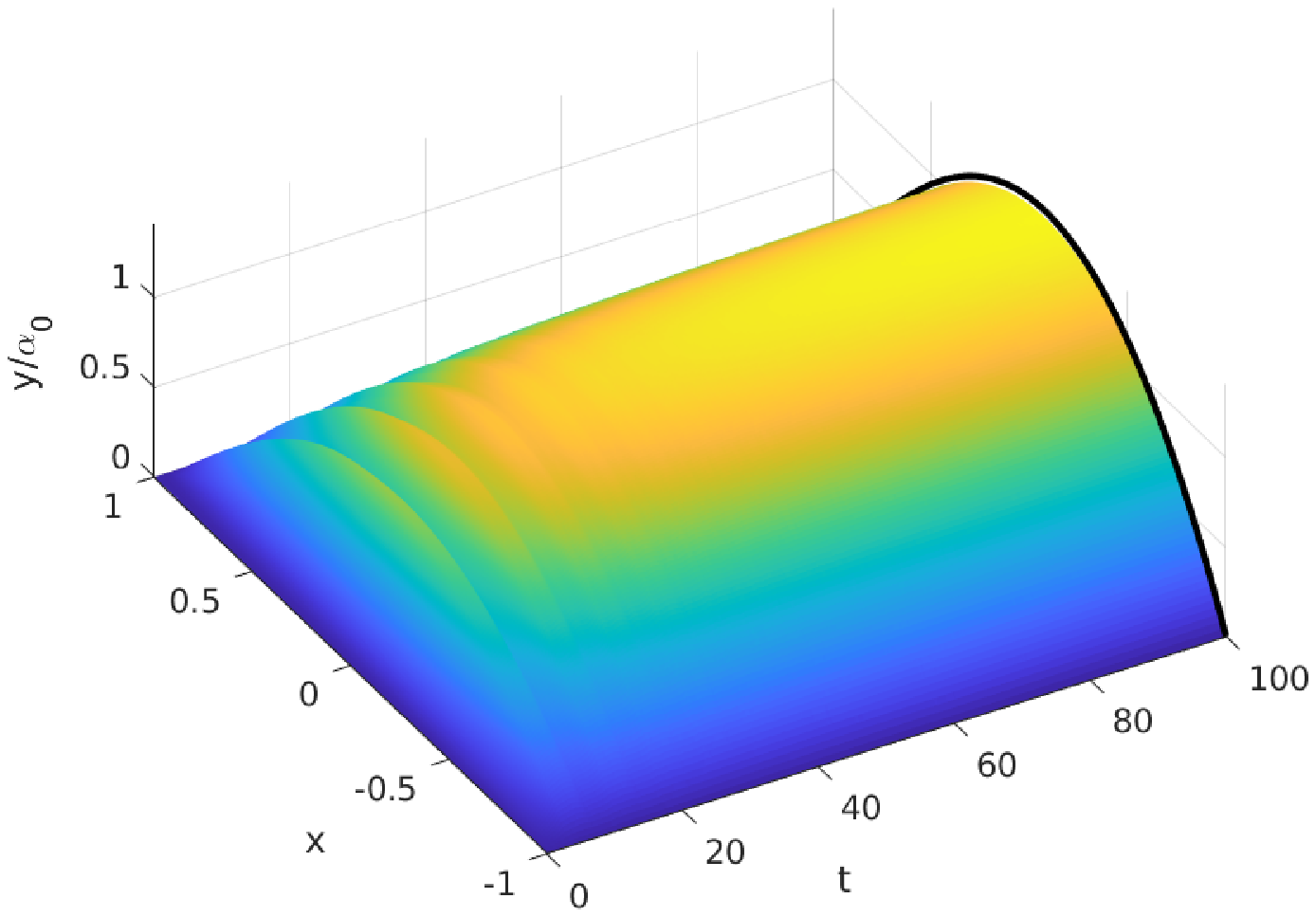}
			\caption{}
			\label{fig:SEG_mem_nom-dynamic}
		\end{subfigure}
		\quad
		\begin{subfigure}[]{0.46\textwidth}
			\includegraphics[width=\textwidth]{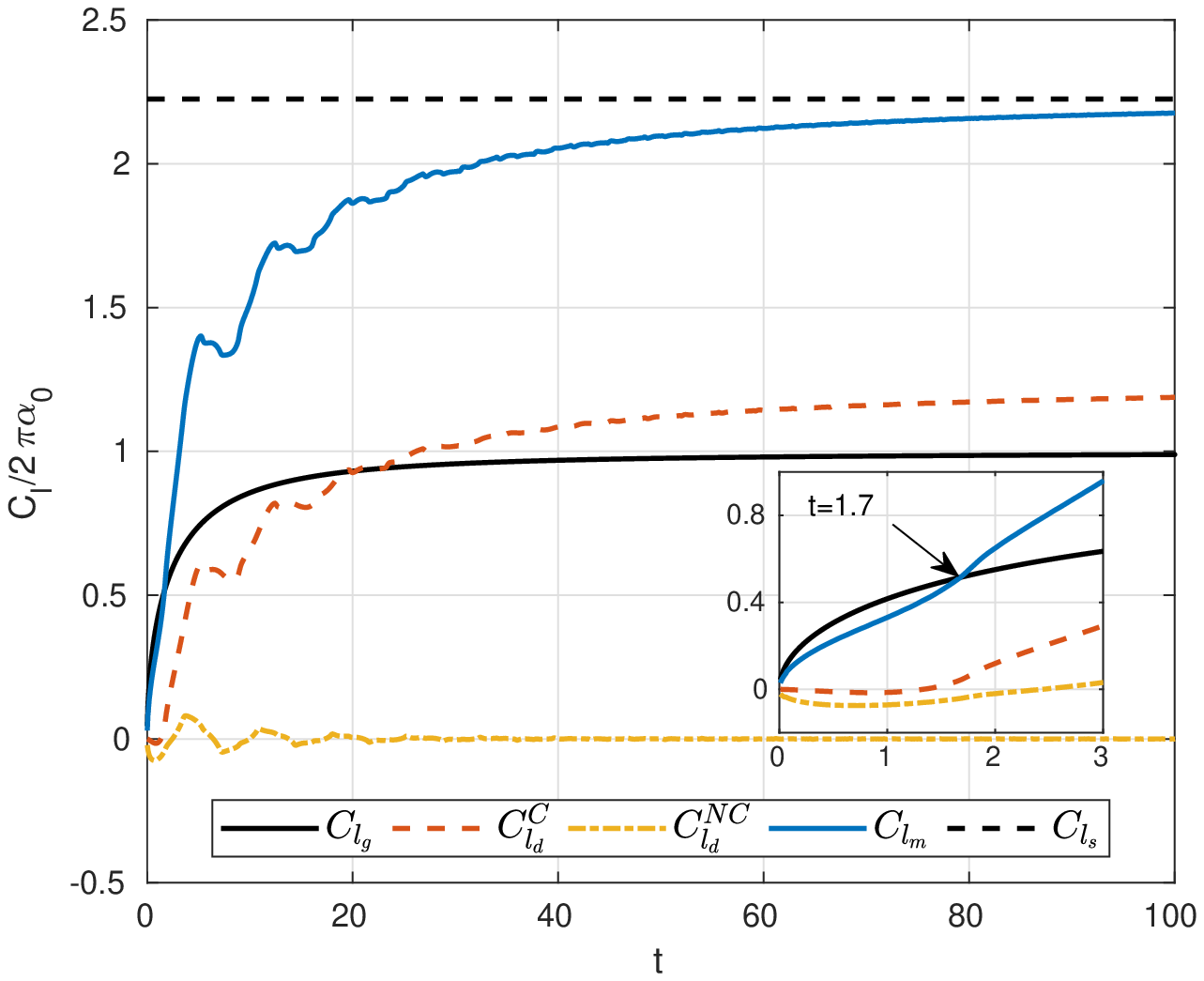}
			\caption{}
			\label{fig:SEG_mem_nom-CL}
		\end{subfigure}
		\caption{Membrane dynamic response (a) and lift response (b) to a sharp edged gust, obtained for a nominal membrane of $C_T=2.5$ and $\mu=1$. Black line in (a) and dashed black line in (b) denote the static solution \citep[][]{Nielsen1963}. The membrane unsteady lift coefficient, $C_{l_m}$, is computed by superposition between the rigid plate indicial lift, $C_{l_g}$, and the lift due to membrane deformation, $C_{l_d}$, which is composed of the circulatory and non-circulatory terms, $C_{l_d}^C$ and $C_{l_d}^\mathit{NC}$, respectively. For $t<1.7$ the membrane deformation yields negative lift that reduces the total membrane lift compared to the rigid plate indicial lift. After time $t=1.7$ the membrane lift surpasses the rigid plate lift due to induced membrane camber and converges to the static solution.}
		\label{fig:SEG_mem_nom}
	\end{center}
\end{figure}

The dynamic response of a nominal membrane wing to a sharp-edged gust is presented in figure~\ref{fig:SEG_mem_nom} in terms of its deformation in time and its lift response. Here we choose to present the actual lift response first (instead of the normalised response represented by the K\"{u}ssner function) to obtain quantitative conclusions on the difference in the lift produced by a flexible membrane wing (blue line) and a rigid flat plate (black line) during sharp-edged gust encounter. While the transient response up to $t\cong30$ introduces oscillations in the membrane deformation, resulting in oscillations in the lift response, at a later time a fully convex shape is obtained converging to the appropriate static solution \citep[][]{Nielsen1963}.
The lift response in figure~\ref{fig:SEG_mem_nom-CL} shows that, as expected, the membrane achieves a much higher lift coefficient than a rigid flat plate, converging to more than double the lift coefficient of the flat plate, as in the case of a step change in angle of attack (cf. figure~\ref{fig:MemAoAstepRes_mu1_CT2_5}). However, zooming in onto the transient lift response during gust penetration reveals that at initial stage, for $t<1.7$, the membrane presents a lower lift coefficient than the rigid plate. 
The inset in figure~\ref{fig:SEG_mem_nom-CL} shows that the transient membrane deformation leads to a negative contribution to the lift coefficient for $t<1.7$, which results in a reduced lift coefficient compared to the rigid plate indicial lift. As the gust front approaches the trailing edge, the circulatory lift due to membrane deformation, $C_{l_d}^C$, increases first until it overcomes the non-circulatory lift, $C_{l_d}^{\mathit{NC}}$, for $t=1.7$. 
We observe that this time is longer than the time required to compensate for the initial lift deficit in the membrane response to a step change in angle of attack (cf. figure~\ref{fig:MemAoAstepRes_mu1_CT2_5-CL}).
From this moment on, the membrane lift coefficient surpasses the rigid-plate lift and slowly converges to the static solution.

\begin{figure}
    \begin{center}
    \includegraphics[width=0.46\textwidth]{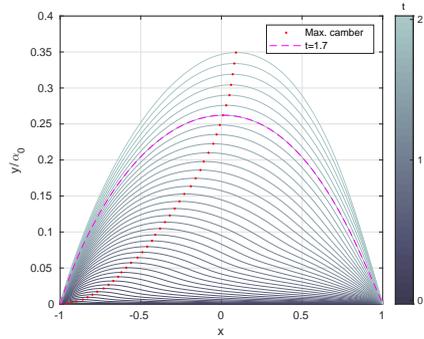}
    \caption{Membrane deformation in response to an encounter with a sharp-edged gust, obtained during gust penetration (when the gust front travels along the chord during ${0\le t\le 2}$) for a nominal membrane of $C_T=2.5, \mu=1$. Red points denote maximum camber point at each time step, and a dashed magenta line is used to identify the membrane profile at time $t=1.7$ (also denoted in figure~\ref{fig:SEG_mem_nom-CL} for the lift response).}
    \label{fig:SEG_mem_nom2-transient}
    \end{center}
\end{figure}

Figure~\ref{fig:SEG_mem_nom2-transient} presents the membrane profiles as obtained for $t\le2$, during gust penetration when the gust front travels along the chord.
The membrane encounters the gust at the leading edge at time $t=0$, in a taut initial position. As the gust front moves downstream, a small hump appears in the membrane profile near the leading edge, which increases in size and moves downstream with the advancement of the gust.
At time $t=1.7$, for which the membrane lift surpasses the rigid plate lift, the gust front has not yet reached the trailing edge, but a fully convex shape has developed with a maximum camber point at the aft part of the aerofoil. Note that the membrane profile obtains a positive camber during gust penetration, which would result with increased static lift compared to a rigid flat plate. 
However, the unsteady response of the membrane, and in particular its acceleration, produces a negative non-circulatory lift at the initial response of the membrane (figure~\ref{fig:SEG_mem_nom-CL}), highlighting the importance of a full unsteady aerodynamic model for predicting the unsteady lift response of the membrane.

\begin{figure}
	\begin{center}
	    \begin{subfigure}[]{0.46\textwidth}
			\includegraphics[width=\textwidth]{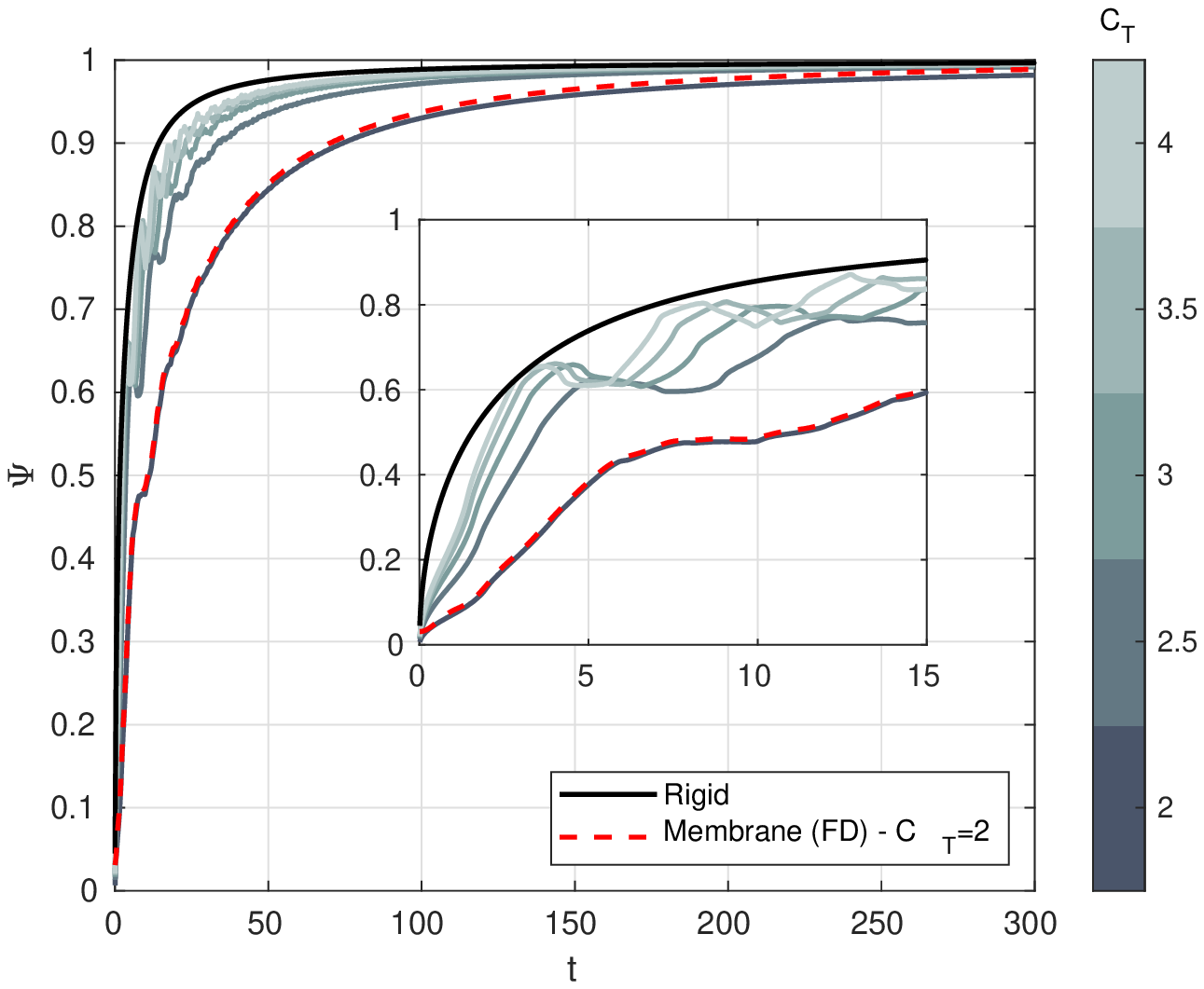}
			\caption{$\mu=1$}
			\label{fig:Kussner_mem-CT_effect}
		\end{subfigure}
		\quad
		\begin{subfigure}[]{0.46\textwidth}
			\includegraphics[width=\textwidth]{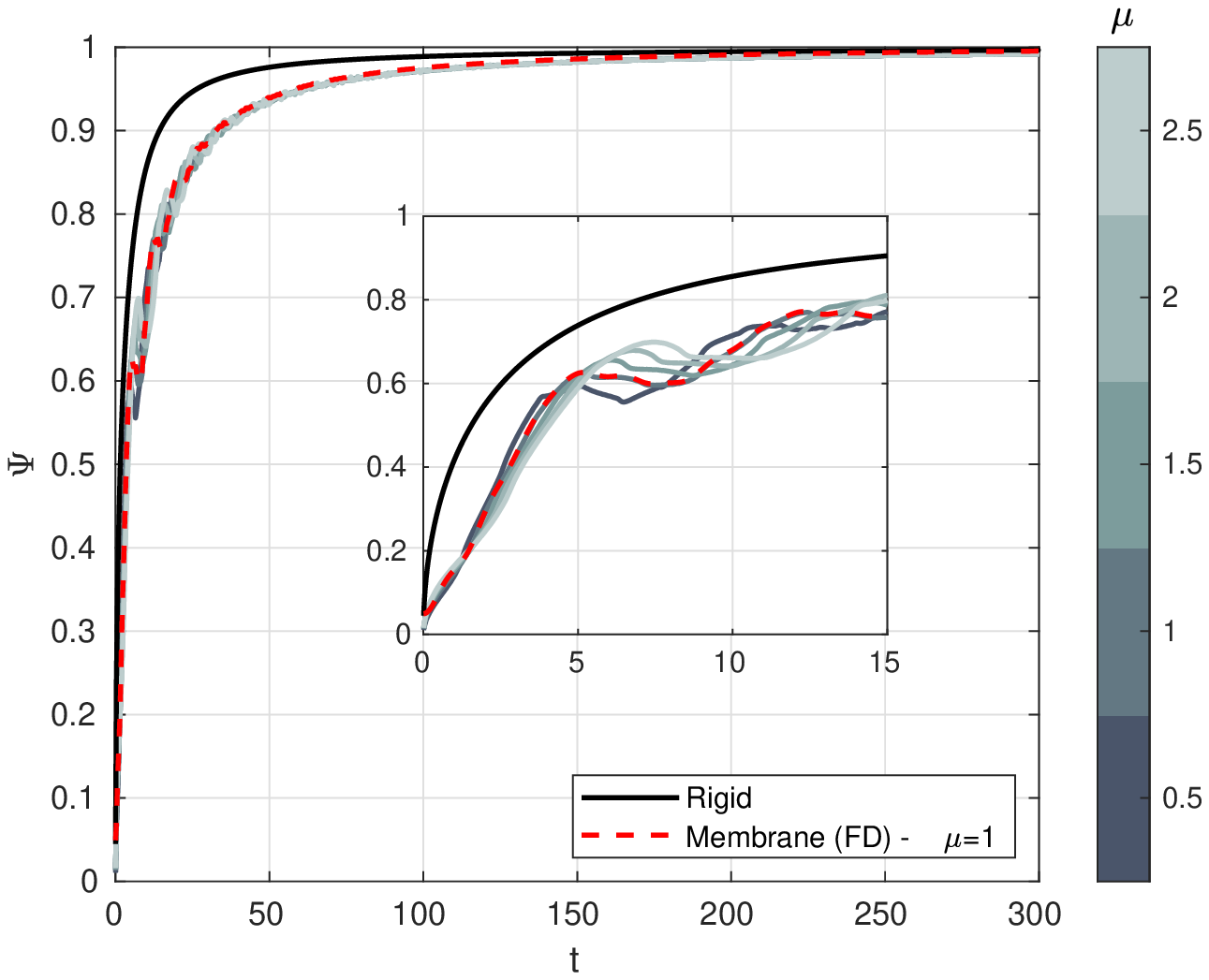}
			\caption{$C_T=2.5$}
			\label{fig:Kussner_mem-mu_effect}
		\end{subfigure}
		\caption{Effect of tension coefficient (a) and mass ratio (b) on the membrane equivalent K\"{u}ssner function, compared to the standard K\"{u}ssner function for a rigid aerofoil (black line). Results are obtained via Laplace domain solution and are validated against the frequency domain solution (red dashed line).}
		\label{fig:Kussner_mem}
	\end{center}
\end{figure}

The membrane equivalent K\"{u}ssner function is presented in figure~\ref{fig:Kussner_mem} to study separately the effect of the tension coefficient and mass ratio. These plots were computed using the Laplace-domain solution \eqref{eq:KussnerMem_exp-LD} and were verified by comparison to the frequency-domain solution \eqref{eq:KussnerMem_def}. In general, the membrane response to a sharp-edged gust is slower than the rigid plate response, similarly to the Wagner function case. As the tension coefficient is reduced, a slower response is obtained since a larger camber profile is achieved at steady state, which takes a longer time to attain. Contrarily, the membrane mass ratio appears to affect only the initial oscillatory stage, while the rest of the response is practically independent of the mass ratio, as expected.

\section{Concluding remarks}\label{sec:conclusions}

Unsteady aerodynamic theory is extended to include the aeroelastic deformations of flexible membrane wings in response to unsteady flow conditions. The pressure loads and membrane deformations due to dynamic fluid-membrane coupling are determined generally and exactly in the Laplace domain for small-amplitude prescribed chord motions or transverse gust encounters, which are also evaluated in the time domain using the numerical inversion scheme of \citet{Valsa1998}. Lift responses computed for the canonical unsteady aerodynamic scenarios of harmonic aerofoil motions or gusts, as well as of a step change in the angle of attack or gust profile, constitute aeroelastic extensions to the classical Theodorsen, Sears, Wagner, and K\"{u}ssner functions, respectively, for a membrane aerofoil.

The membrane lift response to harmonic heave motions and sinusoidal transverse gusts are verified against a separated analysis in the frequency domain. In each scenario, the membrane-equivalent Theodorsen and Sears functions follow the parametric trends of their classical, rigid aerofoil counterparts at low reduced frequency, $k$, albeit with a reduced unsteady lift amplitude and an increased phase lag. As the reduced frequency increases and approaches the first resonance of the fluid-loaded membrane, the membrane-equivalent Theodorsen and Sears functions introduce distinct circular paths in the complex plane, which initiate at different values of $k$ for each function. Closed-form expressions for both functions reveal that these circular paths are related to the membrane dynamic response during oscillations, where each circle corresponds to a different dominant mode, and the inflection points that connect the circles represent the shift in dominance between two consecutive membrane modes.

The model results for these harmonic motions or incoming flow disturbances suggest parametric regions where the aeroelastic response of the membrane could enable performance improvements for flapping flight or gust resilience. The unsteady lift amplitude of the membrane in each of these scenarios is higher than that of the rigid flat plate for a range of reduced frequencies in the neighbourhood of the first resonance. Thus, in this frequency regime the standard Theodorsen and Sears functions underestimate the load on the aerofoil.
This parametric region is controlled by the tension coefficient, whereby the aerodynamic load on the aerofoil may be enhanced (or reduced) through either passive or modest active control of the membrane pretension. 
For reduced frequencies in a regime above the first resonance frequency, the present model predicts that membrane oscillations attenuate the unsteady lift response to sinusoidal gusts or harmonic heave motion.
In addition, by increasing the membrane pretension, the beneficial region of reduced frequency shifts to higher frequencies, which could be exploited as a gust mitigation strategy in practice.

Closed-form expressions derived in the Laplace domain for the membrane equivalent Wagner and K\"{u}ssner functions reveal the direct dependence of these functions on the original rigid-plate functions and on the Fourier coefficients that describe the membrane profile. Each indicial lift response of the membrane aerofoil is slower relative to the rigid plate response but results in a significantly higher lift in the steady state due to aeroelastic membrane deformation. The membrane initial response in short times to an abrupt change in angle of attack or to an encounter with a sharp-edged gust produces a negative non-circulatory lift due to the acceleration of the membrane profile from a still and taut position. Therefore, the overall initial lift response of the membrane is smaller than the rigid plate response due to the gradual elastic reaction of the membrane to the changes in the fluid flow. However, at later times the non-circulatory lift due to the abrupt change in the flow field weakens and the circulatory lift increases, such that the membrane lift quickly overcomes the rigid plate lift response and converges to the static solution.

Results from the present theoretical effort invite computational and experimental companion efforts to elucidate the practical range of validity of the model, including the influence of nonlinear flow effects that occur at large gust ratios \citep[e.g.,][]{andreu2020effect,Jones2020}, which are expected to inform future improvements to the predictive aeroelastic framework.

\section*{Acknowledgement}
This work was supported by the Zuckerman-CHE STEM Leadership Program, with partial support from the National Science Foundation under award 1846852.

\section*{Declaration of interests}
The authors report no conflicts of interest.

\appendix

\section{Mathematical identities\label{a4:Math}}
\be
\label{eq:a4_trigo3}
1=\f{2}{\pi}\sum_{n=1}^\infty \f{1-\left(-1\right)^n}{n}\sin n\theta , \qquad \qquad \qquad 0<\theta<\pi .
\ee
\be
\label{eq:a4_trigo4}
\cos\theta=\f{8}{\pi}\sum_{m=1}^\infty \f{m}{4m^2-1}\sin 2m\theta , \qquad \qquad \; 0<\theta<\pi .
\ee
\be
\label{eq:a4_Lam1}
\Lambda_1\left(x,\xi\right)=\ln\left|\f{\sqrt{(1-x)(1+\xi)}+\sqrt{(1+x)(1-\xi)}} {\sqrt{(1-x)(1+\xi)}-\sqrt{(1+x)(1-\xi)}}\right| .
\ee

\section{Asymptotic analysis of the lift due to membrane deformation for low-frequency regime\label{a2_1:approx_k0}}
\begin{figure}
	\begin{center}
	    \begin{subfigure}[]{0.46\textwidth}
			\includegraphics[width=\textwidth]{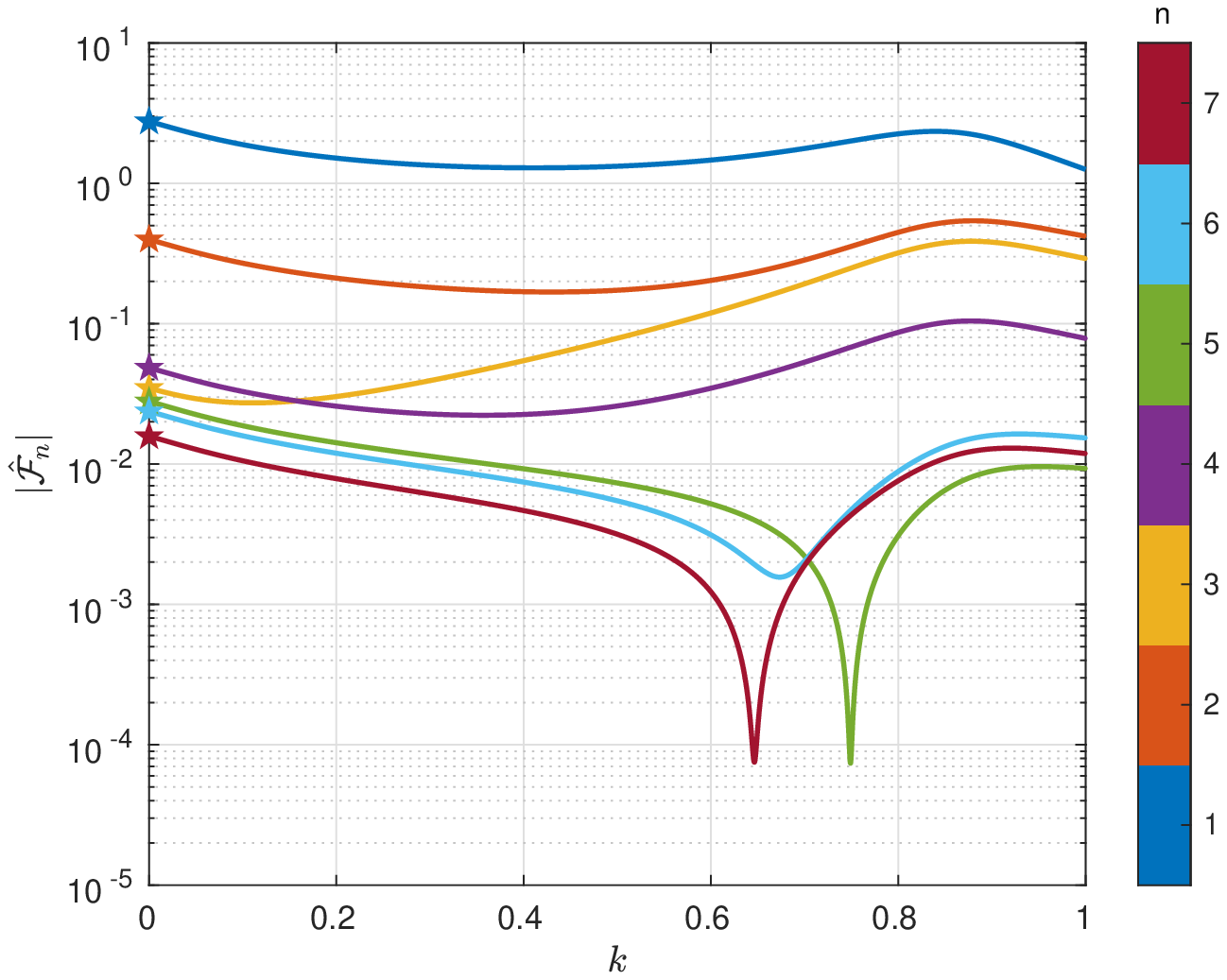}
			\caption{}
			\label{fig:Fn_mem_sears_k0-a}
		\end{subfigure}
		\quad
		\begin{subfigure}[]{0.46\textwidth}
			\includegraphics[width=\textwidth]{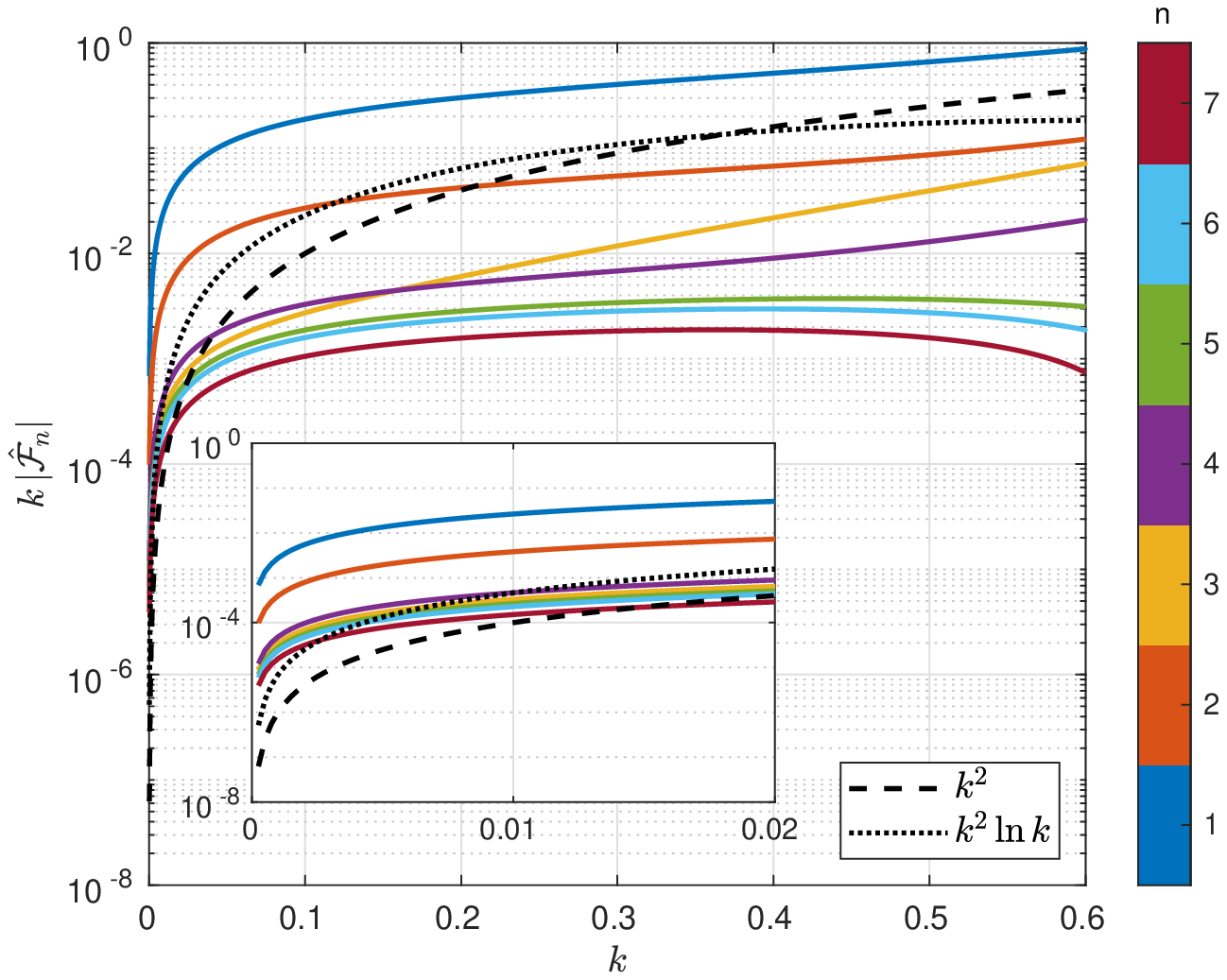}
			\caption{}
			\label{fig:Fn_mem_sears_k0-b}
		\end{subfigure}
		\caption{Fourier coefficients obtained for a nominal membrane wing that encounters a sinusoidal gust of low reduced frequency: (a) modulus, (b) modulus multiplied by the reduced frequency, $k$, and compared against $k^2$ and $k^2\ln{k}$ denoted with a black dashed and dotted lines, respectively. Static solution is denoted in (a) by pentagram markers and is recovered by the unsteady solution for $k\ttz$. Only $7$ of the $24$ coefficients are presented for the sake of clarity, as the magnitude of higher mode coefficients is negligible for low reduced frequencies.}
		\label{fig:Fn_mem_sears_k0}
	\end{center}
\end{figure}

This appendix presents an analysis of the lift due to membrane deformation in the asymptotic regime of low reduced frequencies. 
The analysis begins with an identification of the leading terms in $k$ for the standard Theodorsen function, $C(k)$, and the two auxiliary functions, $\hat{f}(k)$ and $\hat{g}(k)$, as $k\rightarrow0$. These terms are then combined to obtain the leading terms in the lift coefficient, $\hat{C}_{l_d}$, for low reduced frequency.
A series expansion is applied to the standard Theodorsen function, $C(k)$, leading to
\be
\label{eq:a2_1-Theodorsen_limk0}
C(k)\cong 1+k\left[\mi\ln{\f{k}{2}}+\mi \gamma_e-\f{\pi}{2}\right] + \textit{O}\left(k^2\ln{k}\right),
\ee
where $\gamma_e$ is the Euler constant.
For the auxiliary functions, $\hat{f}(k)$ and $\hat{g}(k)$, we derive the leading terms up to $\textit{O}\left(k\hat{\mF}_3 , k\hat{\mF}_4\right)$ and $\textit{O}\left(k^2\right)$, respectively, 
\be
\label{eq:a2_1-fk_lim_k0}
\hat{f}(k) = \f{1}{2}\hat{\mF}_{1}(k)-\f{1}{2}\hat{\mF}_{0}(k)-\f{1}{4}\mi k \hat{\mF}_{0}(k)-\f{1}{4}\mi k \hat{\mF}_{1}(k) +\f{1}{4}\mi k \hat{\mF}_{2}(k) + \textit{O}\left(k\hat{\mF}_3 , k\hat{\mF}_4\right) ,
\ee
\be
\label{eq:a2_1-gk_lim_k0}
\hat{g}(k) = \f{1}{4}\mi k\hat{\mF}_{2}(k)-\f{1}{4}\mi k\hat{\mF}_{0}(k) + \textit{O}\left(k^2\right) .
\ee
We have used the fact that the magnitude of the Fourier coefficients $\hat{\mF_n}(k)$ is strongly reduced for $n>2$ in the low reduced frequency regime, as seen in figure~\ref{fig:Fn_mem_sears_k0-a}, in which the Fourier coefficients obtained for a nominal membrane that encounters a sinusoidal gust of low reduced frequency are presented. For $k\ttz$ the first Fourier coefficient is an order of magnitude larger than the second Fourier coefficient, which is at least one order of magnitude larger than the rest of the coefficients. 
Figure~\ref{fig:Fn_mem_sears_k0-b} illustrates the behaviour of the product $k|\hat{\mF}_n|$ for the first seven Fourier coefficients in the low reduced frequency regime, and compares it to $k^2$ and $k^2\ln{k}$. As $k\ttz$, $k|\hat{\mF}_3|$ and $k|\hat{\mF}_4|$ are comparable and are an order of magnitude larger than $k^2\ln{k}$. However, as $k$ increases this difference diminishes and at $k\cong0.005$ the three functions are comparable.

Further simplification of the above expansion is obtained by recalling that
\be
\label{eq:a2_1-Clsa}
\f{1}{2}\hat{\mF}_{1}(k)-\f{1}{2}\hat{\mF}_{0}(k) \sim  \f{1}{2}\mF_{s1}-\f{1}{2}\mF_{s0} = \f{C_{l_{s\alpha}}}{2\pi}-1  \quad \mbox{as\ }\quad k\ttz ,
\ee
and
\be
\label{eq:a2_1-F0}
\mi k \hat{\mF}_0(k) = \mi k \f{2}{3}\hat{\mF}_2(k) + \textit{O}\left(k\hat{\mF}_4\right) ,
\ee
which when substituted into \eqref{eq:a2_1-fk_lim_k0} and \eqref{eq:a2_1-gk_lim_k0}, respectively, produces
\be
\label{eq:a2_1-fk_lim_k0_1}
f(k) \cong \left( \f{C_{l_{s\alpha}}}{2\pi}-1\right)\left[1-\f{1}{2}\mi k \right] -\f{\mi k}{12}\hat{\mF}_{2}(k) + \textit{O}\left(k\hat{\mF}_3\right) ,
\ee
\be
\label{eq:a2_1-gk_lim_k0_1}
g(k) \cong \f{\mi k}{12}\hat{\mF}_{2}(k) + \textit{O}\left(k\hat{\mF}_4 , k^2 \right) .
\ee
Substitution of \eqref{eq:a2_1-Theodorsen_limk0}, \eqref{eq:a2_1-fk_lim_k0_1}, and \eqref{eq:a2_1-gk_lim_k0_1} into \eqref{eq:Cl_d-Freq} yields
\be
\label{eq:a2_1-Cl_d_k0}
\hat{C}_{l_d}(k) = \alpha_0 \left( C_{l_{s\alpha}}-2\pi\right)\left\{1 + k\left[\mi\ln{\f{k}{2}} -\f{\pi}{2}\right]\right\} + \textit{O}\left( k\hat{\mF}_3 , k\hat{\mF}_4 , k^2\ln{k} , k^2 \right),
\ee
which represents the behaviour of the lift due to membrane deformation in response to a low-frequency flapping motion or gust encounter.
We note that the lift amplitude is controlled by the difference between the static membrane lift-slope and the rigid plate lift slope, $C_{l_{s\alpha}}-2\pi$, as this term describes the static lift due to membrane camber, which is recovered by applying $k=0$ to the unsteady problem.
The tension coefficient on the membrane controls the static membrane lift-slope and as the tension increases the lift slope decreases due to decreased camber (increased stiffness) and approaches $2\pi$. 
In the rigid-membrane limit of $C_T\rightarrow\infty$, the lift due to membrane deformation converges to zero as the reduced frequency approaches zero, as expected.
In addition, it is obvious that the membrane inertia has no notable role in the low-frequency lift response of the membrane.

\bibliographystyle{jfm}
\bibliography{main}

\begin{thebibliography}{48}
\expandafter\ifx\csname natexlab\endcsname\relax\def\natexlab#1{#1}\fi
\def\au#1{#1} \def\ed#1{#1} \def\yr#1{#1}\def\at#1{#1}\def\jt#1{\textit{#1}}
  \def\bt#1{#1}\def\bvol#1{\textbf{#1}} \def\vol#1{#1} \def\pg#1{#1}
  \def\publ#1{#1}\def\arxiv#1{#1}\def\org#1{#1}\def\st#1{\textit{#1}}

\bibitem[Alben(2008)]{Alben2008}
{\sc \au{Alben, S.}} \yr{2008}  \at{Optimal flexibility of a flapping appendage
  in an inviscid fluid}.  \jt{Journal of Fluid Mechanics}  \bvol{614},
  \pg{355--380}.

\bibitem[{Alon Tzezana} \& Breuer(2019)]{Alon2019}
{\sc \au{{Alon Tzezana}, G.} \& \au{Breuer, K.~S.}} \yr{2019}  \at{Thrust, drag
  and wake structure in flapping compliant membrane wings}.  \jt{Journal of
  Fluid Mechanics}  \bvol{862},  \pg{871--888}.

\bibitem[Andreu-Angulo {\em et~al.\/}(2020)Andreu-Angulo, Babinsky, Biler,
  Sedky \& Jones]{andreu2020effect}
{\sc \au{Andreu-Angulo, I.}, \au{Babinsky, H.}, \au{Biler, H.}, \au{Sedky, G.}
  \& \au{Jones, A.~R.}} \yr{2020}  \at{Effect of transverse gust velocity
  profiles}.  \jt{AIAA Journal}  \bvol{58}~(12),  \pg{5123--5133}.

\bibitem[Arbos-Torrent {\em et~al.\/}(2013)Arbos-Torrent, Ganapathisubramani \&
  Palacios]{Arbos-Torrent2013}
{\sc \au{Arbos-Torrent, S.}, \au{Ganapathisubramani, B.} \& \au{Palacios, R.}}
  \yr{2013}  \at{Leading-and trailing-edge effects on the aeromechanics of
  membrane aerofoils}.  \jt{Journal of Fluids and Structures}  \bvol{38},
  \pg{107--126}.

\bibitem[Baddoo {\em et~al.\/}(2021)Baddoo, Hajian \&
  Jaworski]{Baddoo2021unsteady}
{\sc \au{Baddoo, P.~J.}, \au{Hajian, R.} \& \au{Jaworski, J.~W.}} \yr{2021}
  \at{Unsteady aerodynamics of porous aerofoils}.  \jt{Journal of Fluid
  Mechanics}  \bvol{913},  \pg{A16}.

\bibitem[Berci {\em et~al.\/}(2013)Berci, Gaskell, Hewson \&
  Toropov]{Berci2013}
{\sc \au{Berci, M.}, \au{Gaskell, P.~H.}, \au{Hewson, R.~W.} \& \au{Toropov,
  V.~V.}} \yr{2013}  \at{A semi-analytical model for the combined aeroelastic
  behaviour and gust response of a flexible aerofoil}.  \jt{Journal of Fluids
  and Structures}  \bvol{38},  \pg{3--21}.

\bibitem[Bisplinghoff {\em et~al.\/}(1996)Bisplinghoff, Ashley \&
  Halfman]{Bisplinghoff_book1996}
{\sc \au{Bisplinghoff, R.~L.}, \au{Ashley, H.} \& \au{Halfman, R.~L.}}
  \yr{1996} {\em Aeroelasticity\/}.  \publ{Dover}.

\bibitem[Chin \& Lentink(2016)]{Chin2016}
{\sc \au{Chin, D.~D.} \& \au{Lentink, D.}} \yr{2016}  \at{Flapping wing
  aerodynamics: from insects to vertebrates}.  \jt{Journal of Experimental
  Biology}  \bvol{219}~(7),  \pg{920--932}.

\bibitem[Drischler(1956)]{Drischler1956}
{\sc \au{Drischler, J.~A.}} \yr{1956}  \bt{Calculation and compilation of the
  unsteady-lift functions for a rigid wing subjected to sinusoidal gusts and to
  sinusoidal sinking oscillations}. TN 3748.  \org{{National Advisory Committee
  for Aeronautics}}.

\bibitem[Edwards(1979)]{Edwards1979}
{\sc \au{Edwards, J.~W.}} \yr{1979}  \at{Unsteady aerodynamic modeling for
  arbitrary motions}.  \jt{AIAA Journal}  \bvol{17}~(4),  \pg{365--374}.

\bibitem[Elbanhawi {\em et~al.\/}(2017)Elbanhawi, Mohamed, Clothier, Palmer,
  Simic \& Watkins]{elbanhawi2017enabling}
{\sc \au{Elbanhawi, M.}, \au{Mohamed, A.}, \au{Clothier, R.}, \au{Palmer,
  J.~L.}, \au{Simic, M.} \& \au{Watkins, S.}} \yr{2017}  \at{Enabling
  technologies for autonomous {MAV} operations}.  \jt{Progress in Aerospace
  Sciences}  \bvol{91},  \pg{27--52}.

\bibitem[Eldredge \& Jones(2019)]{Eldredge2019ARFM}
{\sc \au{Eldredge, J.~D.} \& \au{Jones, A.~R.}} \yr{2019}  \at{Leading-edge
  vortices: mechanics and modeling}.  \jt{Annual Review of Fluid Mechanics}
  \bvol{51},  \pg{75--104}.

\bibitem[Gopalakrishnan \& Tafti(2010)]{Gopalakrishnan2010}
{\sc \au{Gopalakrishnan, P.} \& \au{Tafti, D.~K.}} \yr{2010}  \at{Effect of
  wing flexibility on lift and thrust production in flapping flight}.  \jt{AIAA
  Journal}  \bvol{48}~(5),  \pg{865--877}.

\bibitem[Gordnier(2009)]{Gordnier2009}
{\sc \au{Gordnier, R.~E.}} \yr{2009}  \at{High-fidelity computational
  simulation of a membrane wing airfoil}.  \jt{Journal of Fluids and
  Structures}  \bvol{25}~(5),  \pg{897--917}.

\bibitem[Hassanalian \& Abdelkefi(2017)]{Hassanalian2017}
{\sc \au{Hassanalian, M.} \& \au{Abdelkefi, A.}} \yr{2017}
  \at{Classifications, applications, and design challenges of drones: A
  review}.  \jt{Progress in Aerospace Sciences}  \bvol{91},  \pg{99--131}.

\bibitem[Hedenstr\"{o}m \& Johansson(2015)]{Hedenstrom2015}
{\sc \au{Hedenstr\"{o}m, A.} \& \au{Johansson, L.~C.}} \yr{2015}  \at{{Bat
  flight: aerodynamics, kinematics and flight morphology}}.  \jt{Journal of
  Experimental Biology}  \bvol{218}~(5),  \pg{653--663}.

\bibitem[Iosilevskii(2007)]{Iosilevskii2007}
{\sc \au{Iosilevskii, G.}} \yr{2007}  \at{Control with trim tabs and
  history-dependent aerodynamic forces}.  \jt{Journal of Fluids and Structures}
   \bvol{23}~(3),  \pg{365--389}.

\bibitem[Jaworski \& Gordnier(2012)]{Jaworski2012}
{\sc \au{Jaworski, J.~W.} \& \au{Gordnier, R.~E.}} \yr{2012}  \at{High-order
  simulations of low {Reynolds} number membrane airfoils under prescribed
  motion}.  \jt{Journal of Fluids and Structures}  \bvol{31},  \pg{49 -- 66}.

\bibitem[Jaworski \& Gordnier(2015)]{Jaworski2015}
{\sc \au{Jaworski, J.~W.} \& \au{Gordnier, R.~E.}} \yr{2015}  \at{Thrust
  augmentation of flapping airfoils in low {Reynolds} number flow using a
  flexible membrane}.  \jt{Journal of Fluids and Structures}  \bvol{52},
  \pg{199--209}.

\bibitem[Jones(2020)]{Jones2020}
{\sc \au{Jones, A.~R.}} \yr{2020}  \at{Gust encounters of rigid wings: Taming
  the parameter space}.  \jt{Physical Review Fluids}  \bvol{5},  \pg{110513}.

\bibitem[Jones {\em et~al.\/}(2022)Jones, Cetiner \& Smith]{Jones2022}
{\sc \au{Jones, A.~R.}, \au{Cetiner, O.} \& \au{Smith, M.~J.}} \yr{2022}
  \at{Physics and modeling of large flow disturbances: Discrete gust encounters
  for modern air vehicles}.  \jt{Annual Review of Fluid Mechanics}  \bvol{54},
  \pg{469--493}.

\bibitem[von K\'{a}rm\'{a}n \& Sears(1938)]{vonKarman1938}
{\sc \au{von K\'{a}rm\'{a}n, T.} \& \au{Sears, W.~R.}} \yr{1938}  \at{Airfoil
  theory for non-uniform motion}.  \jt{Journal of the Aeronautical Sciences}
  \bvol{5}~(10),  \pg{379--390}.

\bibitem[Katz \& Plotkin(2001)]{KatzPlotkin2001}
{\sc \au{Katz, J.} \& \au{Plotkin, A.}} \yr{2001} {\em Low-Speed
  Aerodynamics\/}, 2nd edn.  \publ{Cambridge University Press}.

\bibitem[Kornecki {\em et~al.\/}(1976)Kornecki, Dowell \&
  {O'Brien}]{Kornecki1976}
{\sc \au{Kornecki, A.}, \au{Dowell, E.H.} \& \au{{O'Brien}, J.}} \yr{1976}
  \at{On the aeroelastic instability of two-dimensional panels in uniform
  incompressible flow}.  \jt{Journal of Sound and Vibration}  \bvol{47}~(2),
  \pg{163--178}.

\bibitem[Mavroyiakoumou \& Alben(2020)]{Mavroyiakoumou2020}
{\sc \au{Mavroyiakoumou, C.} \& \au{Alben, S.}} \yr{2020}  \at{Large-amplitude
  membrane flutter in inviscid flow}.  \jt{Journal of Fluid Mechanics}
  \bvol{891},  \pg{A23}.

\bibitem[Mavroyiakoumou \& Alben(2021)]{Mavroyiakoumou2021}
{\sc \au{Mavroyiakoumou, C.} \& \au{Alben, S.}} \yr{2021}  \at{Eigenmode
  analysis of membrane stability in inviscid flow}.  \jt{Physical Review
  Fluids}  \bvol{6},  \pg{043901}.

\bibitem[Minami(1998)]{Minami1998}
{\sc \au{Minami, H.}} \yr{1998}  \at{Added mass of a membrane vibrating at
  finite amplitude}.  \jt{Journal of Fluids and Structures}  \bvol{12}~(7),
  \pg{919--932}.

\bibitem[Mohamed {\em et~al.\/}(2014)Mohamed, Massey, Watkins \&
  Clothier]{Mohamed2014}
{\sc \au{Mohamed, A.}, \au{Massey, K.}, \au{Watkins, S.} \& \au{Clothier, R.}}
  \yr{2014}  \at{The attitude control of fixed-wing mavs in turbulent
  environments}.  \jt{Progress in Aerospace Sciences}  \bvol{66},  \pg{37--48}.

\bibitem[Muijres {\em et~al.\/}(2008)Muijres, Johansson, Barfield, Wolf,
  Spedding \& Hedenstr{\"o}m]{Muijres2008}
{\sc \au{Muijres, F.~T.}, \au{Johansson, L.~C.}, \au{Barfield, R.}, \au{Wolf,
  M.}, \au{Spedding, G.~R.} \& \au{Hedenstr{\"o}m, A.}} \yr{2008}
  \at{Leading-edge vortex improves lift in slow-flying bats}.  \jt{Science}
  \bvol{319}~(5867),  \pg{1250--1253}.

\bibitem[Nielsen(1963)]{Nielsen1963}
{\sc \au{Nielsen, J.~N.}} \yr{1963}  \at{Theory of flexible aerodynamic
  surfaces}.  \jt{Journal of Applied Mechanics}  \bvol{30},  \pg{435--442}.

\bibitem[Rao(2007)]{Rao_book2007}
{\sc \au{Rao, S.~S.}} \yr{2007} {\em Vibration of continuous systems\/}.
  \publ{John Wiley \& Sons, Ltd}.

\bibitem[Rojratsirikul {\em et~al.\/}(2009)Rojratsirikul, Wang \&
  Gursul]{Rojratsirikul2009}
{\sc \au{Rojratsirikul, P.}, \au{Wang, Z.} \& \au{Gursul, I.}} \yr{2009}
  \at{Unsteady fluid-structure interactions of membrane airfoils at low
  {Reynolds} numbers}.  \jt{Experiments in Fluids}  \bvol{46},  \pg{859--872}.

\bibitem[Rojratsirikul {\em et~al.\/}(2010)Rojratsirikul, Wang \&
  Gursul]{Rojratsirikul2010}
{\sc \au{Rojratsirikul, P.}, \au{Wang, Z.} \& \au{Gursul, I.}} \yr{2010}
  \at{Effect of pre-strain and excess length on unsteady fluid-structure
  interactions of membrane airfoils}.  \jt{Journal of Fluids and Structures}
  \bvol{26},  \pg{359--376}.

\bibitem[Schwarz(1940)]{Schwarz1940}
{\sc \au{Schwarz, L.}} \yr{1940}  \at{{Berechnung der Druckverteilung einer
  harmonisch sich verformenden Tragfl{\"a}che in ebener Str{\"o}mung}}.
  \jt{Luftfahrtforschung}  \bvol{17},  \pg{379--386}.

\bibitem[Sears(1940)]{Sears1940}
{\sc \au{Sears, W.~R.}} \yr{1940}  \at{Operational methods in the theory of
  airfoils in non-uniform motion}.  \jt{Journal of the Franklin Institute}
  \bvol{230}~(1),  \pg{95--111}.

\bibitem[Serrano-Galiano {\em et~al.\/}(2018)Serrano-Galiano, Sandham \&
  Sandberg]{Serrano-Galiano2018}
{\sc \au{Serrano-Galiano, S.}, \au{Sandham, N.~D.} \& \au{Sandberg, R.~D.}}
  \yr{2018}  \at{Fluid{\textendash}structure coupling mechanism and its
  aerodynamic effect on membrane aerofoils}.  \jt{Journal of Fluid Mechanics}
  \bvol{848},  \pg{1127--1156}.

\bibitem[Shyy {\em et~al.\/}(2013)Shyy, Aono, Kang \& Liu]{Shyy2013book}
{\sc \au{Shyy, W.}, \au{Aono, H.}, \au{Kang, C.} \& \au{Liu, H.}} \yr{2013}
  {\em An introduction to flapping wing aerodynamics\/}.  \publ{Cambridge
  University Press}.

\bibitem[Shyy {\em et~al.\/}(2016)Shyy, Kang, Chirarattananon, Ravi \&
  Liu]{Shyy2016}
{\sc \au{Shyy, W.}, \au{Kang, C.}, \au{Chirarattananon, P.}, \au{Ravi, S.} \&
  \au{Liu, H.}} \yr{2016}  \at{Aerodynamics, sensing and control of
  insect-scale flapping-wing flight}.  \jt{Proceedings of the Royal Society A}
  \bvol{472}~(2186),  \pg{20150712}.

\bibitem[S\"{o}hngen(1939)]{Sohngen1939}
{\sc \au{S\"{o}hngen, H.}} \yr{1939}  \at{{Die L\"{o}sungen der
  Integralgleichung und deren Anwendung in der Tragfl\"{u}geltheorie}}.
  \jt{Mathematische Zeitschrift}  \bvol{45},  \pg{245--264}.

\bibitem[Song {\em et~al.\/}(2008)Song, Tian, Israeli, Galvao, Bishop, Swartz
  \& Breuer]{Song2008}
{\sc \au{Song, A.}, \au{Tian, X.}, \au{Israeli, E.}, \au{Galvao, R.},
  \au{Bishop, K.}, \au{Swartz, S.} \& \au{Breuer, K.}} \yr{2008}
  \at{Aeromechanics of membrane wings with implications for animal flight}.
  \jt{AIAA Journal}  \bvol{46}~(8),  \pg{2096--2106}.

\bibitem[Sygulski(2007)]{Sygulski2007}
{\sc \au{Sygulski, R.}} \yr{2007}  \at{Stability of membrane in low subsonic
  flow}.  \jt{International Journal of Non-Linear Mechanics}  \bvol{42}~(1),
  \pg{196--202}.

\bibitem[Tiomkin \& Raveh(2017)]{Tiomkin2017}
{\sc \au{Tiomkin, S.} \& \au{Raveh, D.~E.}} \yr{2017}  \at{On the stability of
  two-dimensional membrane wings}.  \jt{Journal of Fluids and Structures}
  \bvol{71},  \pg{143--163}.

\bibitem[Tiomkin \& Raveh(2021)]{Tiomkin2021}
{\sc \au{Tiomkin, S.} \& \au{Raveh, D.~E.}} \yr{2021}  \at{A review of
  membrane-wing aeroelasticity}.  \jt{Progress in Aerospace Sciences}
  \bvol{126},  \pg{100738}.

\bibitem[Tregidgo {\em et~al.\/}(2013)Tregidgo, Wang \& Gursul]{Tregidgo2013}
{\sc \au{Tregidgo, L.}, \au{Wang, Z.} \& \au{Gursul, I.}} \yr{2013}
  \at{Unsteady fluid--structure interactions of a pitching membrane wing}.
  \jt{Aerospace Science and Technology}  \bvol{28}~(1),  \pg{79--90}.

\bibitem[Valsa \& Bran\u{c}ik(1998)]{Valsa1998}
{\sc \au{Valsa, J.} \& \au{Bran\u{c}ik, L.}} \yr{1998}  \at{Approximate
  formulae for numerical inversion of {Laplace} transforms}.  \jt{International
  Journal of Numerical Modelling}  \bvol{11}~(3),  \pg{153--166}.

\bibitem[Wagner(1925)]{wagner1925}
{\sc \au{Wagner, H.}} \yr{1925}  \at{{\"{U}ber die Entstehung des dynamishen
  Auftriebes von Tragfl\"{u}geln}}.  \jt{Zeitschrift f\"{u}r Angewandte
  Mathematik und Mechanik}  \bvol{5},  \pg{17--35}.

\bibitem[Watkins {\em et~al.\/}(2006)Watkins, Milbank, Loxton \&
  Melbourne]{Watkins2006atmospheric}
{\sc \au{Watkins, S.}, \au{Milbank, J.}, \au{Loxton, B.~J.} \& \au{Melbourne,
  W.~H.}} \yr{2006}  \at{Atmospheric winds and their implications for microair
  vehicles}.  \jt{AIAA Journal}  \bvol{44}~(11),  \pg{2591--2600}.

\bibitem[Yadykin {\em et~al.\/}(2003)Yadykin, Tenetov \& Levin]{Yadkyn2003}
{\sc \au{Yadykin, Y.}, \au{Tenetov, V.} \& \au{Levin, D.}} \yr{2003}  \at{The
  added mass of a flexible plate oscillating in a fluid}.  \jt{Journal of
  Fluids and Structures}  \bvol{17}~(1),  \pg{115--123}.

\end{thebibliography}

\end{document}